\documentclass[twocolumn]{aastex631}
\usepackage{lineno}

\usepackage{booktabs} 

\usepackage{natbib}
\usepackage{hyperref} 
\usepackage{soul} 

\usepackage{array}
\usepackage{booktabs}  
\usepackage{graphicx}  
\newcolumntype{C}[1]{>{\centering\arraybackslash}p{#1}}

\newcommand{\UCB}{University of California, Berkeley, 501 Campbell Hall 3411, Berkeley, CA 94720, USA}
\newcommand{\seti}{SETI Institute, 339 Bernardo Ave, Suite 200, Mountain View, CA 94043, USA}
\newcommand{\oxon}{Breakthrough Listen, University of Oxford, Department of Physics, Denys Wilkinson Building, Keble Road, Oxford, OX1 3RH, UK}
\newcommand{\oxford}{Astrophysics, Department of Physics, University of Oxford, Denys Wilkinson Building, Keble Road, Oxford OX1 3RH, UK}
\newcommand{\KZA}{University of Malta, Institute of Space Sciences and Astronomy}
\newcommand{\ska}{SKA Observatory, 26 Dick Perry Avenue, Kensington, WA 6151, Australia}

\newcommand{\turboseti}{\texttt{turboSETI}}

\begin{document}

\title{A Novel Technosignature Search in the Breakthrough Listen Green Bank Telescope Archive}

\author[0009-0003-6274-657X]{Caleb Painter}
\affiliation{Department of Astronomy, Harvard University $|$ Cambridge, MA 02138}

\author[0000-0003-4823-129X]{Steve Croft}
\affiliation{\oxon}
\affiliation{\UCB}
\affiliation{\seti}

\author[0000-0002-7042-7566]{Matthew Lebofsky}
\affiliation{\UCB}

\author[0009-0008-0662-1293]{Alex Andersson}
\affiliation{\oxford}

\author[0009-0008-0662-1293]{Carmen Choza}
\affiliation{\oxon}
\affiliation{\seti}

\author[0009-0008-0662-1293]{Vishal Gajjar}
\affiliation{\UCB}
\affiliation{\seti}

\author[0009-0008-0662-1293]{Danny Price}
\affiliation{\oxon}
\affiliation{\ska}

\author[0000-0003-2828-7720]{Andrew P.\ V.\ Siemion}
\affiliation{\oxon}
\affiliation{\seti}
\affiliation{\KZA}
\affiliation{\UCB}

\begin{abstract}

The Breakthrough Listen program is, to date, the most extensive search for technological life beyond Earth. Over the past nine years it has surveyed thousands of nearby stars and close to 100 nearby galaxies with telescopes around the world, including the Robert C. Byrd Green Bank Telescope (GBT) in West Virginia. The goal is to find evidence of technosignatures of other civilizations, such as narrowband Doppler drifting radio signals. Despite the GBT's location in a radio-quiet zone, the primary challenge of this search continues to be the ability to pick out genuine candidates from the high quantities of human-generated radio-frequency interference (RFI). Here we present a novel search method aimed at finding these `needle-in-a-haystack' type signals, applied to 6,630 observation cadences of 2,623 stars (each observed with one or more of the L, S, C, and X band receivers) from the GBT archive. We implement a low-complexity statistical process to vet out RFI and highlight signals that, upon visual inspection, are less evidently RFI than those from previous analyses. Our work returns candidate signals found previously using both traditional and machine learning algorithms, as well as many not previously identified. This analysis represents the largest dataset searched for technosignatures to date, and highlights the efficacy that traditional algorithms continue to have in these types of technosignature searches. We find that less than 1\%\ of stars host transmitters brighter than $\sim 0.3$ Arecibo radar equivalents broadcasting in our direction over the frequency band covered. 
\end{abstract}




\section{Introduction} \label{sec:intro}

The quest to determine whether or not we are alone in the universe continues to drive the Search for Extraterrestrial Intelligence (SETI), aimed at detecting signs of technologically advanced life beyond Earth. These searches for `technosignatures' are often carried out in the radio domain due to the low extinction of radio waves as they travel through interstellar space, as well as the assumption that the prevalence of radio technology on Earth might also hold for other civilizations. \citet{Cocconi:1959} laid the theoretical groundwork for modern SETI searches in the late 1950s, which was followed by the first such observational program, Project Ozma \citep{project_ozma}. Early searches were restricted to narrow bandwidths and a small range of stars, whereas subsequent searches have sampled increasingly wider bandwidths and a higher number of targets, making them a powerful avenue for SETI research. Successful detection of unambiguous, artificial radio signals would provide the most definitive answer to date on whether there is other life out in the cosmos.   

The largest program currently driving the field of SETI research is the Breakthrough Listen (BL) Initiative \citep{Worden:2017}, which has as part of its core mission a survey of 1 million nearby stars. Over the past nine years, BL has been using the 100-m Robert C. Byrd Green Bank Telescope (GBT) in West Virginia to survey thousands of stars (\citealt{macmahon:18}, \citealt{Price:2020}), close to a hundred galaxies \citep{choza:2024}, and a variety of solar system objects \citep{oumuamua} at frequencies of 1--12\,GHz. BL also observes at other facilities around the globe, including the Murriyang telescope at Parkes in Australia \citep{price:2018}. BL's projects on GBT and Parkes have resulted in the aggregation of over 20\,PB of radio data.

Technosignature searches of the kind BL carries out face many challenges, the most fundamental one being that the signals being searched for have never been found, and may be intrinsically rare. It is difficult to say what a signal would be like, where it would be most likely to come from, or at what frequencies it would be emitted. As such, attempts have been made to cover as wide a frequency range as possible, including in optical lightcurves \citep{cabrales2024searching} and spectra \citep{2023AJ....165..114Z}, and employing a variety of techniques in radio searches \citep[e.g.,][]{Ma_2023, gajjar2022searching, tusay2022search}.

Several technical challenges remain after carrying out radio SETI observations, perhaps the largest of which is Radio Frequency Interference (RFI) created by human technology, which poses a significant problem even in radio-quiet zones. This emission contaminates observations and confuses search algorithms trying to pick up extraterrestrial signals. To help distinguish candidate signals from clear RFI, BL applies a spatial filtering sequence on the GBT, using a ``cadence" of observations that switch between the primary target and secondary targets. This ON target / OFF target pattern is repeated three times, to produce a total of six 5-minute observations, three of which are pointed at the source of interest. Human-generated signals entering the telescope through the antenna side lobes would most likely appear in both the ON and OFF observations, while genuine ETI signals coming from the primary target would appear in only the ON observations. The other primary criterion for a quality candidate signal is that it should be Doppler drifting, due to the motion between the source and the receiver. Together, these two requirements make up the base of many of the algorithms used by BL, including \turboseti\ \citep{turboSETI} and \texttt{seticore}\footnote{\url{https://github.com/lacker/seticore}} which use a “tree de-Doppler” approach to find these narrowband drifting signals that show up in only the ON observations, referred to as ``events'' \citep{Enriquez}.

While \turboseti\ has helped provide some of the best constraints on extraterrestrial life to date, it has some limitations. Some signals (including some narrowband Doppler drifting signals, as well as the broader class of signals with other morphologies) are missed by \turboseti\ \citep{Ma_2023}. Additionally, in each GBT cadence \turboseti\ typically finds thousands of hits (signals detected by \turboseti's narrowband Doppler drifting search), and several events \citep{choza:2024}, requiring tedious visual inspection.  In response, other authors have come up with their own algorithms to find higher-quality signals, including both traditional \citep{margot2023search} as well as machine learning approaches \citep{Ma_2023,brzycki:2020,pinchuk:2022,gajjar:2022,jb:2025}. Some examples of machine learning approaches include the use of convolutional neural networks and random forest decision trees to recognize and classify specific kinds of data, as well as generative adversarial networks to pick out anomalies. These have proven successful in handling the large datasets involved in technosignature searches, but come with their own respective problems. Specifically, these are the high computational costs required to train the neural networks, the dependence on training data, and the lack of interpretability in the decision-making process. Deep learning models also only make predictions based on the data they have been trained on, and cannot extrapolate to novel signal types that may not have been encompassed in the training set. This is one advantage of an unsupervised learning approach (see \citealt{jacobson2025anomaly} for example).   

Here we present a novel algorithm that aims to find a wider range of possible signals without making assumptions about signal morphology, while still being sensitive to those detected through both machine learning and traditional pipelines.  We search through a dataset that encompasses the GBT L, C, S, and X bands, as described in Section~\ref{sec:dataset}. We describe the algorithm in Section~\ref{sec:methods}, present the results of our search in Section~\ref{sec:results}, and discuss these results in section \ref{sec:discussion}.
 
\begin{table}[t]
\centering
\caption{Survey Parameters\label{tab:dataset}}
\begin{tabular}{cccccc}
\hline
\hline
Receiver & Frequency & Cadences & Targets$^a$ & Stars & Time \\
& [GHz] & & & & [Hr]\\
\hline
L & 1.10--1.90 & 2131 & 1627 & 1522 & 1066 \\
S & 1.80--2.80 & 1524 & 1225 & 1092 & 762  \\
C & 4.00--7.80 & 1122 & 928  & 845 & 561\\
X & 7.80--11.20 & 1553 & 1383 & 1275 & 776 \\
\hline
Total & 1.10--11.20 & 6330 & 2841 & 2623 & 3165\\
\hline
\end{tabular}
\begin{flushleft} 
$^a$: Includes both stellar and non-stellar targets (galaxies, dwarf galaxies, globular clusters)\\
\end{flushleft}
\end{table}

\begin{figure*}[t]
    \centering
    \includegraphics[width=0.8\textwidth]{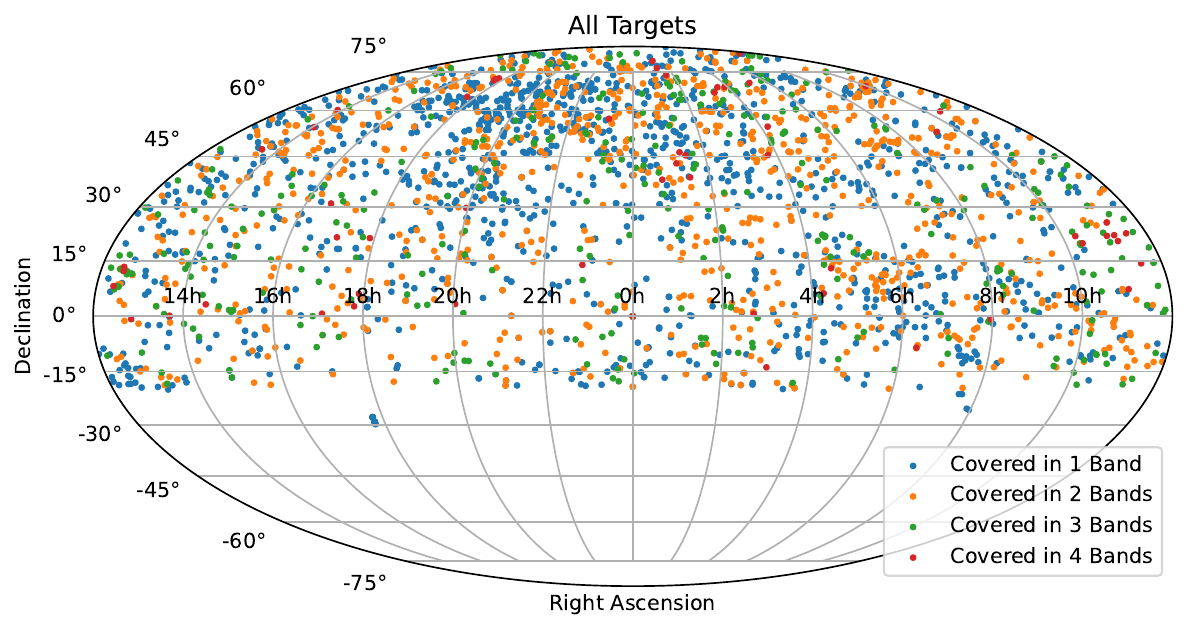}
    \caption{Distribution of sources in our sample in equatorial coordinates. Sources with data available for all four GBT receivers used are plotted in blue, sources observed in three bands are in yellow, sources observed in two bands are in green, and sources observed in only one band are in red.}
    \label{fig:source_positions}
\end{figure*}

\section{Dataset} \label{sec:dataset}

\begin{figure}[t]
    \centering
    \includegraphics[width=0.45\textwidth]{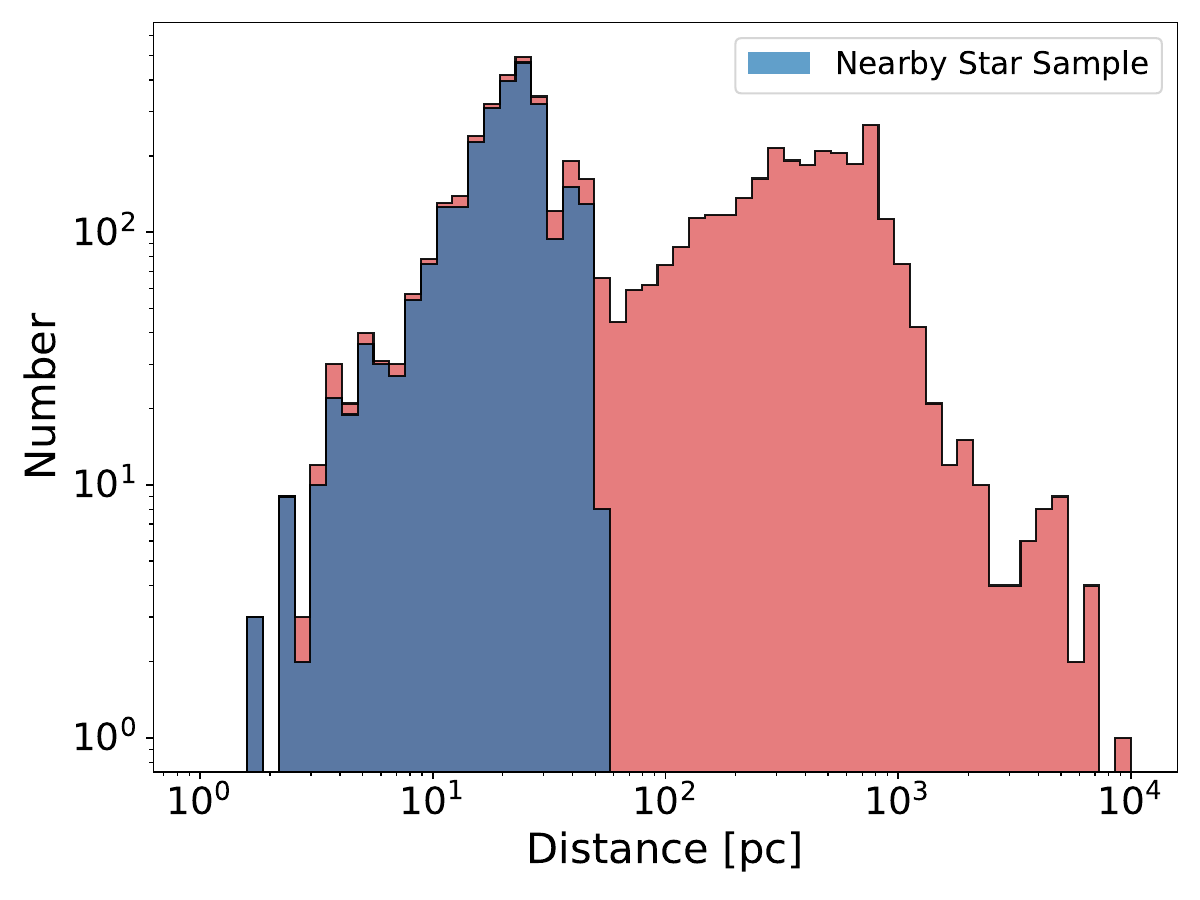}
    \caption{Histograms of the distances for the stars in our sample. Sources in blue, are part of the Breakthrough Listen nearby star sample compiled by  \citet{isaacson:2017}, while sources in red are all stars observed. Distances to galaxies and globular clusters in our sample are not shown.}
    \label{fig:source_distances}
\end{figure}

We deployed our search algorithm on a large fraction of the entire GBT dataset in the Breakthrough Listen data center at the University of California, Berkeley as of 2023 July; we did not analyze additional GBT data stored onsite at the Green Bank Observatory. This comprises 3,165 hours of of on-sky data across 6,630 cadences and 2,841 unique pointings (of which 2,623 are stars, and the rest are galaxies, globular clusters, and solar system objects)\footnote{The sample size has been corrected from the original version of this paper, and the affected figures, tables, and transmitter rates have been updated.}. The position of each source is shown in Figure \ref{fig:source_positions}, along with a histogram of all their distances in Figure \ref{fig:source_distances}.  This makes this study one of the largest SETI searches to date in terms of number of targets and observations. The observations come from hundreds of different observing sessions over the past eight years. The dataset used covers all four of the main bands used to date in BL's GBT searches, namely the 1.1--1.9\,GHz L band, the 1.8--2.7\,GHz S band, the 4.0--7.8\,GHz C band, and the 7.8--11.2\,GHz X band. The full number of cadences and unique targets in each band can be found in Table~\ref{tab:dataset}. Each cadence in this dataset is composed of six 5-minute observations, three on-source and three off-source. These data are recorded in HDF5 format, with a time resolution of $\sim 18$\,s and a fine frequency resolution of 2.79\,Hz for 96.4\% of the data and 2.84\,Hz for the other 3.6\%\ \citep{lebofsky:19}. 

Each BL compute node captures a portion of the total bandwidth during observing with a maximum of 187.5\,MHz. Of the 6,630 cadences in our sample, 67.8\% have been stored as individual 187.5\,MHz data products. These have been grouped with the other five 187.5\,MHz data products spanning the same frequency range, but corresponding to the other ON and OFF scans in the cadence. These groups of six 187.5\,MHz data products form what we label a `sub-cadence', and we split these sub-cadences into subsamples that we process in parallel. The other 32.2\% of our observations have been spliced together in frequency into one large data product spanning the full band (as much as 4\,GHz). Due to the differences in size, we combine all of these spliced data products into their own sub-sample. This leaves us with 62,308 sub-cadences, which we group into 63 sub-samples each containing 1,000 sub-cadences (except for the last subsample which contains slightly fewer).

Some frequency ranges are affected by strong RFI and are blocked by notch filters, superconducting filters permanently installed immediately after the first amplifiers in the GBT. These block the 1200–-1340\,GHz range in L band, to suppress interference from a nearby Air Surveillance Radar, and the 2300–-2360\,MHz range in S band, to reduce interference from Sirius and XM satellite radio transmissions.

\begin{figure*}
    \includegraphics[width=1\textwidth]{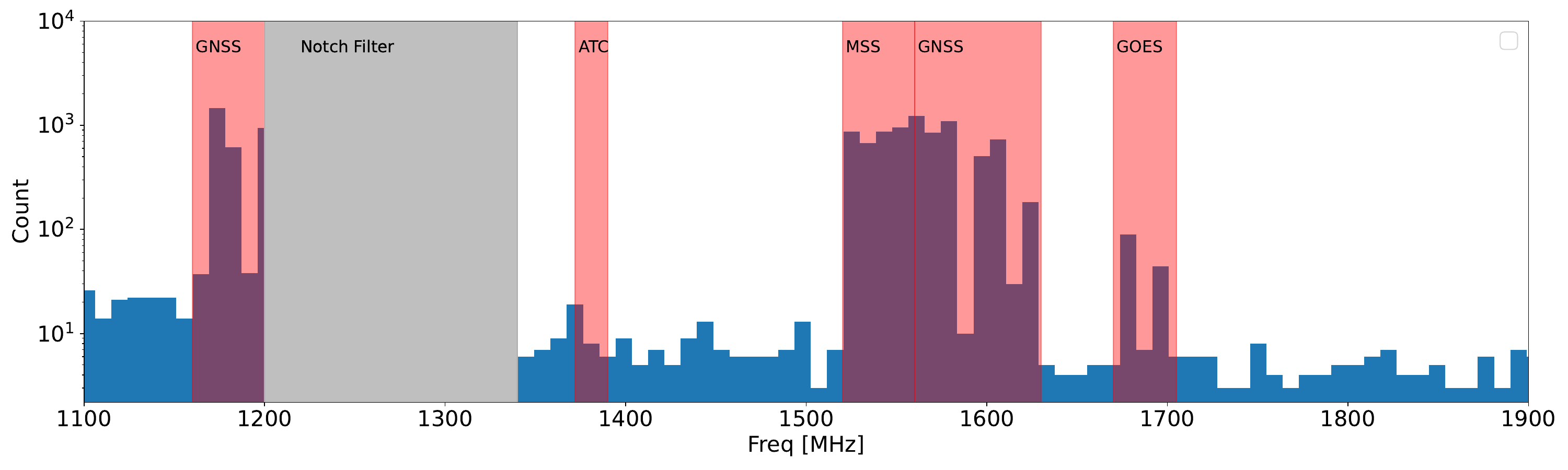}
    \caption{The distribution (blue histogram) of hotspot regions (2.86\,kHz frequency blocks from individual observations) across L band, with kurtoses above our threshold but before any further filtering is performed. Red rectangles highlight heavy RFI regions, and contain 97\% of the frequency blocks. }
    \label{fig:red_cutouts}
\end{figure*}

While all cadences used alternate between ON-source and OFF-source observations, two different patterns are used. Earlier cadences (prior to 2016 April 26) follow an ABABAB source pattern, with the same offset position (B) being used on all three OFF scans between the primary source (A) scans. All other cadences use an ABACAD pattern, with a different source (B,C,D) used in each OFF scan. For the ABABAB observations, the angular offset between ON and OFF is 2\degr\ in declination (for comparison, the GBT FWHM is $11\farcm 5$ at 1.1\,GHz). For the ABACAD observations, the secondary targets are at least five beam-widths away from the primary target. For the purposes of our search we treat both type of cadence scheme equally.

Some of the data analyzed were included in previously published work (see Section \ref{sec:comparisons}), so we are also able to benchmark the results from this search against earlier ones. This gives some reference for our algorithm's performance against both other classical algorithms (such as the standard de-Doppler technique used by \turboseti), as well as machine learning techniques.

\section{Methodology and Algorithm} \label{sec:methods}

\subsection{Pre-Processing}

Our pipeline runs on the sub-cadences described in Section \ref{sec:dataset}, using two of the compute nodes at the Breakthrough Listen data center (Table \ref{tab:computing_specs}).  

Prior to passing in the data for analysis, we cut out parts of the frequency band that were particularly heavily affected by RFI. These frequency regions are based on work by \citet{choza:2024}, and are listed in Table \ref{tab:rfi}. These regions comprised a total of 1768\,MHz, or 20\% of our total coverage, and are shown in red in Figure \ref{fig:red_cutouts}. While this is a significant percentage of the total frequency space, removing these regions greatly reduces the computational time needed to process each cadence, which scales linearly with the number of strong signals present in the data. 

\begin{table}[t]
\centering
\caption{Computing Specifications \label{tab:computing_specs}}
\begin{tabular}{ccc}

\hline 
\hline
& Compute Node 1 & Compute Node 2 \\
\hline 
CPU & Intel Xeon E5-2630 & Intel Xeon Silver 4210 \\
GPU-0 & NVIDIA TITAN X & NVIDIA TITAN X \\
GPU-1 & NVIDIA TITAN XP & NVIDIA GTX 1080 \\
GPU-2 & NVIDIA GTX 1080 & NVIDIA GTX 1080 Ti \\
GPU-3 & NVIDIA TITAN XP & NVIDIA GTX 1080 Ti \\
RAM & 256 GB & 196 GB \\
\hline
\end{tabular}
\end{table}

\subsection{Algorithm}\label{sec:algorithm}

The novel algorithm we present in this study is dubbed \texttt{pickles}\footnote{\url{https://github.com/UCBerkeleySETI/pickles}}, Pipeline for Identification of Candidates using Kurtosis to Locate Extraterrestrial Signals. The core idea is that a narrow subset of the frequency space without any signal present will be almost entirely noise, while such a frequency subset with a signal present will have a power spectrum that deviates significantly from the background noise distribution. We label these subsets of the frequency space as `frequency blocks' or `blocks' for short, and they correspond to a sliver of a cadence's frequency coverage. They consist of a set number of frequency channels from all six observations in a cadence, all centered on the same frequency. 

If this frequency block does not contain any signal, its spectra will be primarily noise and follow a Chi-squared distribution \citep{Brzycki_2022}. Following the Central Limit Theorem, this noise will tend towards a Gaussian distribution, which has a Fisher kurtosis of zero. The Fisher kurtosis can be calculated as

\begin{equation}
{\rm Kurt}_{\rm Fisher}([x_1, ...., x_N]) = \frac{\sum_{i=1}^N\left(x_i-\bar{x}\right)^4 / N}{\sigma^4}-3, 
\end{equation}

where $\sigma$ is the standard deviation of the spectra in the frequency block, $x_i$ are the individual intensity values, $\bar{x}$ the mean intensity, and N the number of data points. On the other hand, a frequency block with a signal in it could have a much higher kurtosis (we will define what `high' could mean in Section \ref{sec:filtering}). We note that this use of kurtosis is different than previous uses of kurtosis for RFI flagging \citep{nita2007PASP..119..805N}, where a `spectral' kurtosis is instead calculated on the time series of each channel, yielding multiple kurtosis values. Here, we instead compute a `block' kurtosis, where we flatten our entire frequency block across multiple time channels into a 1D dataset. We refer to this `block' kurtosis as kurtosis for simplicity. 

A strong candidate ET signal would manifest only in the ON scans and not in the OFF scans of a cadence. The frequency block this signal falls in would thus have a high kurtosis in all of its ON scans and a very low kurtosis in all of its OFF scans. By dividing up the cadence's full frequency range into narrow frequency blocks, we can thus test each block against this condition. While the requirement for a low kurtosis in the OFF observations means that any genuine ET signal in the same frequency block as RFI will not be found, this is a trade-off we accept in exchange for keeping our algorithm low in complexity and reducing the number of false positives.

The size of the frequency block that we split each cadence into is an important choice. A smaller block will produce a higher kurtosis compared to the same signal in a larger block, which enables the detection of weaker signals. It also reduces the probability that RFI falls into the same block as a genuine signal. However, if the block is too small then a high-drift signal might only appear in one of the ON scans. We opt for a frequency block of 1024 fine channels, equivalent to $2.86$\,kHz, dividing up each cadence into $n = N_{\rm chan}/1024$ separate blocks where $N_{\rm chan}$ is the total number of fine channels in the file. We then create an additional $n-1$ frequency blocks of the same size, each centered on the divide between the original regions. This gives a total of $2n-1$ overlapping frequency blocks. We then remove all frequency blocks falling into the high RFI regions and notch-filter regions defined above. 

Loading arrays of length power of 2 into memory uses this available memory better as the hardware and the operating system manage memory in units of power of two. As a result we initially considered four potential block sizes, of 512, 1024, 2048, and 4096 channels. We opt to divide our observations into frequency blocks of 1024 fine channels for two main reasons:

\begin{enumerate}
\item The quantity of noise in the frequency block we choose must not overpower potential signals. Figure \ref{fig:kurt_snr} shows the kurtosis of an injected signal as a function of its signal-to-noise ratio for varying block sizes. For a signal with SNR=10, only blocks of size 512 and 1024 frequency channels have a kurtosis greater than 1. The kurtosis of blocks wider in frequency are dominated by the noise in the rest of the observation, which drowns out the signal.

\begin{figure}
    \centering
    \includegraphics[width=0.45\textwidth]{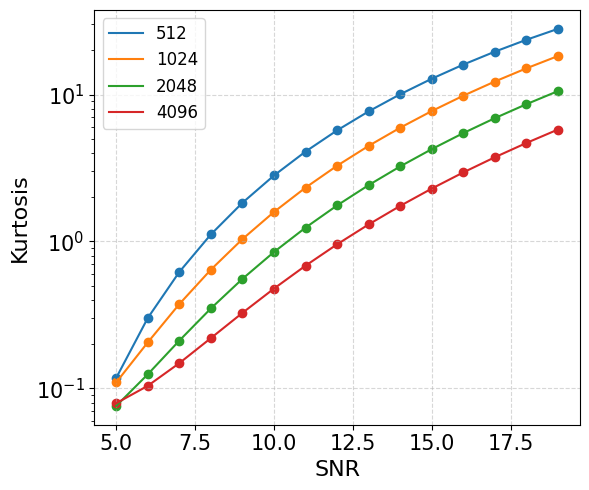}
    \caption{Kurtosis as a function of signal strength for a narrowband signal present in each time bin of an observation. Only frequency blocks of size 512 and 1024 have an expected kurtosis $>1$ for a signal with SNR=10. This is true for L, S, C, and X bands.}
    \label{fig:kurt_snr}
\end{figure}

\begin{figure}
    \includegraphics[width=0.45\textwidth]{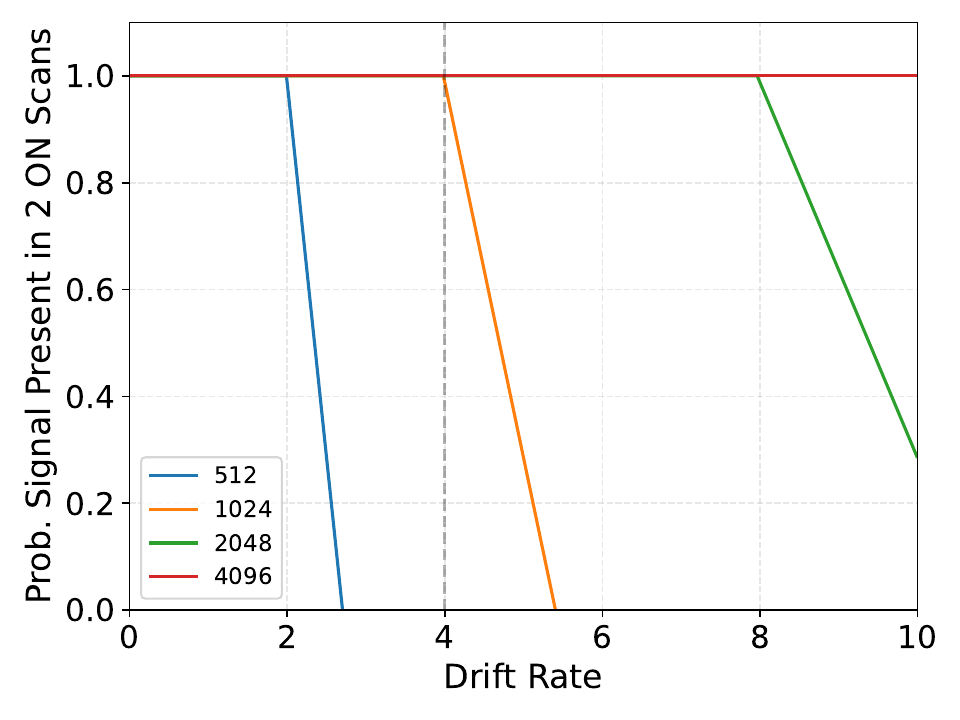}
    \caption{The probability that a signal is present in at least two ON scans within a frequency block, as a function of drift rate. A frequency block of at least 1024 frequency bins (2.86\,kHz) is needed to guarantee that a signal with a drift rate of $\pm$4\,Hz/s falls within the bounds of at least two of the ON scans.}

    \label{fig:block_size}
\end{figure}

\item We want the size of our blocks to capture drifting signals. A signal might drift out of frame if a block size is too small. Figure \ref{fig:block_size} shows the probability of a signal being present in at least 2 of the 3 ON scans of a cadence for blocks of varying size, as a function of the drift rate of a signal. We aim to be sensitive to signals drifting up to $\pm 4$ Hz/s, consistent with past Breakthrough Listen searches. The 512 channel block has a 0\% probability of capturing a signal drifting $\geq 3$ Hz/s. Blocks of size 1024 channels have a 98.7\% chance of capturing some portion of a signal drifting at $\pm 4$ Hz/s in at least two of the three ON scans. We thus choose to use blocks of size 1024 channels, as they have a near 100\% chance of capturing signals drifting $\pm 4$ Hz/s and will produce kurtoses $>1$ if they contain signals with SNR $\approx$ 10. We define the SNR of a signal as the signal amplitude divided by the median absolute deviation of the noise in its block.  
\end{enumerate}

The next step is to remove all blocks that we believe are unlikely to contain a signal. Since input/output (I/O) read time is the most time-intensive part of this algorithm, we would like to avoid analyzing blocks that contain only noise in all of their ON observations. To do this we load the entire frequency data of one time slice in each ON observation (the choice of time slice is somewhat arbitrary, we use the eighth), and then split this array into the frequency blocks that were created earlier. This results in a set of length 1024 arrays, each representing one time slice of a 2.86\,kHz frequency block, which are each individually checked to see if they contain any peaks with a median absolute deviation $>5\sigma$, and the frequency blocks that do are recorded as `hotspots'. This is repeated for the equivalent time slices of the second and third ON scans, with all hotspots recorded for each. All frequency blocks that contain at least two ON observations labeled as hotspots are deemed likely to have a signal in at least two of the ON observations, and are kept for full analysis. These are deemed Level 0 frequency blocks, and get passed on to the rest of our pipeline for full analysis \textbf{(see Section \ref{sec:pipeline_over}  for more details)}.

The most time intensive step in our algorithm is the I/O time of loading individual blocks of an observation from the BL data archive into memory. To speed this up, we divide each full frequency band into 8 sub-bands. We load the sub-bands of each observation in a cadence into memory at once, and analyze all frequency blocks present in that sub-band before moving on to the next sub-band. This allows us to load observations faster into memory compared to reading in and processing each 1024-channel frequency block one by one, because of the switching cost during I/O vs processing.


\begin{table}
\begin{center}
\caption{High-Density RFI Removed Frequencies}\label{tab:rfi}
\begin{tabular}{ll}
\hline \hline
Frequency & Federal Allocation \\
$(\mathrm{MHz})$ &  \\
\midrule
$1164-1200$ & Global Navigation Satellite System (GNSS) \\
$1200-1340$ & Notch Filter \\
$1370-1390$ & Air traffic control (ATC) \\
$1520-1560$ & Mobile-satellite service (MSS) \\
$1560-1630$ & GNSS \\
$1670-1705$ & Geostationary operational \\
& environmental satellite (GOES) \\
$1915-2000$ & Personal communications services (PCS) \\
$2025-2035$ & GOES \\
$2100-2120$ & NASA Deep Space Network use (DSN) \\
$2180-2200$ & MSS \\
$2200-2280$ & Earth exploration satellite (EES) \\
$2300-2360$ & Notch Filter \\
$2485-2500$ & MSS \\
$4680-4800$ & Fixed Satellite Service (FSS) \\
$8150-8350$ & EES \\
$9550 - 9650$ & EES \\
$10700-11200$ & FSS \\
\hline
\end{tabular}
\end{center}
\begin{flushleft}
Notes. Modified and extended from \citet{Price:2020},
 (
\hyperlink{ref[https://transition.fcc.gov/oet/spectrum/table/fcctable.pdf}{FCC Table of Frequency Allocations})
\end{flushleft}
\end{table}

\begin{table*} 
\begin{center}
\begin{tabular}{|l|l|l|}
\hline \textbf{Column} &  \textbf{Data Type} & \textbf{Content}  \\
\hline 
\hline 
Batch Info & Int & The batch that the cadence came from and the cadence number within that batch.\\
All Files & Double & List of the paths of the .h5 files making up the respective cadence. \\
Index & Int & HDF5 index of the 2.86\,kHz block corresponding to the frequency. \\
Freq & Double & Frequency in the observation where the signal was found. \\
ON-1 Maxes & Array &The maximum values of each time slice in the first ON observation.\\
ON-2 Maxes & Array &The maximum values of each time slice in the second ON observation.\\
ON-3 Maxes & Array &The maximum values of each time slice in the third ON observation. \\
k1 & Double &Kurtosis of observation 1. \\
k2 & Double &Kurtosis of observation 2. \\
k3 & Double & Kurtosis of observation 3. \\
k4 &Double & Kurtosis of observation 4. \\
k5 &Double &Kurtosis of observation 5. \\
k6 &Double &Kurtosis of observation 6. \\
Drift &Boolean & Whether or not the signal present in the region has a drift of zero. \\
\hline
\end{tabular}
\caption{\label{tab:stats_recorded} Properties of each frequency block recorded by our pipeline.}
\end{center}
\end{table*}

For each block, the properties in Table \ref{tab:stats_recorded} are recorded. In addition to the kurtosis of each observation, the maximum value of each time slice in each ON observation is recorded. These are used later to filter out small `blips' and broadband contamination, which is discussed in Section \ref{sec:filtering}. We also record identifying information for each region, notably the batch and cadence number that it was part of, the list of files that make up the cadence, the index of the frequency block, and the frequency of the block itself. An estimate of the drift rate is also recorded, to filter out zero drift signals later on. All of these properties for each Level 0 block are recorded and passed into the filtering segment of the pipeline.

\subsection{Filtering}\label{sec:filtering}

The properties recorded above are used to select strong candidates from the set of frequency blocks analyzed. The kurtosis statistic is the primary metric. The goal is to find frequency blocks with high kurtoses in the ON scans, and low kurtoses in the OFF scans. We impose a cutoff of $k_{\rm ON, med} \geq 0.5$, where $k_{\rm ON, med}$ is the median of the three ON scan kurtoses.  This is to take into account that a rapidly drifting signal (close to $\pm$4\,Hz/s) may not appear in all 3 ON observations, but should still appear in at least 2. The median kurtosis value of the ON scans should thus be $>1$ if a strong (SNR $> 10$) signal is present. For the block size we use here, a narrowband signal with a SNR of 10 would produce an associated kurtosis of $\gtrsim1$ when injected into a background of sampled GBT receiver noise (see Section \ref{sec:algorithm}). In theory this means we are sensitive to signals with even lower SNR, closer to SNR $\approx$ 8. We require that the kurtosis of each OFF scan is $\leq 0.25$, so that there is little or no signal in each OFF scan. This $k_{\rm OFF, max}$ cutoff can be lowered to decrease the number of false positives.  

After imposing the kurtosis cuts detailed above, we are left with $175,160$ frequency blocks from the original $24,368,715$ analyzed. This is about a $0.7\%$ detection rate, or equivalently a $0.7\%$ false positive rate if we assume essentially all detected signals are RFI. These are still too many frequency blocks to inspect visually, so additional filters must be applied. We use two: a requirement that any signal present in the frequency block have non-zero drift; and that the signal present is not a broadband streak present in only a small number of time bins. 

Our drift-filter checks a signal's drift rate by comparing the location of the signal in each ON scan. We integrate each ON scan along the time dimension. A signal with zero drift will appear in the same frequency bin in each time slice, and thus show up clearly when the ON scan is time-integrated. We record the frequency location of these signals by running a peak-finding algorithm on the time-integrated ON scans, and compare the frequency locations of the peaks between each ON scan using the Python package \texttt{scipy}'s \texttt{findpeaks} routine. If the peaks in adjacent ON scans are less than 2 frequency bins apart (indicating a drift rate of $< 0.01$\,Hz/s) then we mark the signal as having zero drift.

\begin{figure}
    \centering
    \hspace{-.6cm}
    \includegraphics[width=0.5\textwidth]{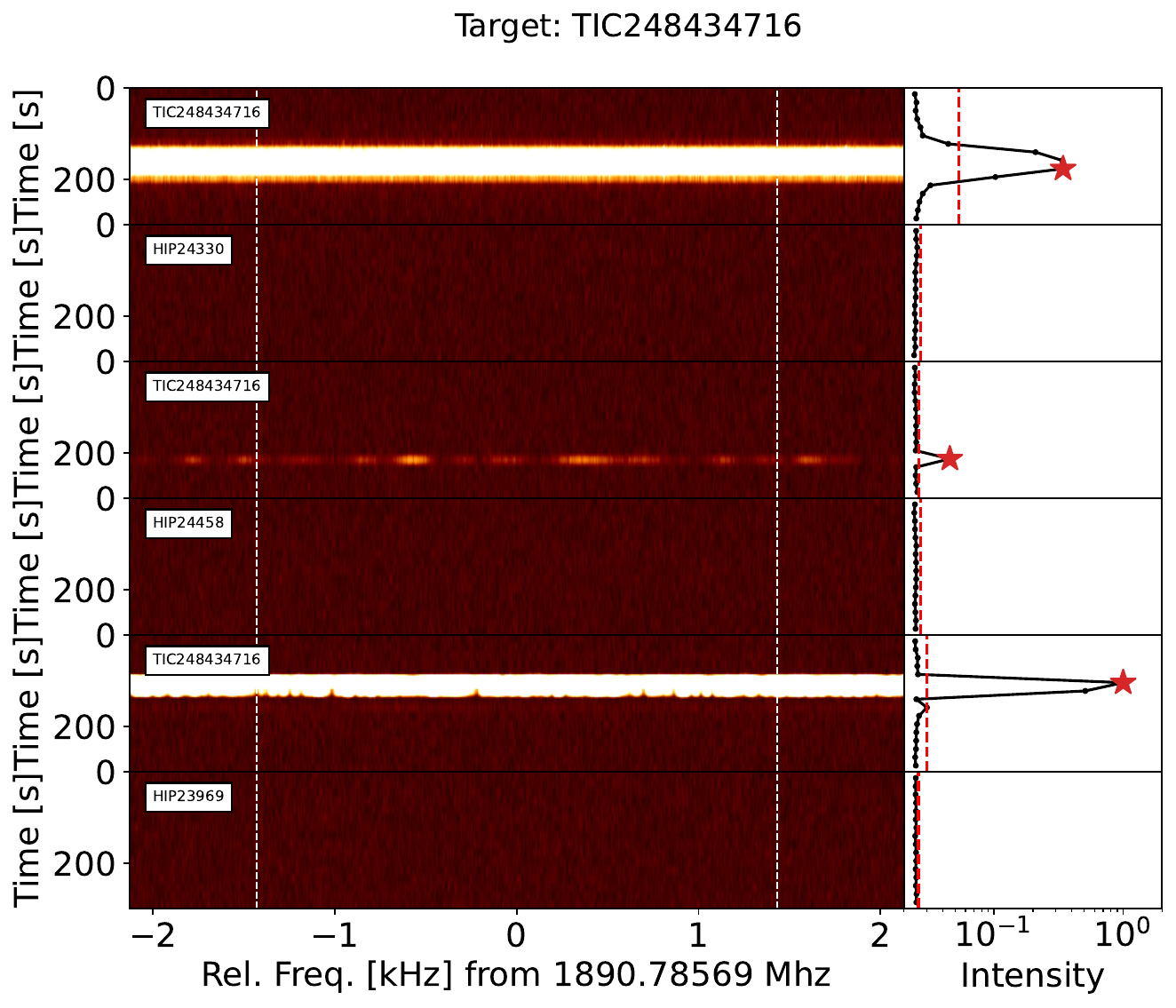}
    \caption{An example signal that would be rejected by our broadband filter. The left panels contain the waterfall plot of the signal, with frequency on the x-axis and time on the y-axis, while the right panels show the frequency-integrated observations, with normalized intensity on the x-axis and time on the y-axis.  The dashed red line is $10\sigma$ above the median, where $\sigma$ is the median absolute deviation. The red star highlights time bins identified by a peak finding algorithm above this level. The white dashed lines on the left panel show the extent of the 1024 channel block that our pipeline analyzes.}
    \label{fig:broadband_filter}
\end{figure}

\begin{figure*}
    \centering
\centering
    \includegraphics[width=1\textwidth]{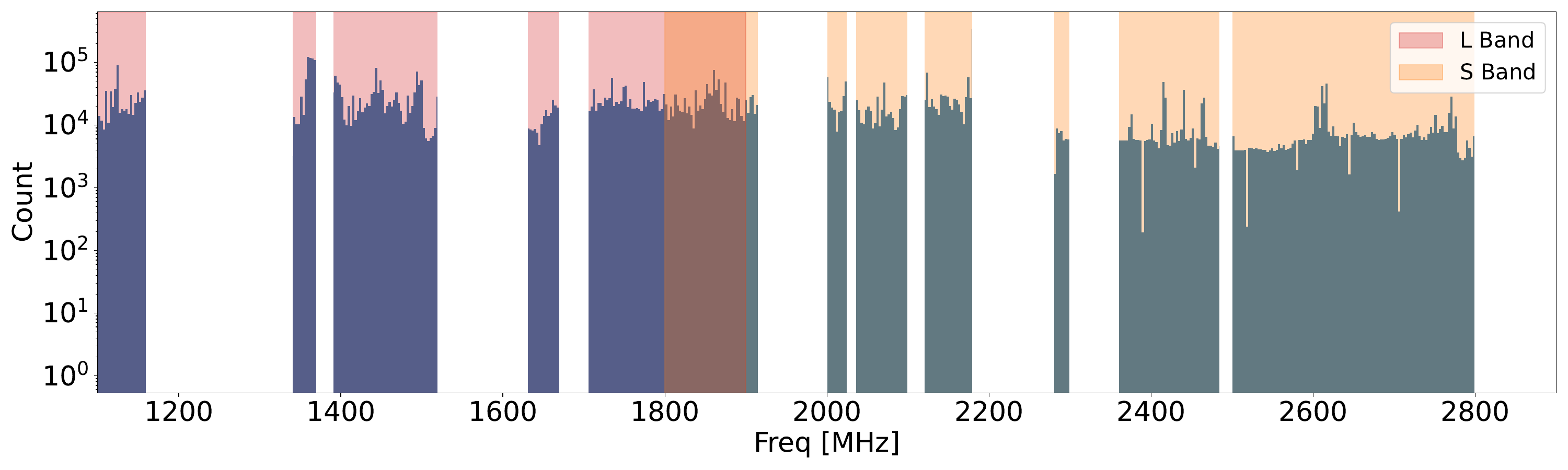}
    \includegraphics[width=1\textwidth]{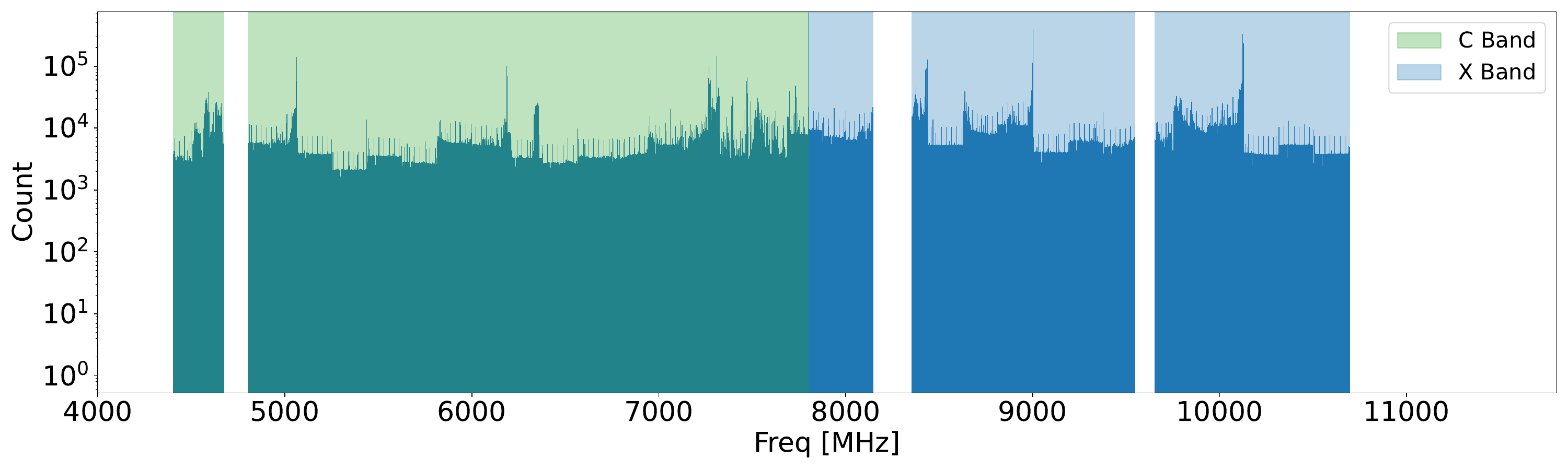}

    \caption{The frequency distribution of all Level 0 blocks.}
    \label{fig:frequency_block_distribution}
\centering
\end{figure*}

Our broadband filter functions by checking if there are large changes in intensity over time in an observation. Broadband interference often presents itself as horizontal streaks across the entire frequency range of a block, but confined to just one or two time bins. To identify such contamination, we average across the frequency of a block's ON scan. The broadband interference will show up as a spike in intensity at a certain time over the course of the observation. If the peak has a median absolute deviation $\geq10$ from baseline, we classify it as likely being produced by broadband streaks and reject the signal. Of course, we might also discard some real pulsed technosignatures using this method, but only if they have pernicious duty cycles that happen to line up with the duration of our ON observations. In general, these pulsed signals belong to a different kind of transmitter than this search is optimized for, as our data have a limited time resolution of $\sim18$s. There have been other search approaches built to target pulse beacons (\citealt{gajjar2022searching}, \citealt{suresh20234}). Our broadband filter is illustrated in Figure \ref{fig:broadband_filter}.

A final common false positive is small `blips' of high intensity, which show up in a small number of frequency and time bins in an otherwise empty block. If such blips show up in two or more ON scans and are high enough intensity to significantly skew the kurtosis, then their respective blocks will pass our kurtosis filter. Since these small blips only occupy a few frequency bins, they are averaged out when the entire scan is frequency integrated, and our broadband filter does not pick them up. To reduce the number of such false positives that make it through our pipeline we calculate the maximum of each time slice, and check if there are  significant outliers, defined as $>10\sigma$, within these local maxima. Signals that are fairly constant in intensity over time will not be eliminated. However, this method may eliminate true signals that have a significant intensity spike in one time bin.

\subsection{Pipeline Overview}\label{sec:pipeline_over}

Our pipeline takes as input six observations, corresponding to three ON scans of a target and three OFF scans of a different target. It then performs the following steps.

\begin{enumerate}
    \item Collates the spectrograms of the observations in a cadence by time. This creates one large waterfall plot composed of three ON scans and three OFF scans.   
    \item Divides this large waterfall plot in its frequency space into overlapping blocks of 1024 frequency channels. 
    \item Determines which blocks are likely to have a signal in at least two of the three ON scans, and record these if they do not fall in dense RFI regions. These are deemed \textbf{Level 0 blocks}.
    \item Determines which blocks have no OFF scans with kurtosis greater than $0.25$, and in which the majority of ON scans have kurtosis greater than $0.5$. These are deemed \textbf{Level 1 blocks}. 
    \item Determines which blocks do not contain narrowband signals drifting at $<0.01$ Hz/s, and do not show signs of broadband or blips in the majority of their ON scans. These are deemed \textbf{Level 2} blocks. 
    \item Inspect the remaining blocks visually to determine final candidates for further analysis.
\end{enumerate}

 The number of remaining frequency blocks after each step can be found in Tables \ref{tab:pipeline_outputs} and \ref{tab:densities_hits}.

\begin{figure*}[h!]
    \centering
\centering
    \includegraphics[width=1.02\textwidth]{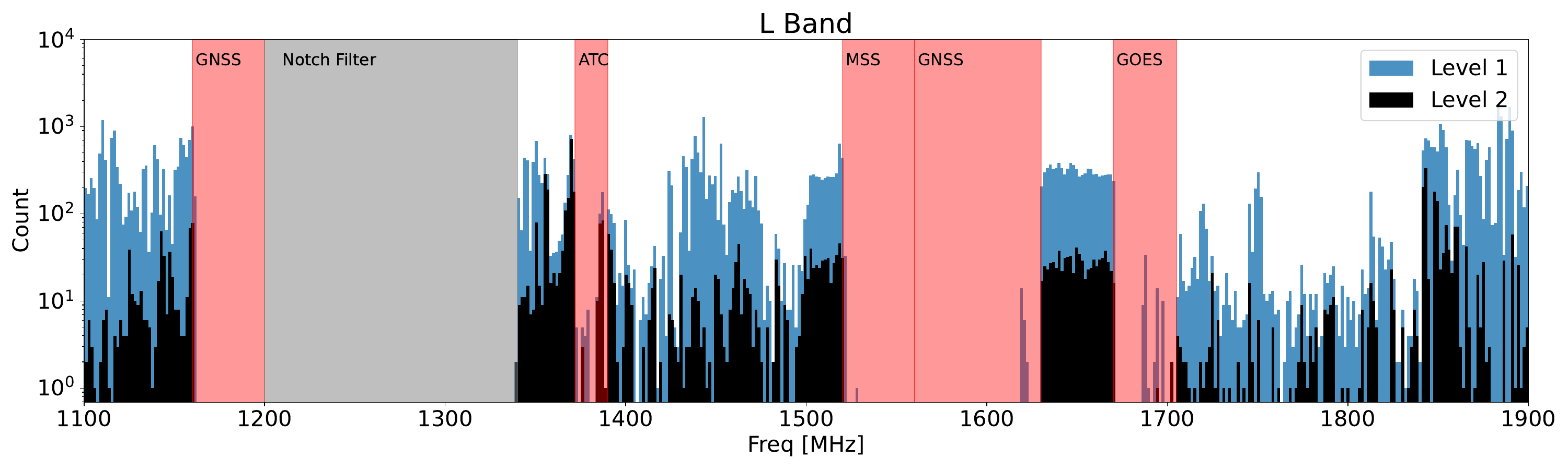}
    \includegraphics[width=1.02\textwidth]{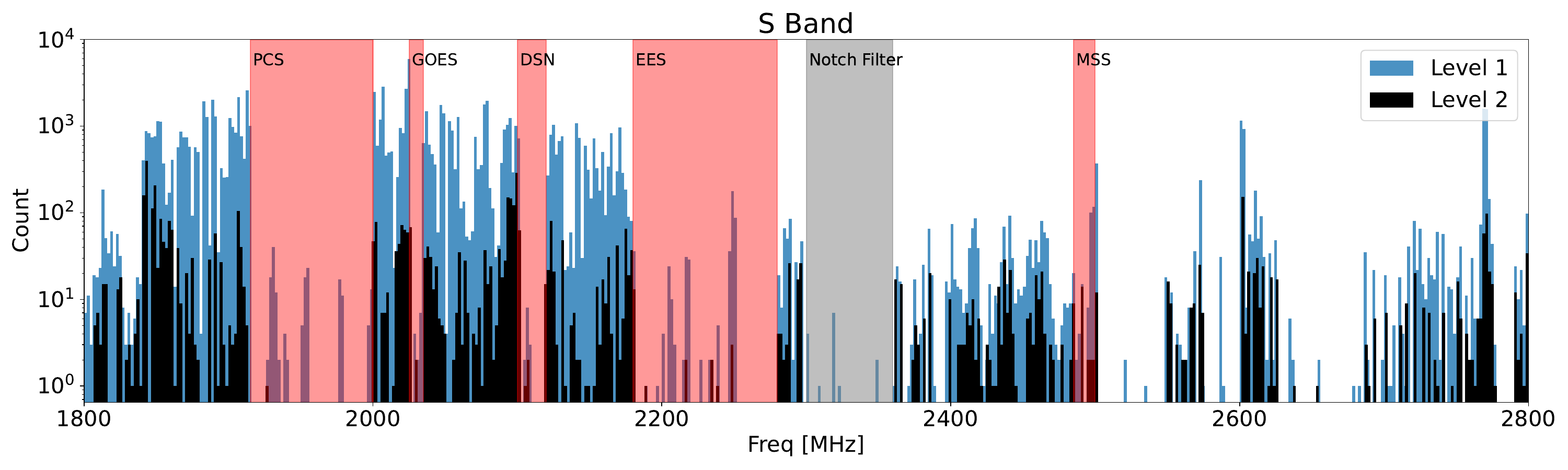}
    \includegraphics[width=1\textwidth]{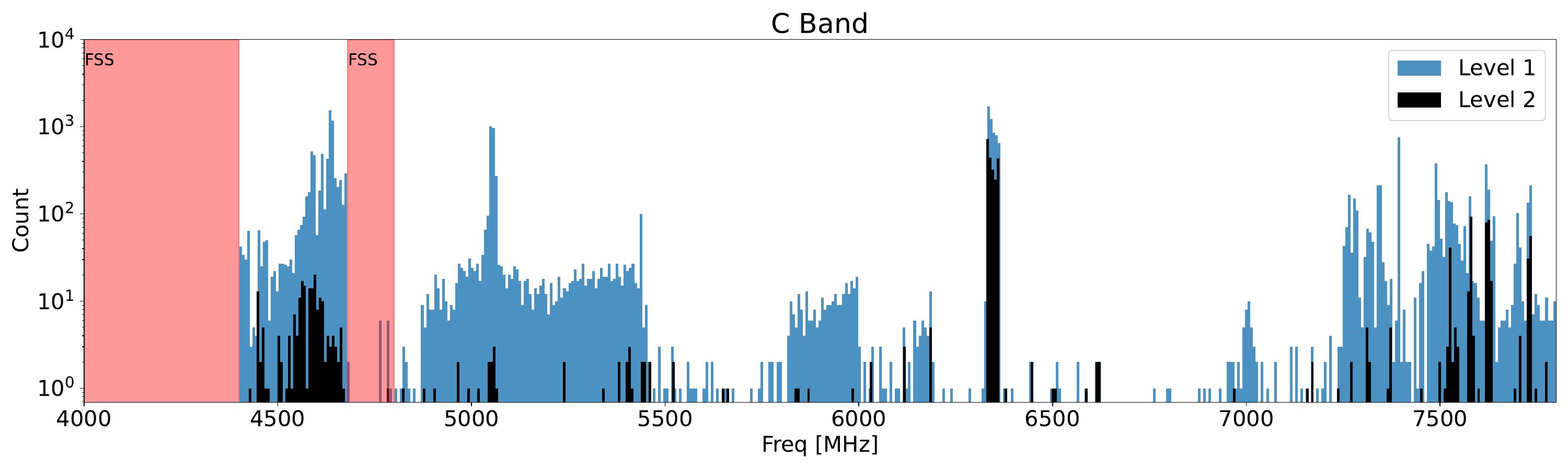}
    \includegraphics[width=1\textwidth]{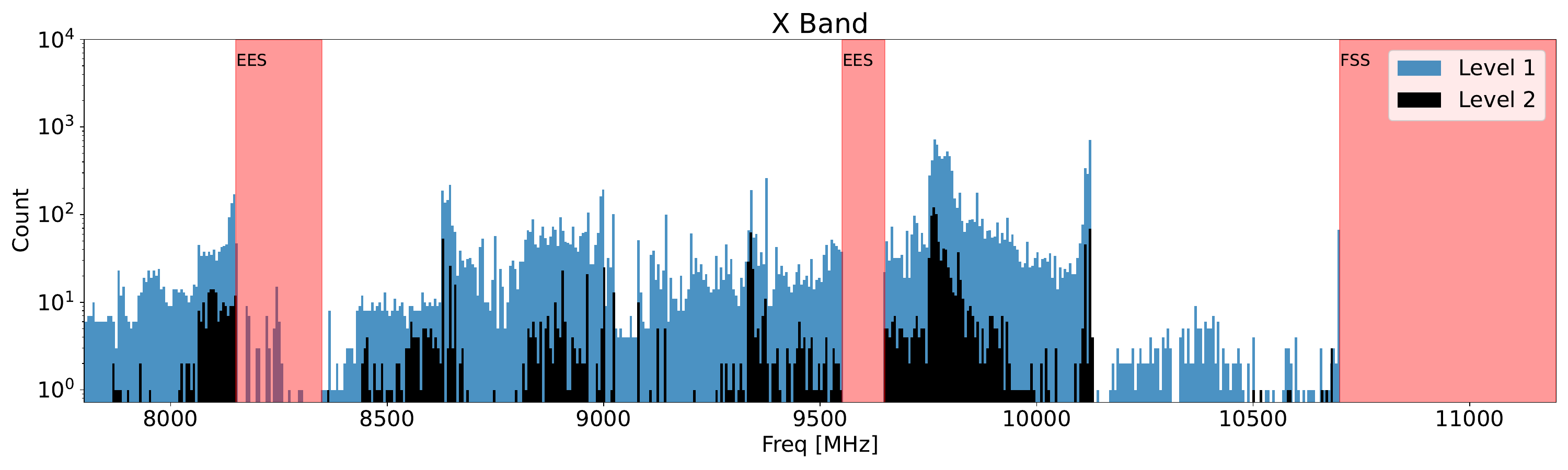}

    \caption{The frequency distribution of Level 1 and Level 2 blocks for each band. Red regions are frequencies with high RFI densities that are excluded from our analysis. Grey regions represent the GBT notch filters and are also excluded.}
    \label{fig:hit_dist_level12}
\end{figure*}

\begin{figure*}
    \includegraphics[width=0.33\textwidth]{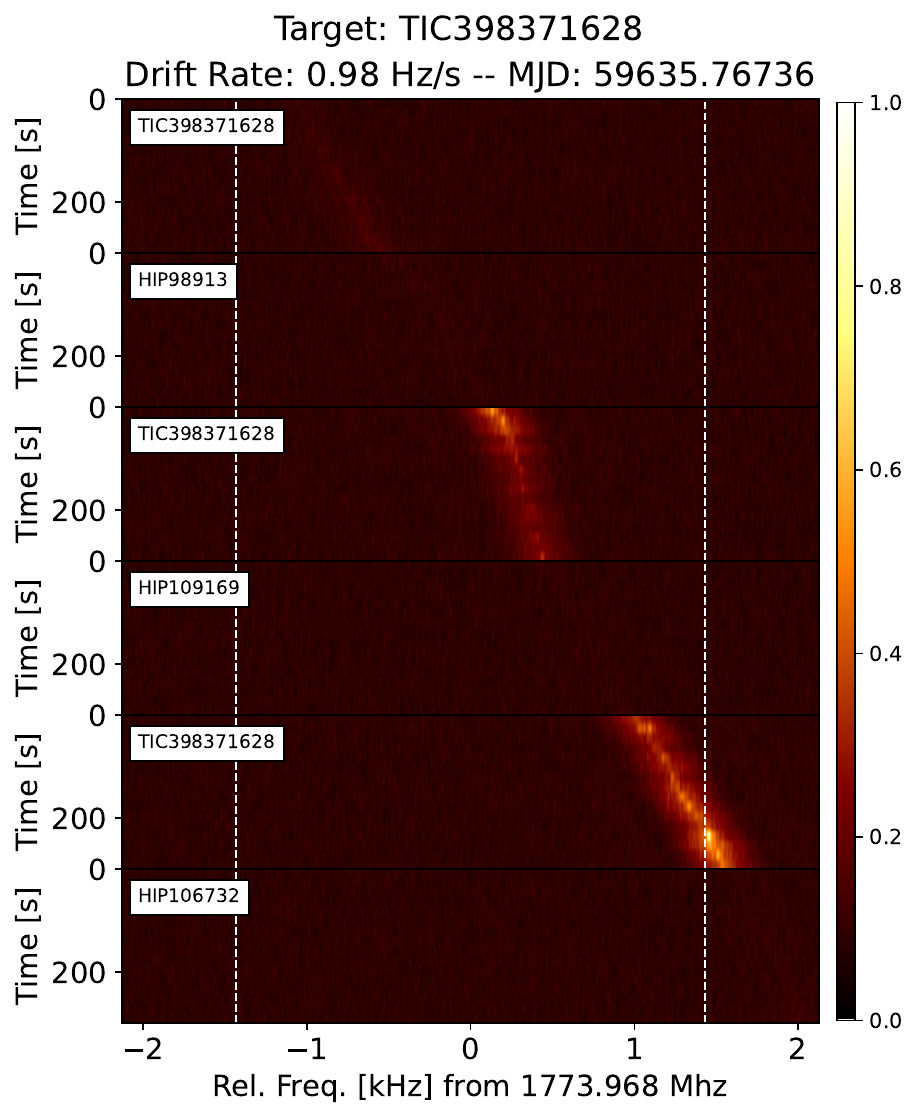}
    \includegraphics[width=0.33\textwidth]{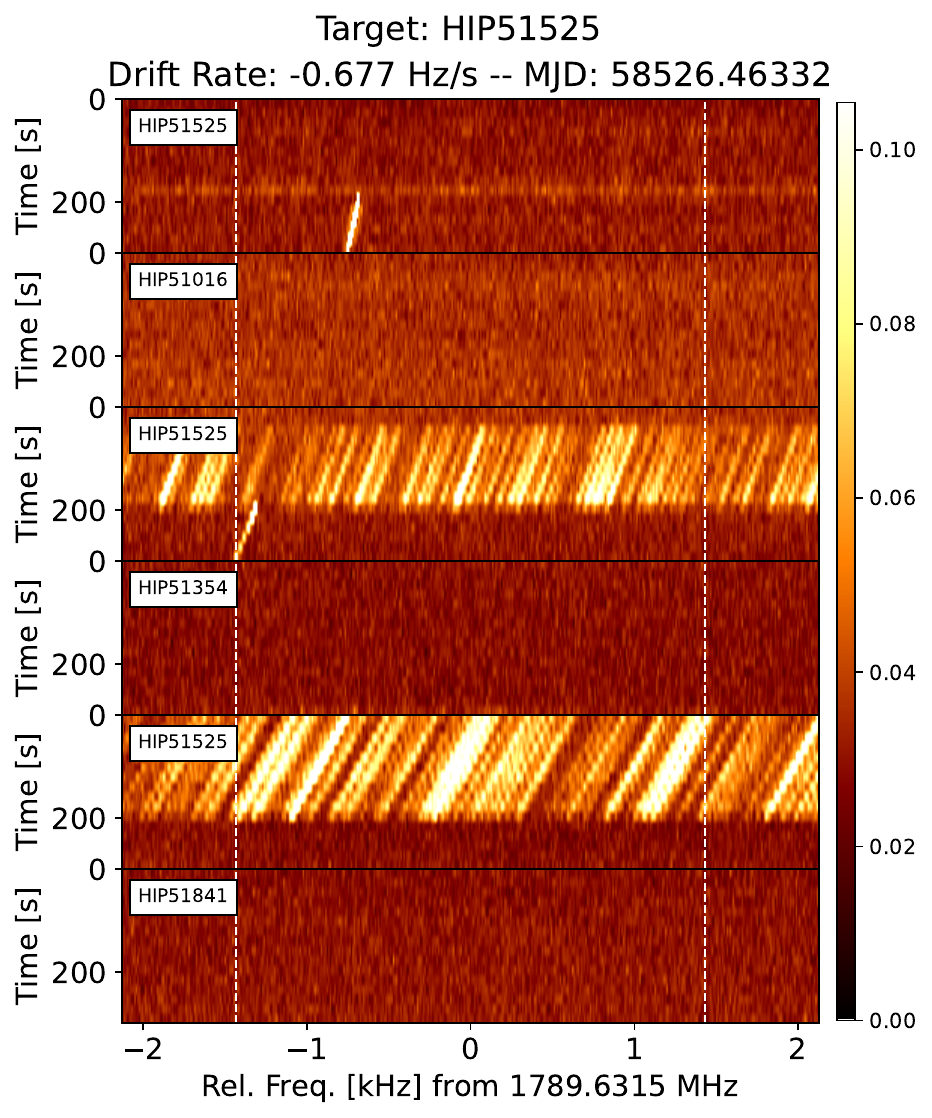}
    \includegraphics[width=0.33\textwidth]{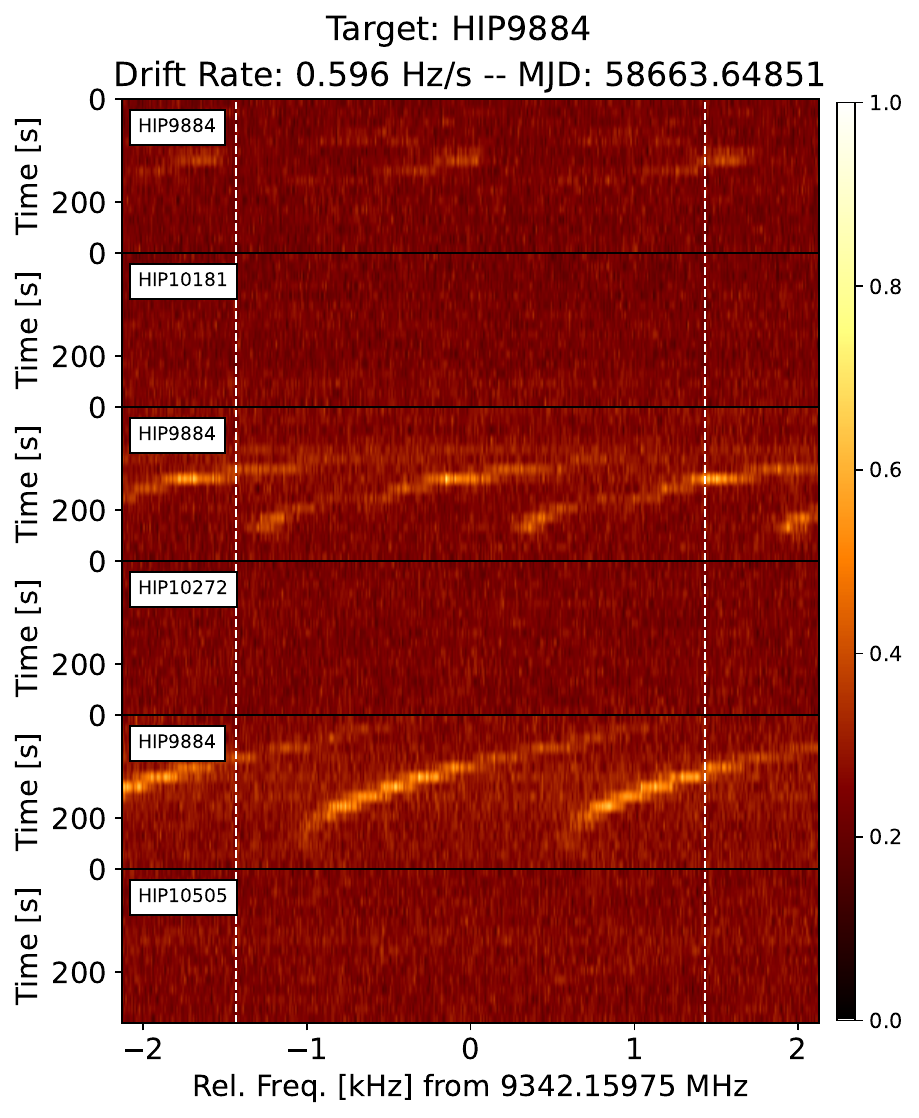}

    \includegraphics[width=0.33\textwidth]{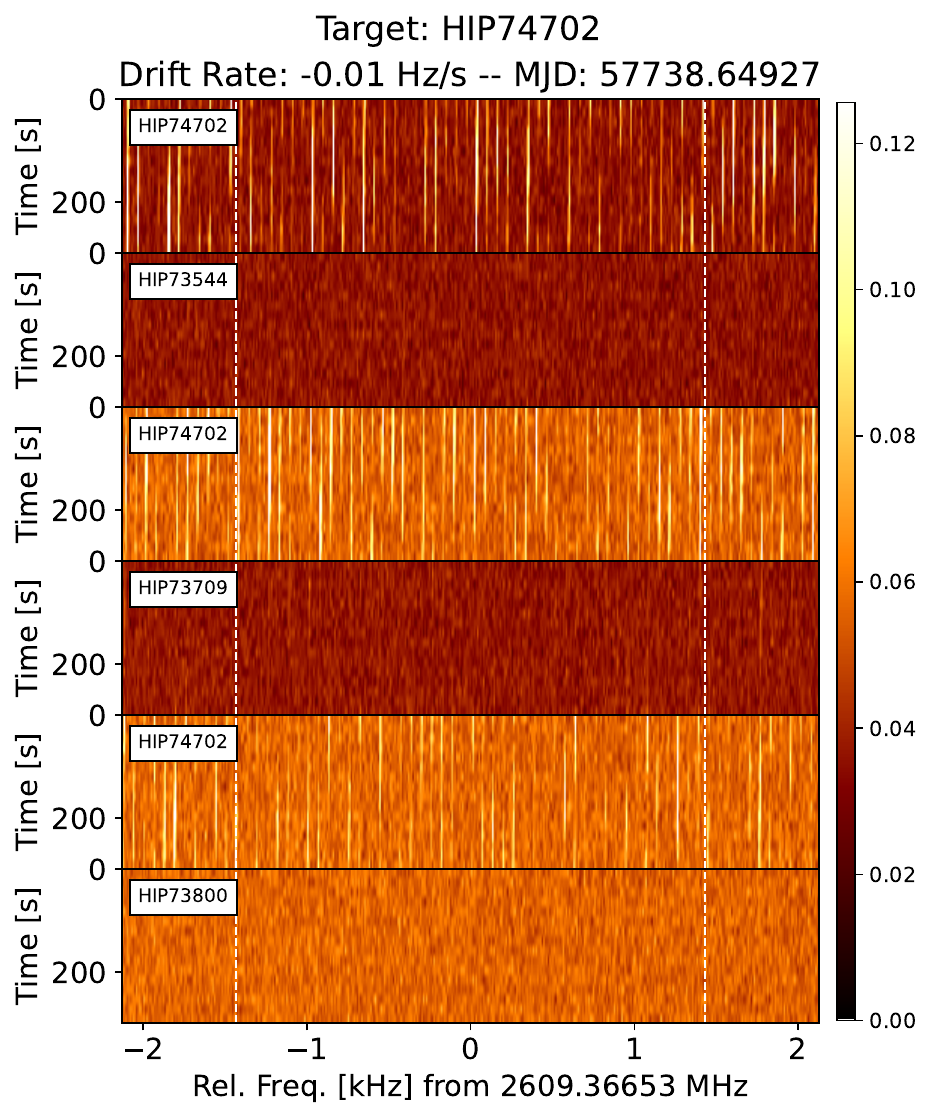}
    \includegraphics[width=0.33\textwidth]{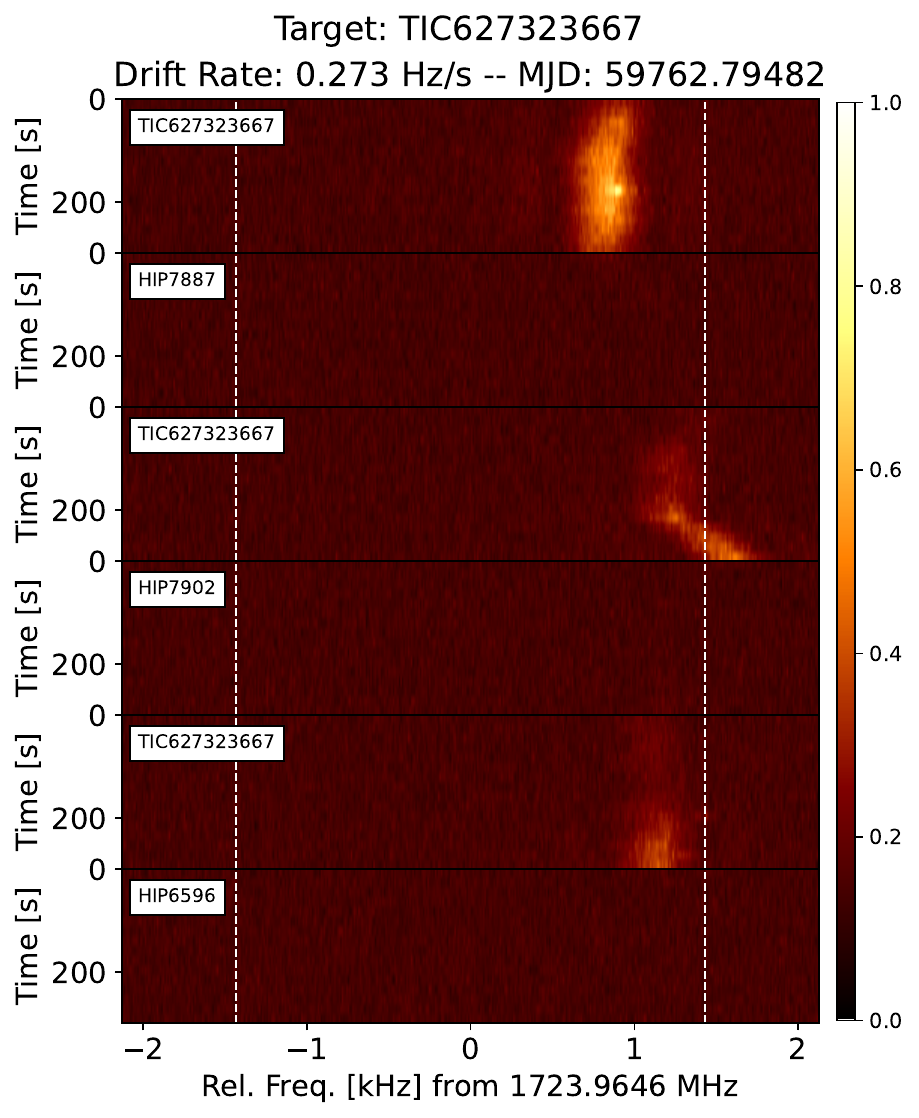}
    \includegraphics[width=0.33\textwidth]{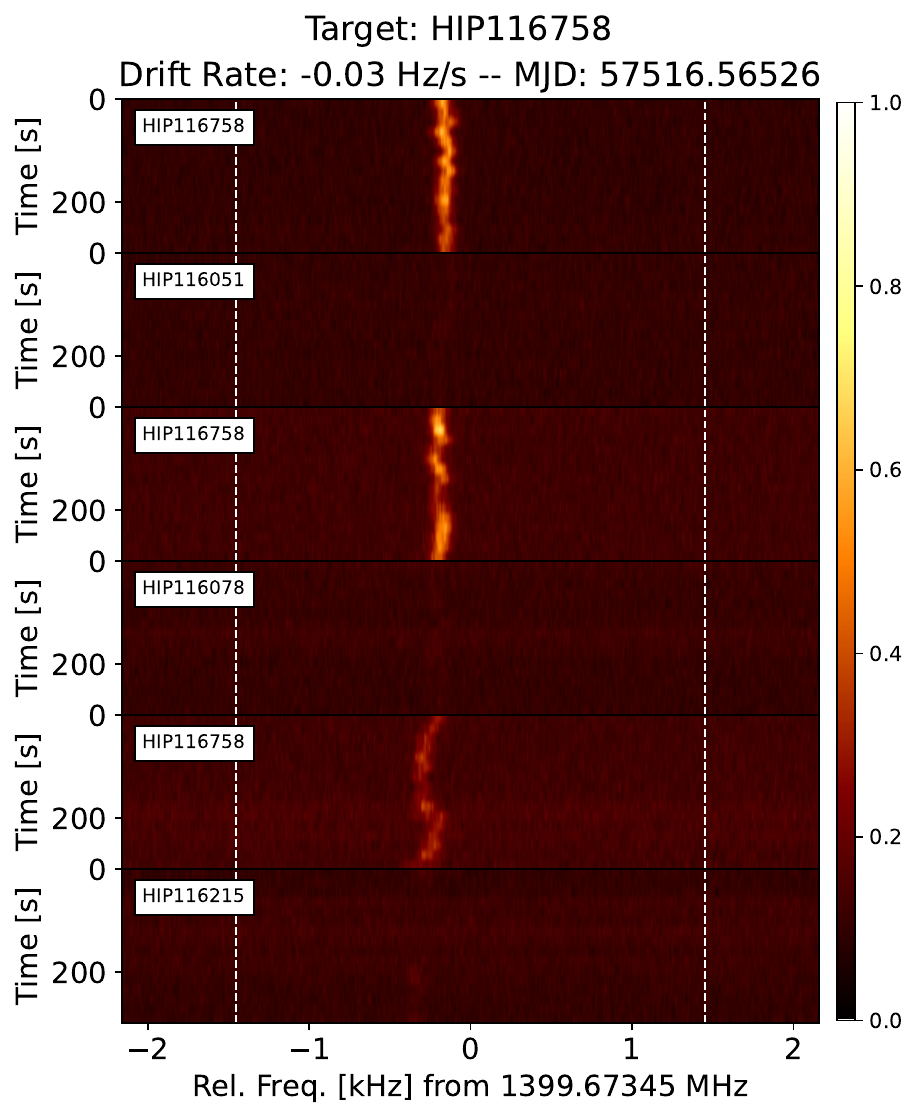}
    \caption{Blocks qualified as Level 2 by our pipeline that do not contain singular, narrowband, linearly drifting signals.}
    \label{fig:funky}

\end{figure*}

\section{Results} \label{sec:results}

We applied our detection algorithm to the bulk of the Breakthrough Listen Green Bank Telescope archive, consisting of $6,630$ cadences, each a collection of six 5-min observations. These observations spanned in frequency from L band to X band. Of this archive, 2,138 were `spliced' files consisting of a observations across a range of nodes stitched together. 

Our pipeline found a total of $24,368,715$ Level 0 frequency blocks. Of these, $175,160$ met our kurtosis criterion and were deemed Level 1 blocks. The distribution of the Level 0 blocks as a function of observing frequency can be seen in Figure \ref{fig:frequency_block_distribution}. While some sections of the observing band have a higher number of frequency blocks than others, for the most part they are fairly evenly distributed across the frequency range. This is largely due to a pre-filtering of the especially dense RFI regions (primarily present in the L and S bands). 




$14,168$ blocks survived filtering for broadband, blips, and zero-drift signals, and made it to Level 2. The distribution of the Level 2 blocks as a function of observing frequency can be seen in Figure \ref{fig:hit_dist_level12}. These $14,168$ Level 2 blocks that passed all of our filters came from a total of 838 unique targets, and 126 of these targets contained more than 10 Level 2 blocks. Our pipeline picks up a wide variety of frequency blocks containing signals strong in the ON scans and weak in the OFF scans. Many of these signals do not display narrowband morphologies that traditional pipelines and machine learning approaches are designed to find. Figure \ref{fig:funky} displays some examples of these kinds of signals. While not all of these signals are compelling candidates, they highlight the flexibility of our approach and the ability it has to pick up on a wide variety of signals. These might include signals with non-linear drift rates, signals that are not narrowband, or multiple signals in close proximity to each other.   

We inspect each of our Level 2 blocks visually to determine which contain signals worthy of further analysis, which we label as our `candidates'. While this process could likely be automated to some degree in future work, as this is the first major application of this pipeline we do it manually to understand the kinds of signals that make it to Level 2.

\begin{figure*}

    \includegraphics[width=0.33\textwidth]{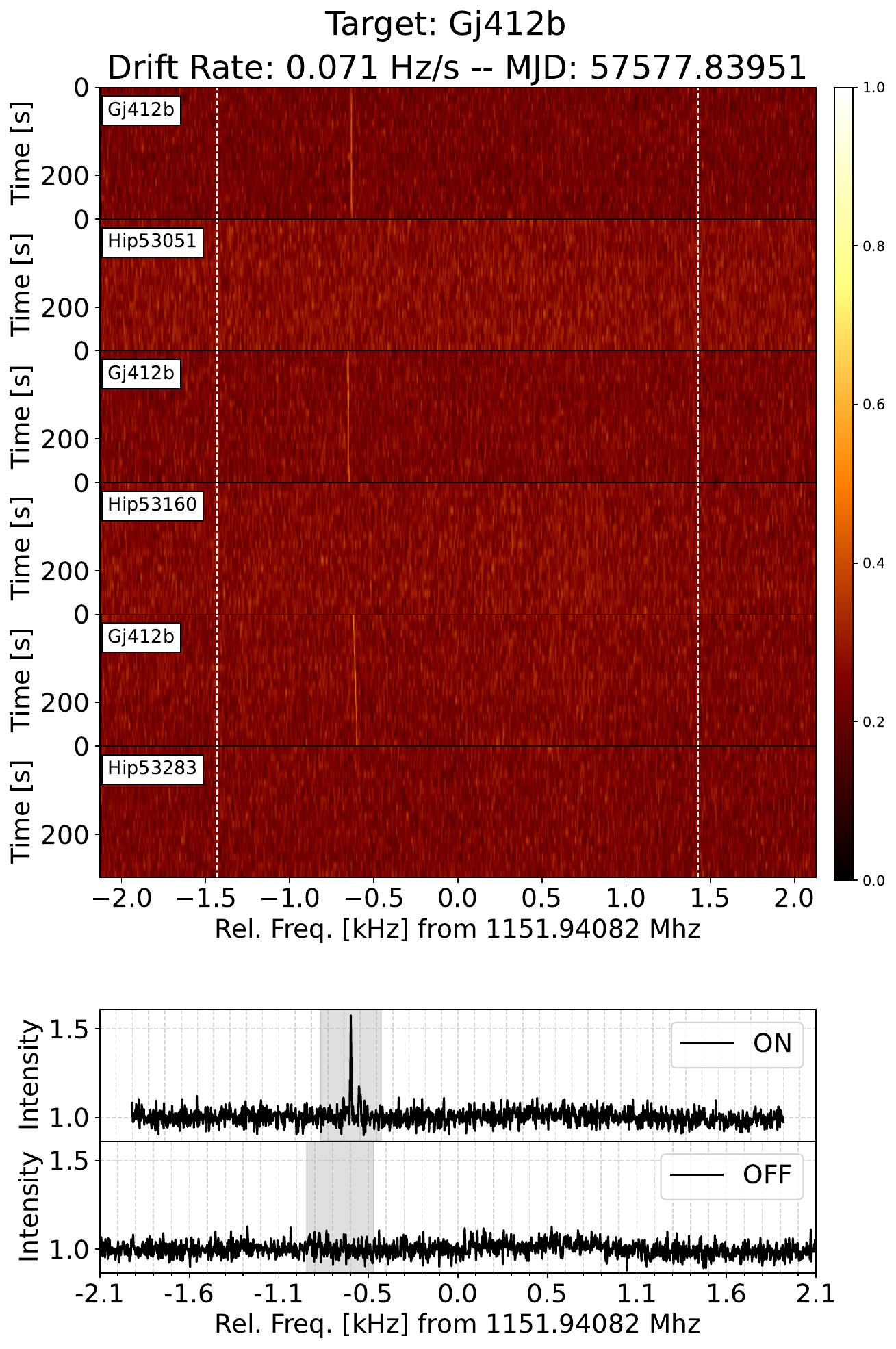}
    \includegraphics[width=0.33\textwidth]{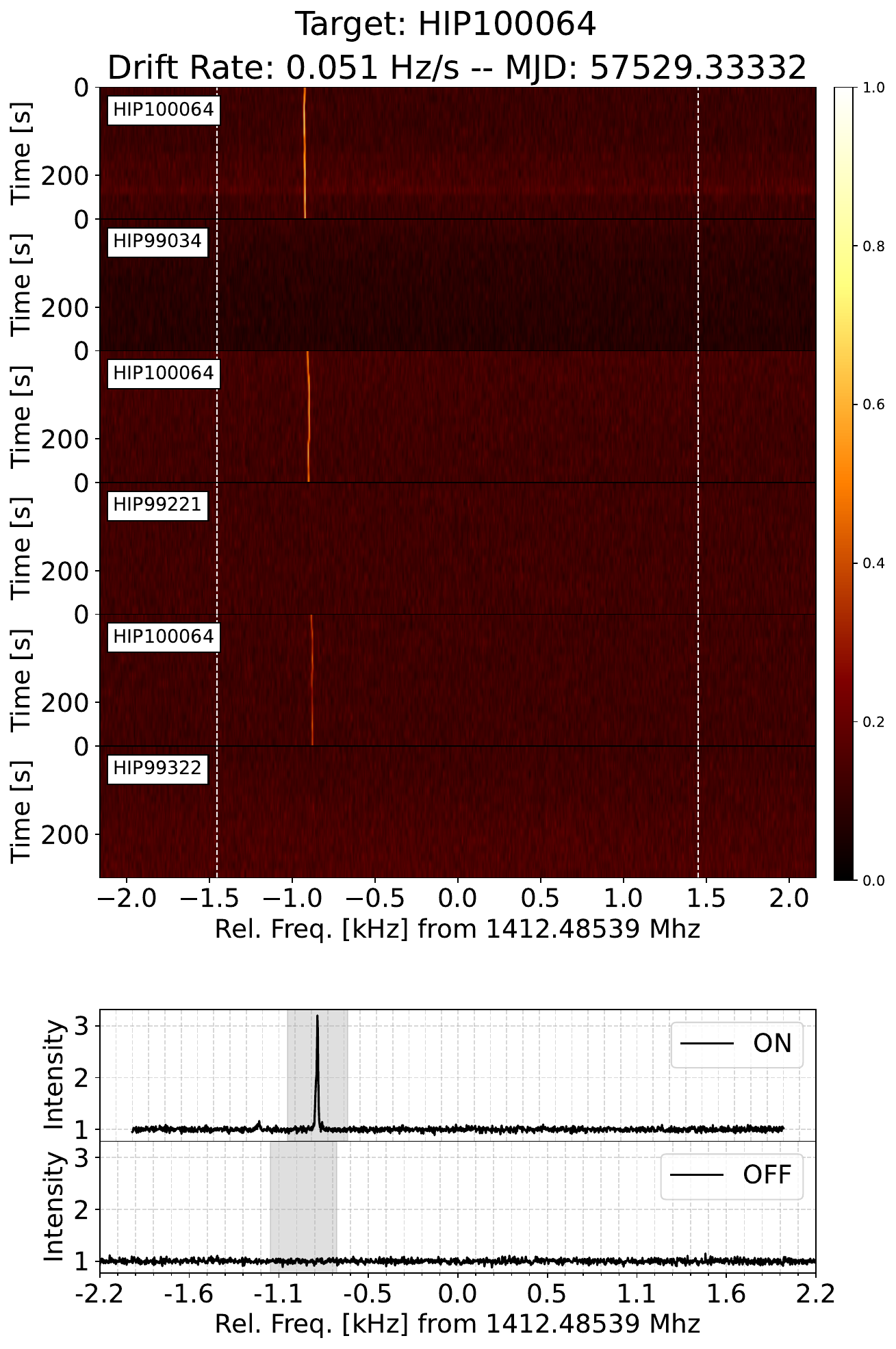}
    \includegraphics[width=0.33\textwidth]{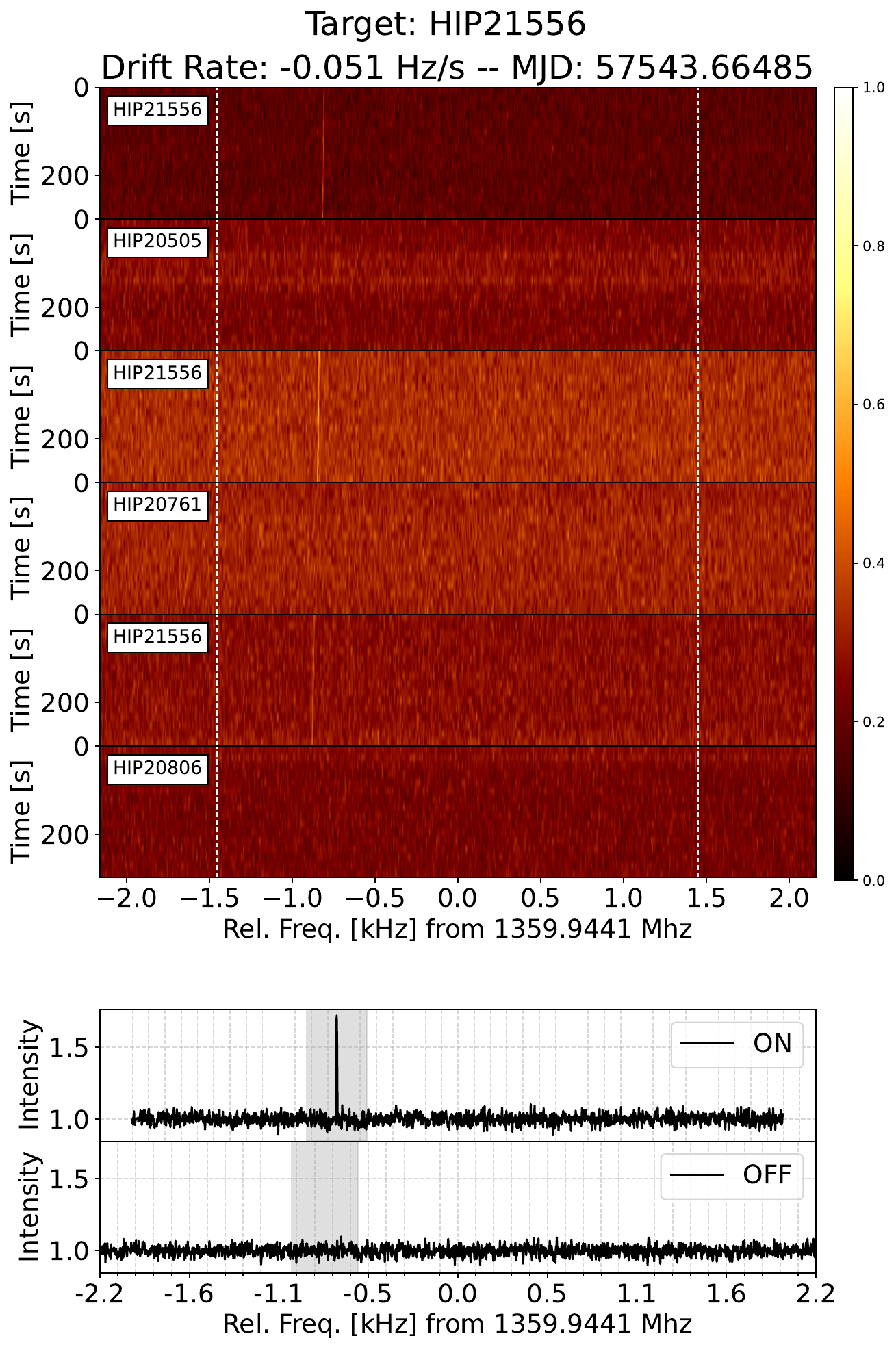}
      \vspace{.1cm}

    \includegraphics[width=0.33\textwidth]{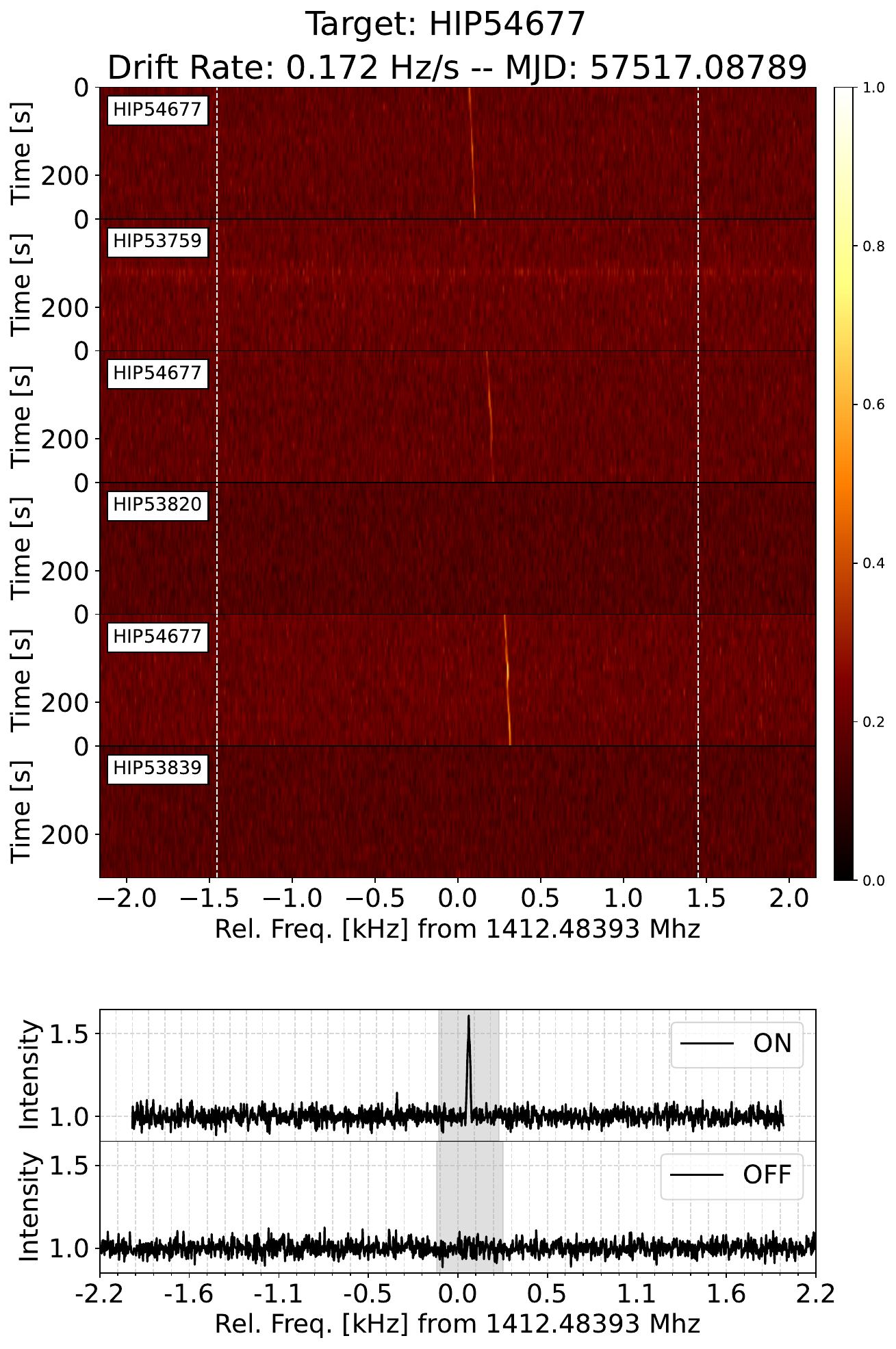}
    \includegraphics[width=0.33\textwidth]{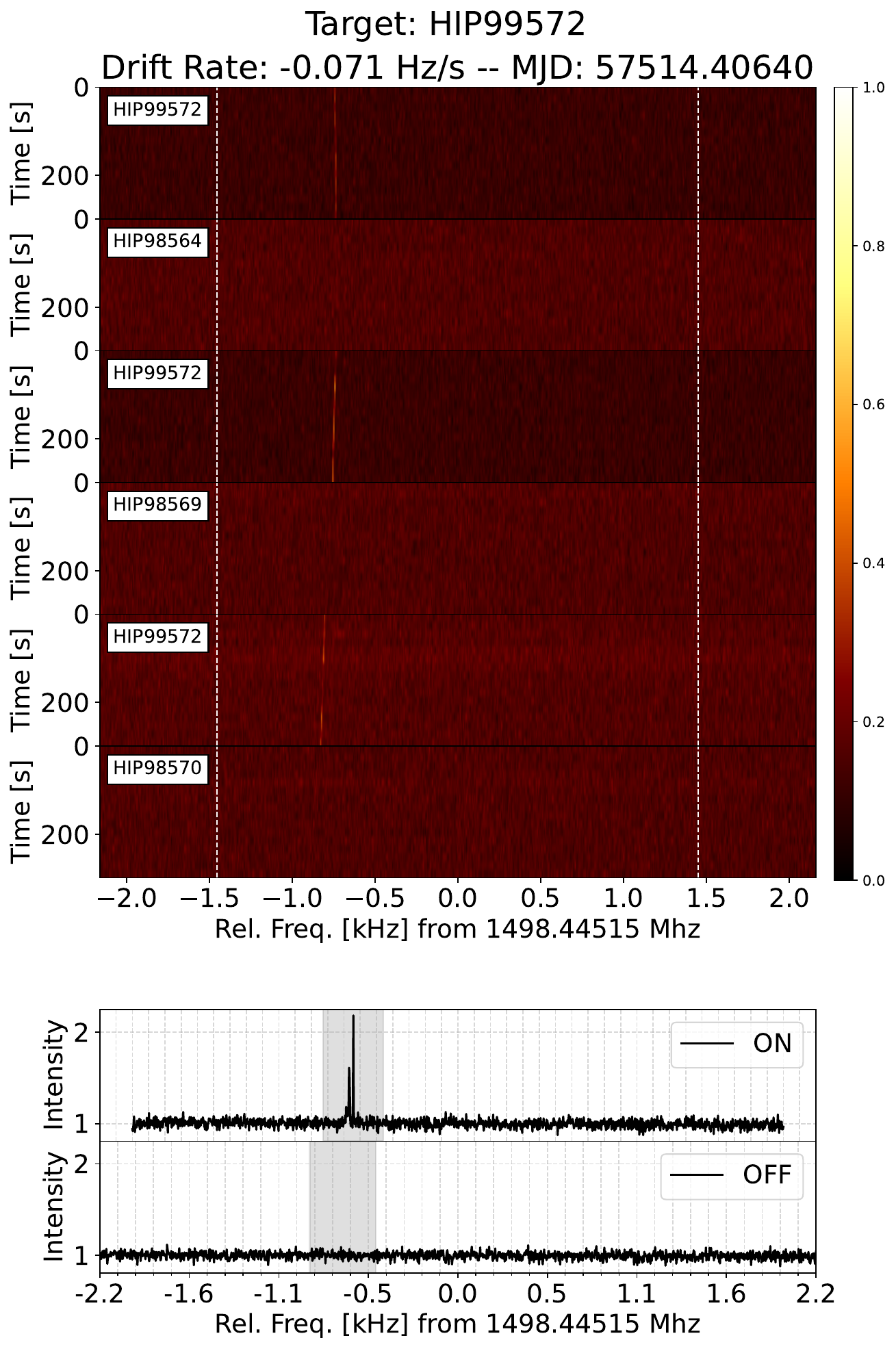}
    \includegraphics[width=0.33\textwidth]{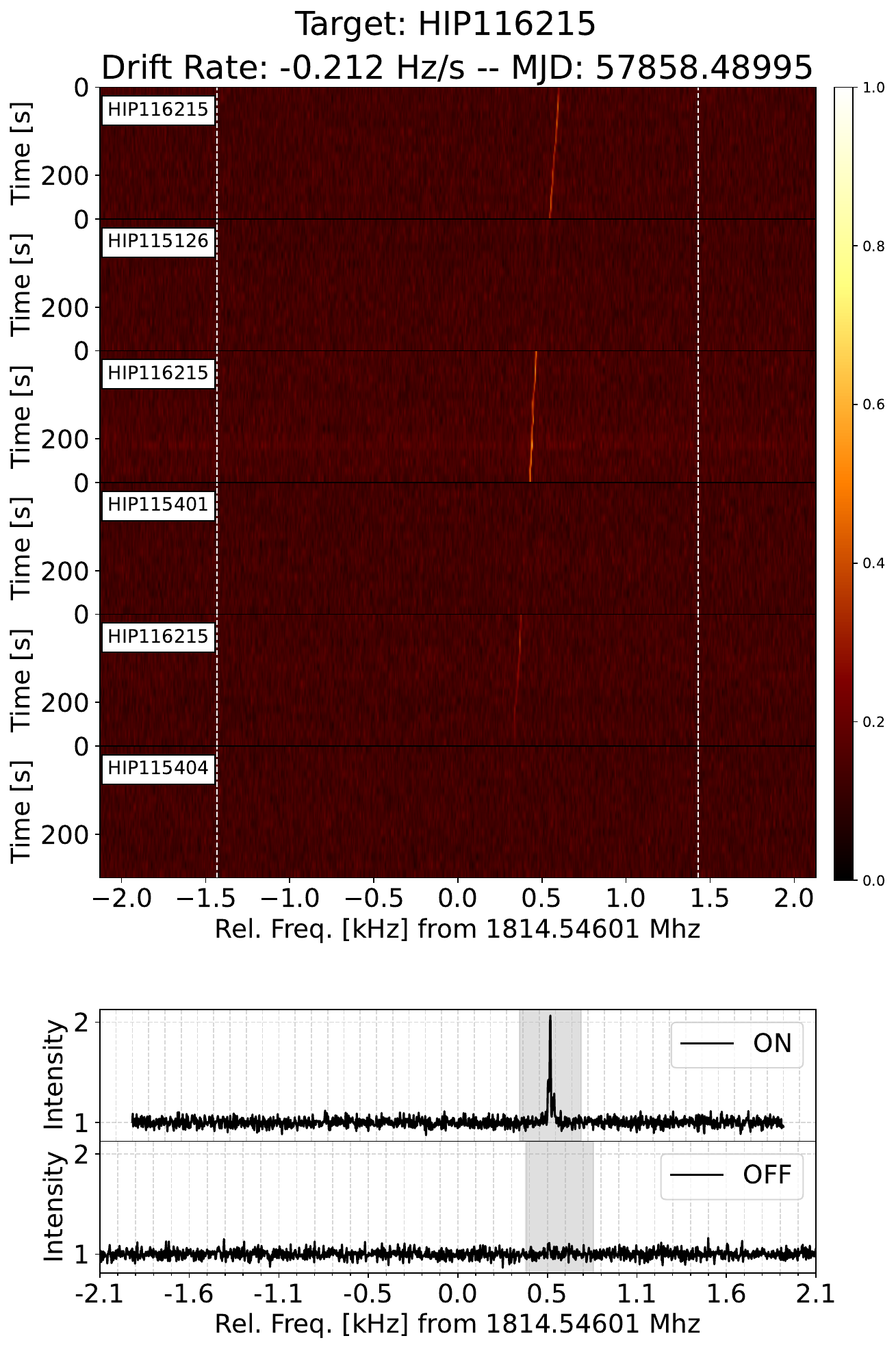}

    \caption{Six candidate signals, with their de-drifted, time-integrated spectra plotted.}
    \label{fig:dedrifted}
\end{figure*}

The majority of our Level 2 blocks contain signals that are weakly present in the OFF scans, but which do not produce high enough kurtosis values to be vetted out by our pipeline. We eliminate these by eye. For the blocks that do not clearly contain signs of a signal continuing into their OFF scans, we estimate a drift rate if the signal is narrowband and then de-drift and integrate the ON and OFF scans respectively. This can reveal signs of signals that are very weak in the OFF scans. For signals that do not have any sign of continuation into the OFF scans, we examine the larger frequency range the block comes from. Oftentimes a signal is part of a larger structure that bleeds into the OFF scan at frequencies outside the extent of the block. These signals are also eliminated. Finally, if a signal is only present in two of the three ON scans even after the larger frequency range is examined, we also do not raise it to the level of candidate. In short, we are looking for signals that are present in all three ON scans and not in any of the OFF scans, including after extending the frequency range beyond just the block's extent. If a signal meets all of these requirements, we deem it a candidate and perform a more in depth study on it. We present examples of some common false positives captured by our pipeline in Section \ref{sec:false_positives}.

\begin{figure*}
\gridline{
\fig{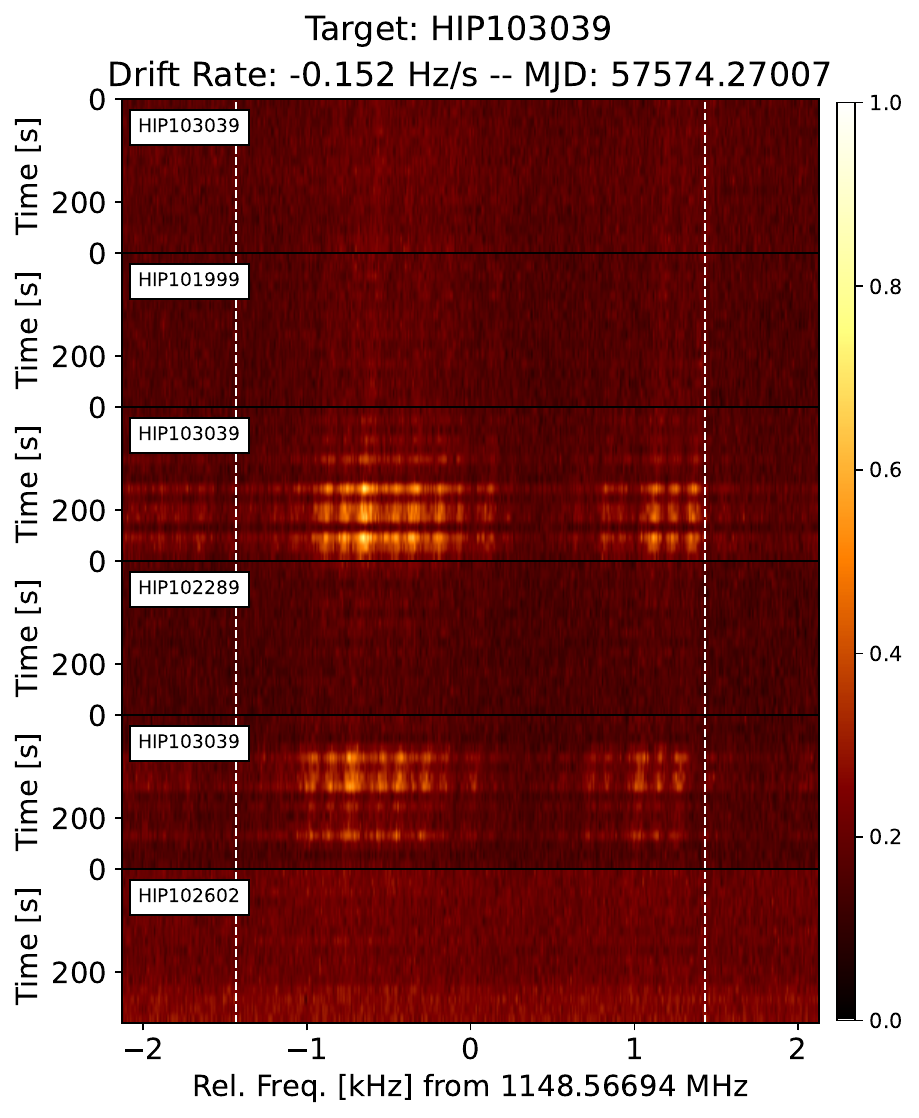}{0.31\textwidth}{(a)}
\fig{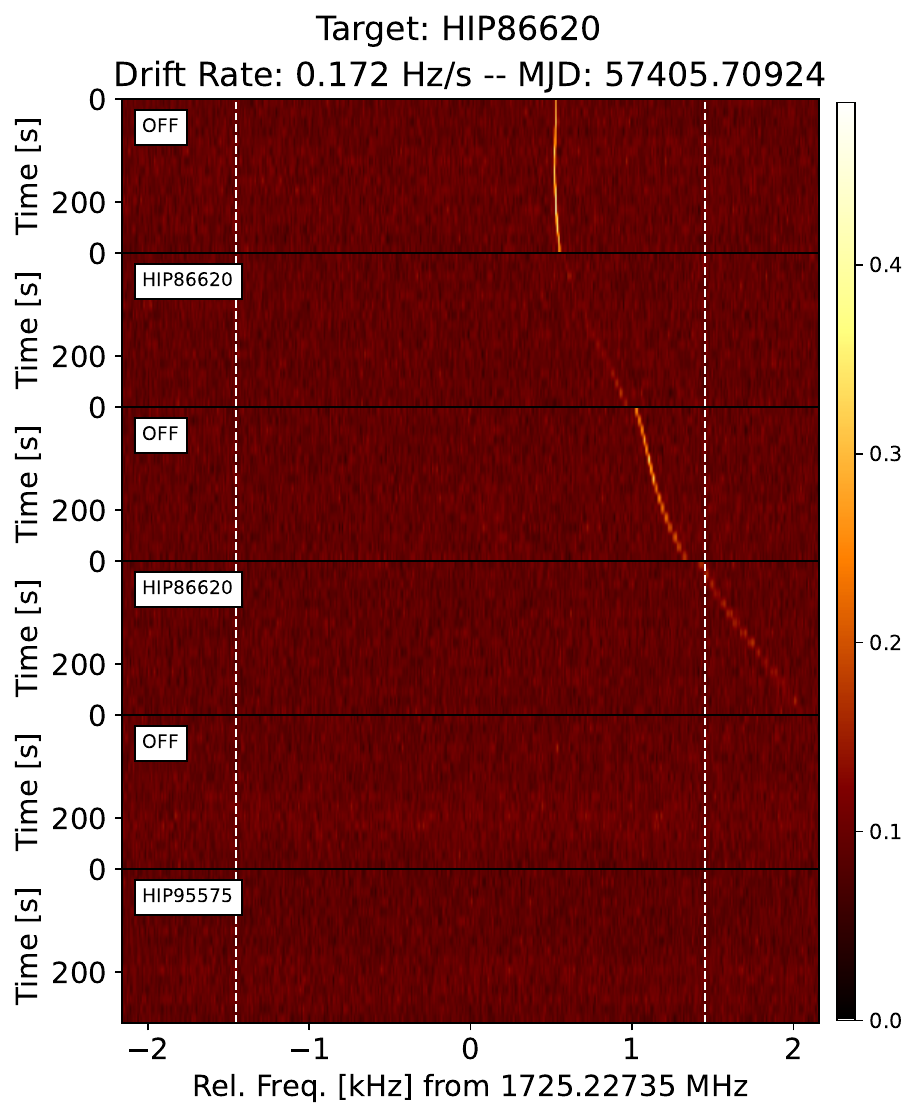}{0.31\textwidth}{(b)}
\fig{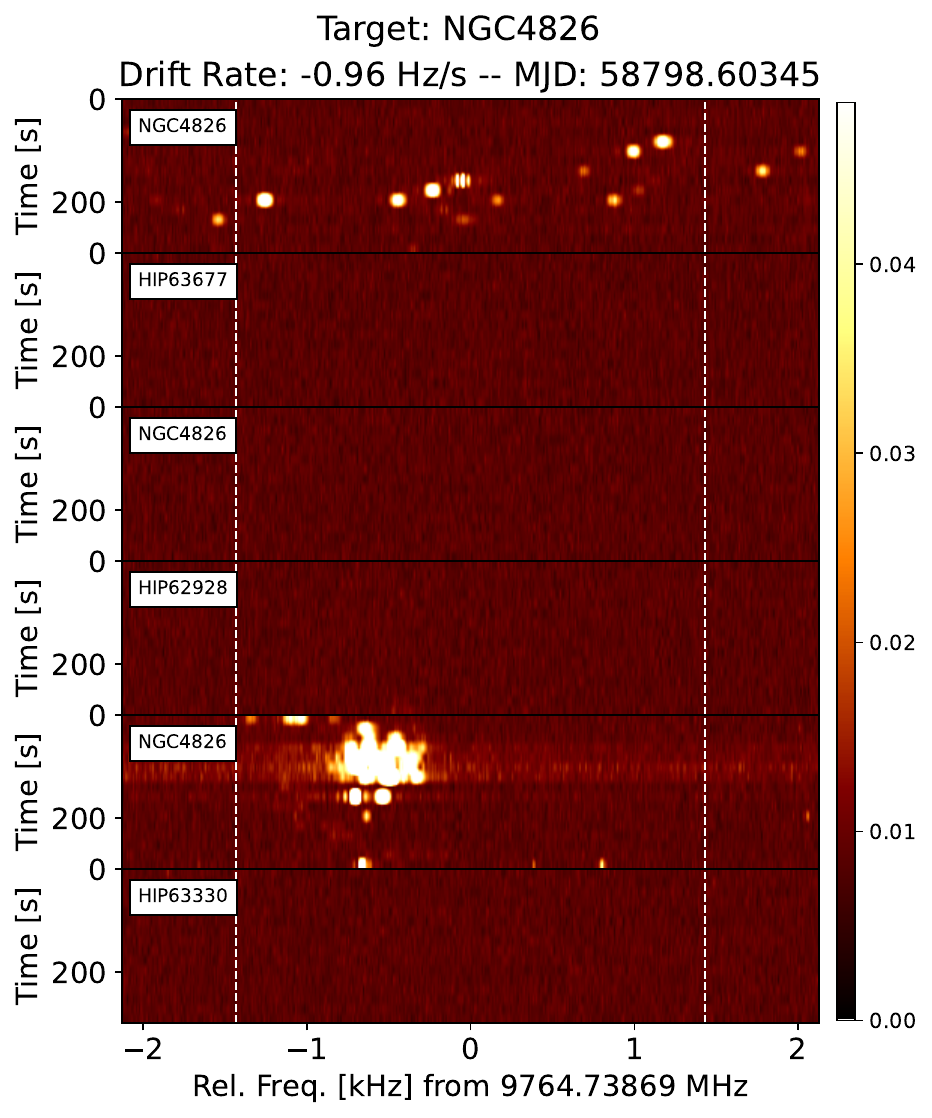}{0.31\textwidth}{(c)}}
\gridline{
\fig{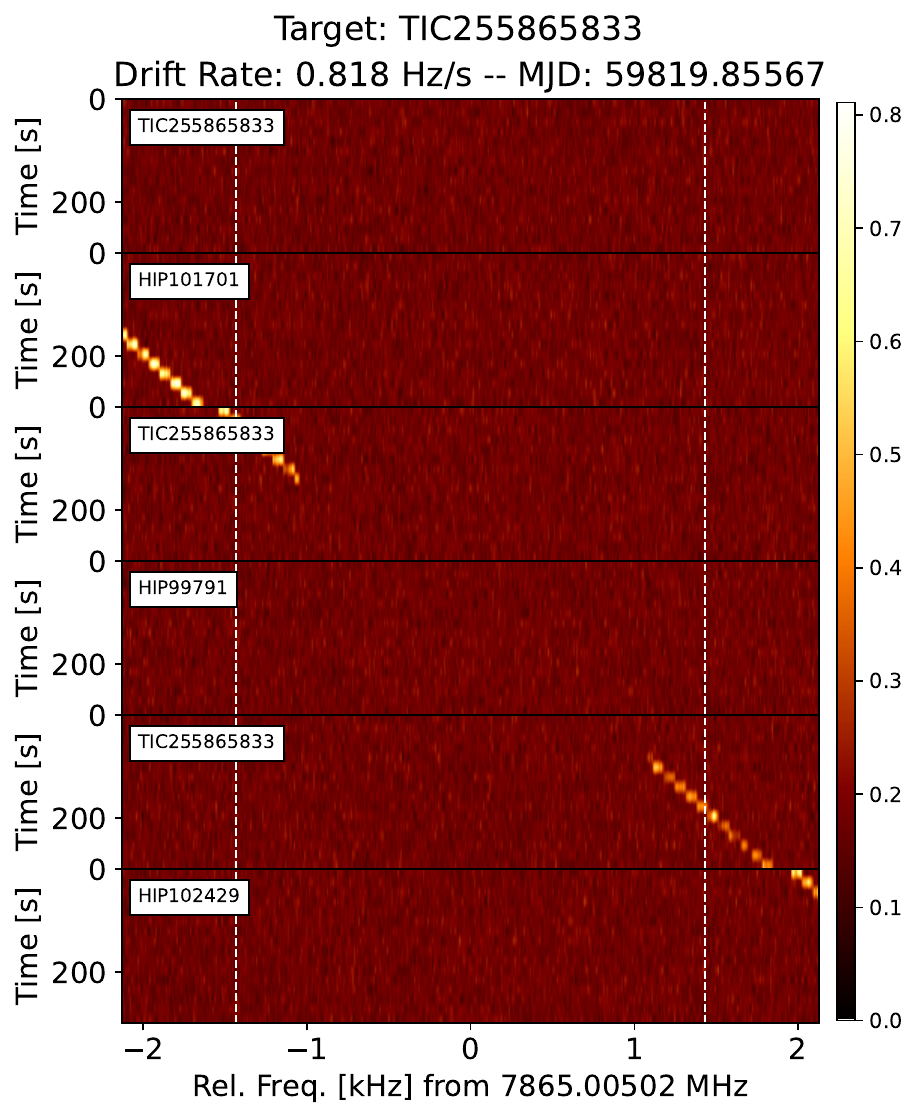}{0.31\textwidth}{(d)}
\fig{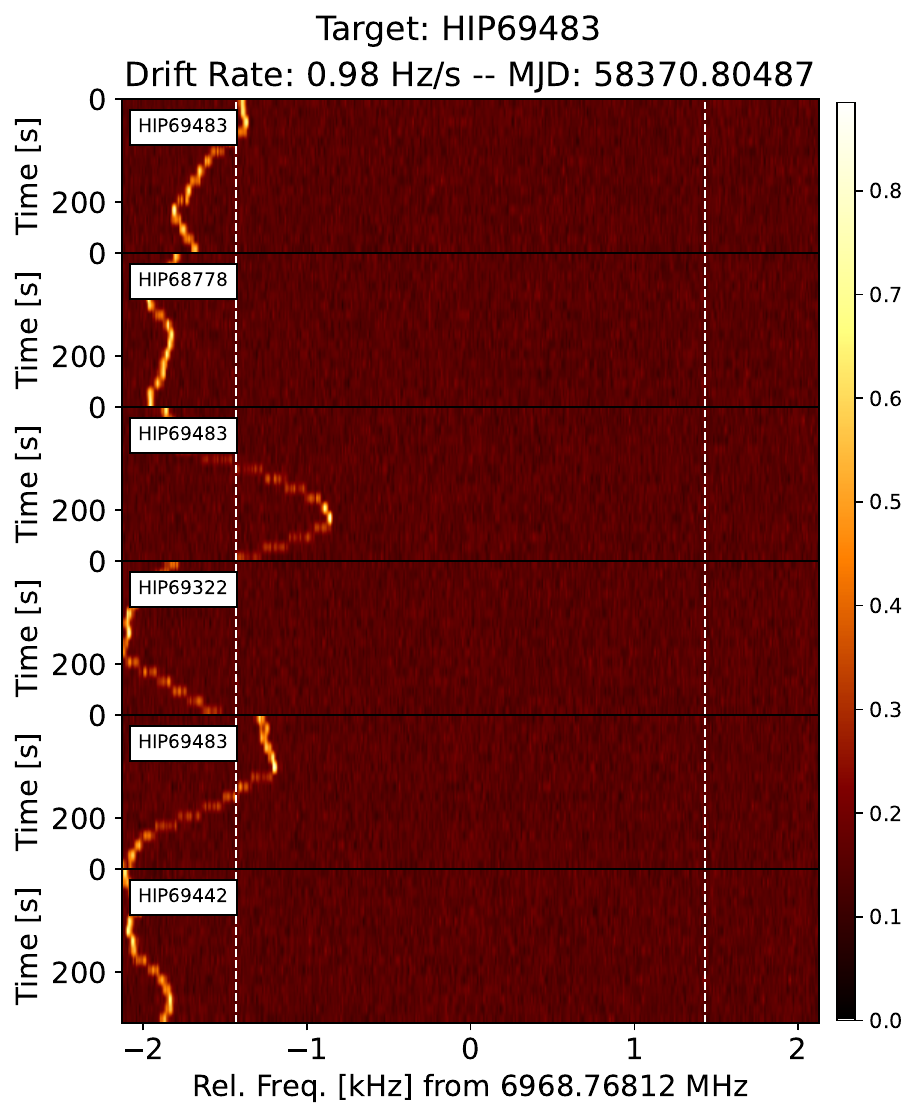}{0.31\textwidth}{(e)}
\fig{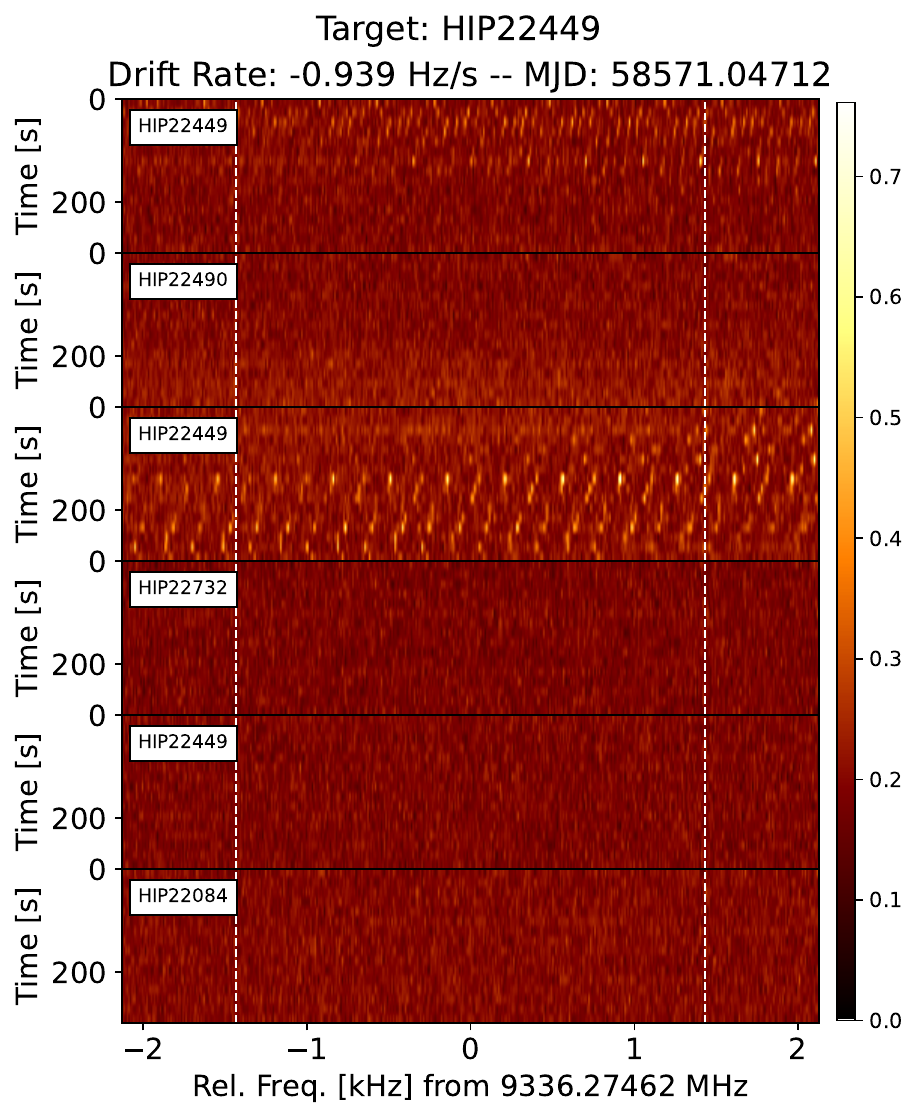}{0.31\textwidth}{(f)}
}
\caption{Common types of false positives that our pipeline deems Level 2 blocks. The drift rates in the heading of each figure are estimates produced by the pipeline, and may not reflect the true drift rate of the signal.}
\label{fig:false_pos}
\end{figure*}

    


\subsection{Candidates Surviving Visual Inspection}\label{sec:candidates}

We find a total of 75 narrowband drifting signals that appear strongly in the ON scans of an observation and are not clearly present in the OFF scans. We highlight six of these candidates in Figure \ref{fig:dedrifted}. These signals showed no sign of any signal in the OFF scans even after de-drifting them. These candidates serve as an example of what our follow up study entails for our candidates. Additional candidates can be found in Figure \ref{fig:candidates} and Figure \ref{fig:candidates2} in Appendix \ref{app:B}.  

GJ\,412B has a candidate signal at 1151.94082\,MHz, and closer inspection reveals no sign of a signal in the OFF scans even after the observations were de-drifted and time-integrated. The closest other signal to this candidate is 38\,MHz away, and is present in all scans. We deem this candidate signal likely to be RFI as a similar signal is found 12\,kHz away from 1151.94082\,MHz in observations of HIP\,62207 taken a month earlier. This other signal drifts with exactly the same drift rate as our candidate, and is also significantly stronger in the ON scans than the OFF scans, although in this case it is weakly visible in the OFF scans. We do not expect that a technosignature would come from two different stars with exactly the same drift rate relative to Earth. 

HIP\,100064 and HIP\,54677 also have candidate signals, which did not appear at all in the OFF scans. These signals were also picked up by \citet{Price:2020}. They were deemed to be RFI due to the presence of similar signals with near identical drift rates at almost the same frequency in observations of other stars. We reach the same conclusion, and the candidate signals in both of these stars occur at exactly 1412.48393\,MHz. 

HIP\,21556 has a candidate signal at 1359.9441\,MHz. The closest signal neighbor to this is 15\,MHz away, and can be attributed to RFI as it appears strongly in the OFF scans. We deem this candidate as likely RFI due to a similar signal with near-identical drift rate is found 85\,Hz away in observations of another target, HIP18280, which also weakens significantly in OFF scans. 

HIP\,99572 has a candidate signal at 1498.44515\,MHz. This signal is in an RFI sparse environment, with no other signals present within 100\,kHz. We deem it attributable to RFI due to similar signals appearing less than 10\,kHz away in multiple other stars, including HIP\,84012, HIP\,54426, HIP\,66207, and HIP\,54677. All of these signals also are strong in the ON scans and weak in the OFF scans of their respective observations, and drift at similar rates.

HIP\,116215 contains a signal at 1814.54601\,MHz, drifting at -0.212 Hz/s. The closest RFI to this signal is 5\,kHz away. Again, we attribute this candidate signal to RFI, as a signal drifting at the same rate is present in observations of HIP\,22044 at the same frequency.  

We carry out a similar vetting process on all of our candidates, including those shown in Figures \ref{fig:candidates} and \ref{fig:candidates2}. Overall, we find that all of our candidates can likely be attributed to RFI.

\subsection{Common False Positives}\label{sec:false_positives}

While we attribute all of the signals we pick up on to RFI, some are more clearly RFI than others. We highlight some of these common false positives in Figure \ref{fig:false_pos}. 

Figures a) and b) display two signal morphologies that our pipeline struggles to vet out. The signal displayed in a) spans multiple frequency channels and is part of a larger structure. The signal in b) is a narrowband signal, with a non-linear drift. Both of these are signals that manifest in both ON and OFF scans, but weaken significantly in the OFF scans or strengthen significantly in the ON scans. As a result, two of the ON scans have very high kurtosis, while all of the OFF scans have low kurtosis. Neither of these signals is excised by our drift, broadband, or blip filters. 

Figures c) and f) displays another common false positive, which manifests as a signal in two of the three ON scans. These signals are not constant in time, and are usually part of a larger structure. Examining a wider frequency range usually reveals parts of this structure to appear in the OFF scans.  

Figures d) and e) show signals whose drift rates cause them to appear only in the ON scans of a given frequency block, while the signals themselves may appear strongly in the OFF scans outside of this frequency block. In figure d) this manifests as a high drift rate and non time-constant signal that can be seen extending into the OFF scans outside of the white dashed line which represent the frequency block our pipeline analyzes. In figure e) the signal's drift rate is extremely variable, and it drifts in and out of the block. 

\begin{figure*}
    \includegraphics[width=0.48\textwidth]{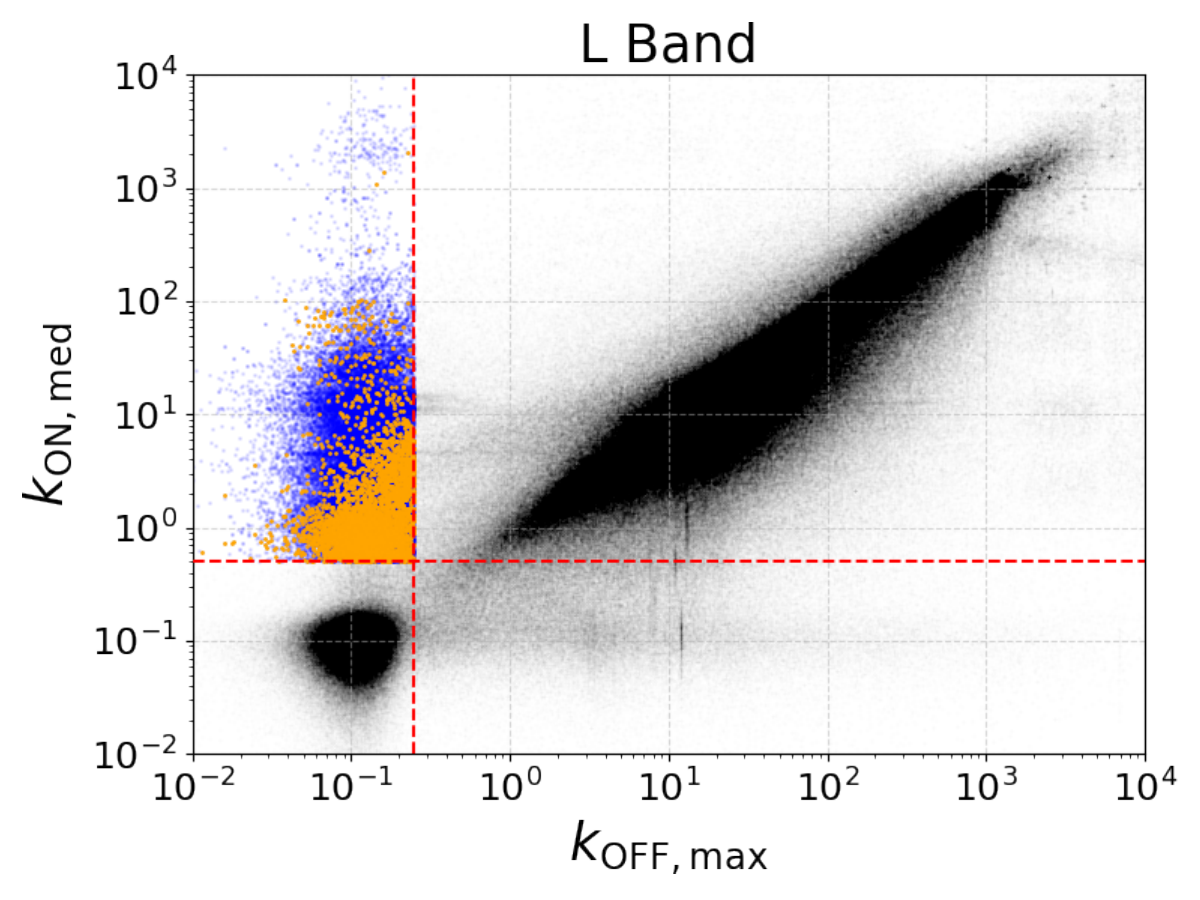}
    \includegraphics[width=0.48\textwidth]{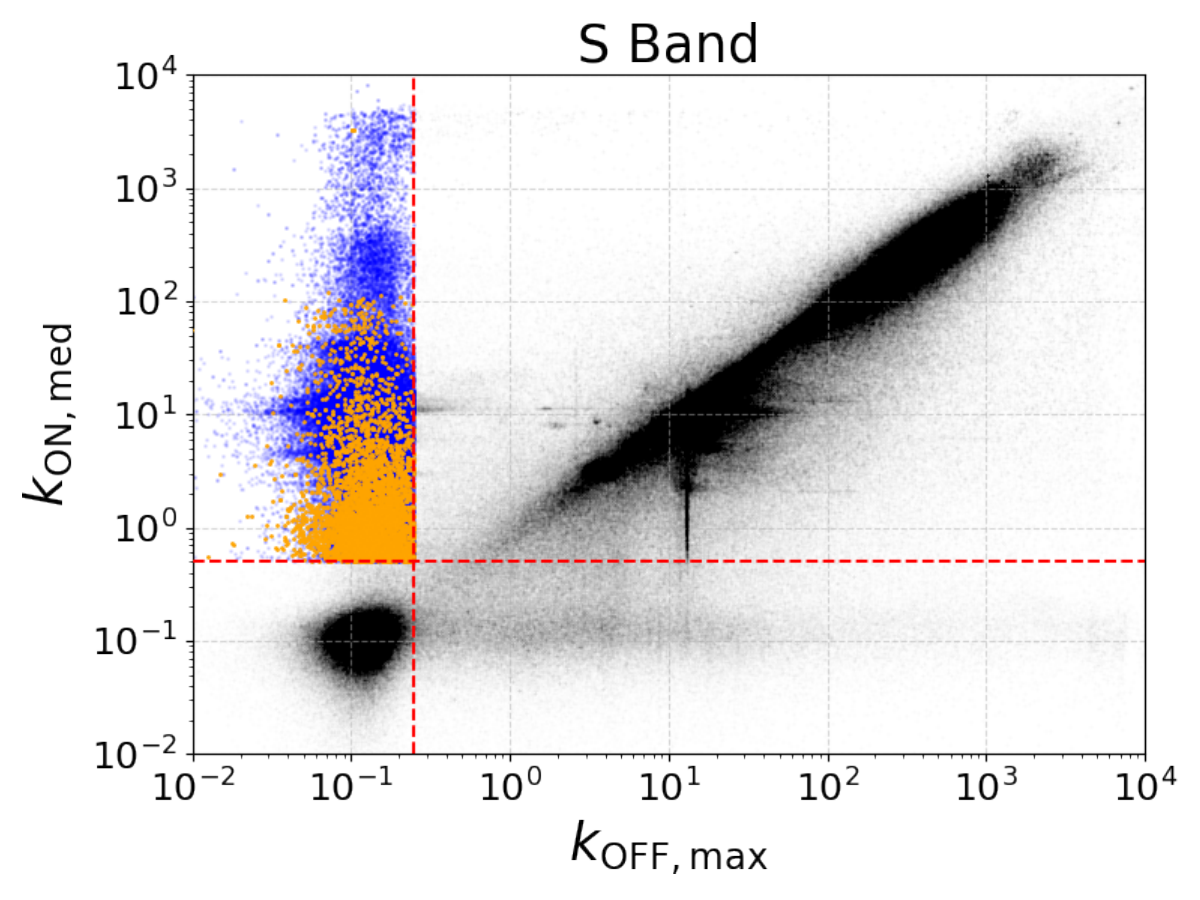}
    
    \includegraphics[width=0.48\textwidth]{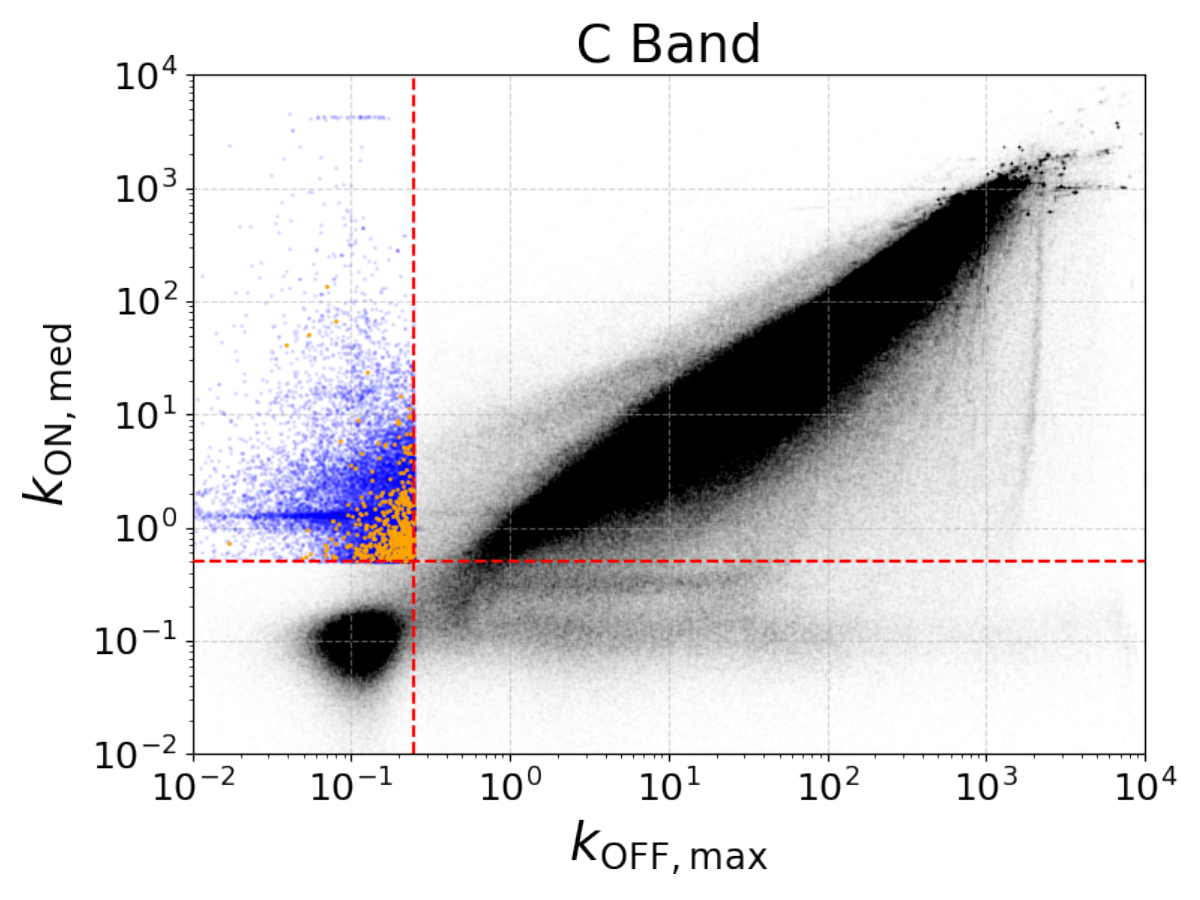}
    \includegraphics[width=0.48\textwidth]{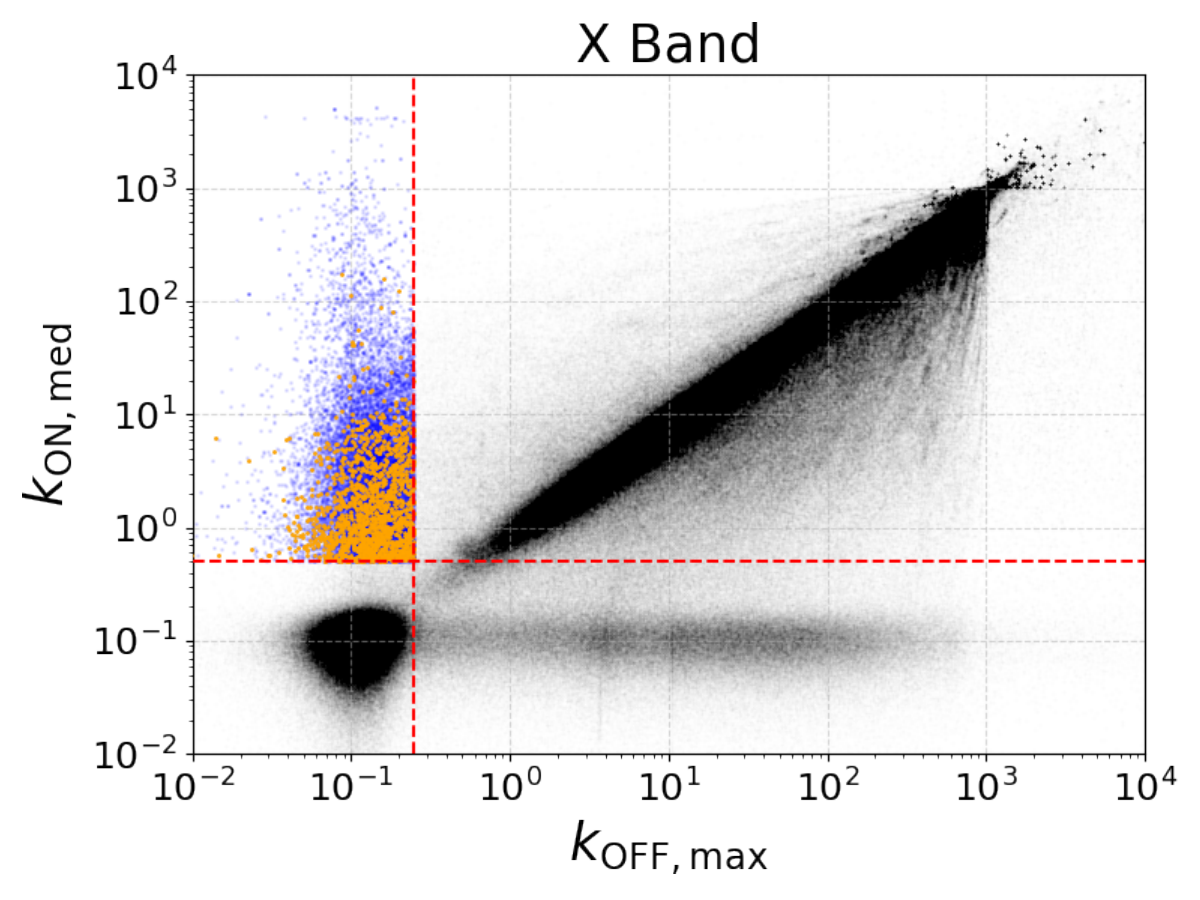}
    \caption{Scatterplot of all blocks analyzed by our pipeline according to their median ON scan kurtosis and maximum OFF scan kurtosis. Level 0 blocks are shown in black, Level 1 blocks in blue, and Level 2 blocks in orange. The dashed red lines represent the kurtosis cut offs described in Section \ref{sec:pipeline_over}. All Level 1 and Level 2 blocks thus fall in the upper left quadrant, which contains blocks with signals that appear strongly in the ON scans and weakly in the OFF scans. The vast majority of blocks fall along the $K_{\rm OFF, max} = k_{\rm ON, med}$ line, as most blocks contain RFI signals that appear equally strongly in all ON and OFF scans.}
    \label{fig:all_ks}
\end{figure*}

\begin{figure*}
    \centering
\centering
    \includegraphics[width=.45\textwidth]{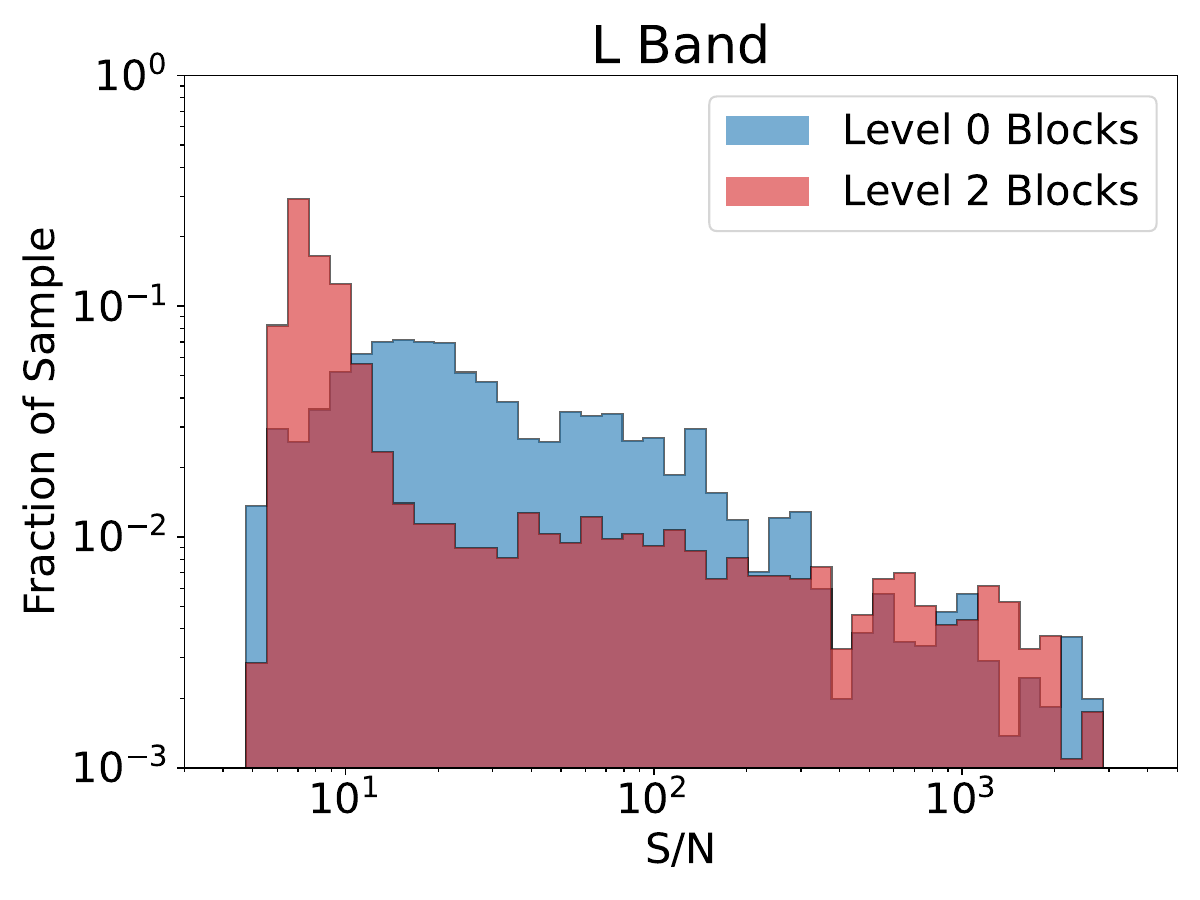}
    \includegraphics[width=.45\textwidth]{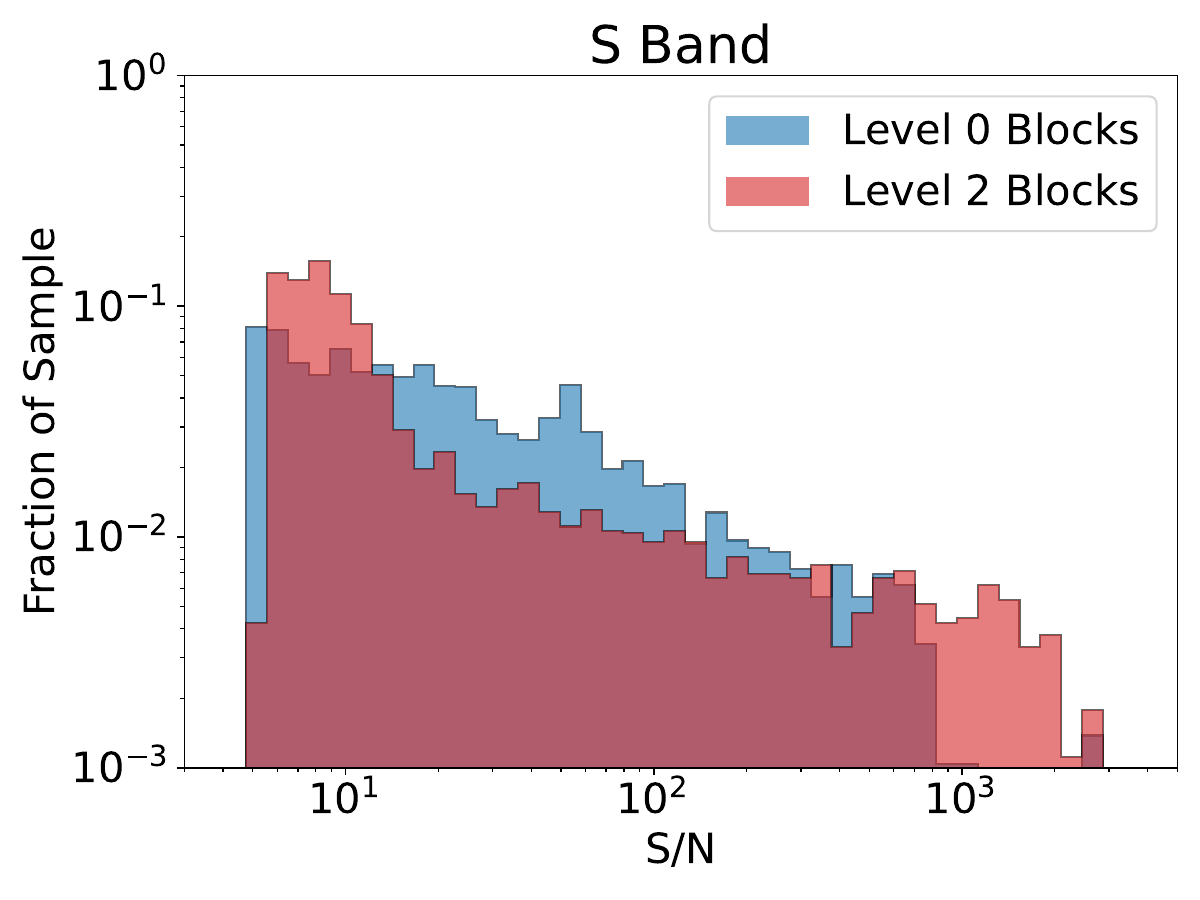}
    \includegraphics[width=.45\textwidth]{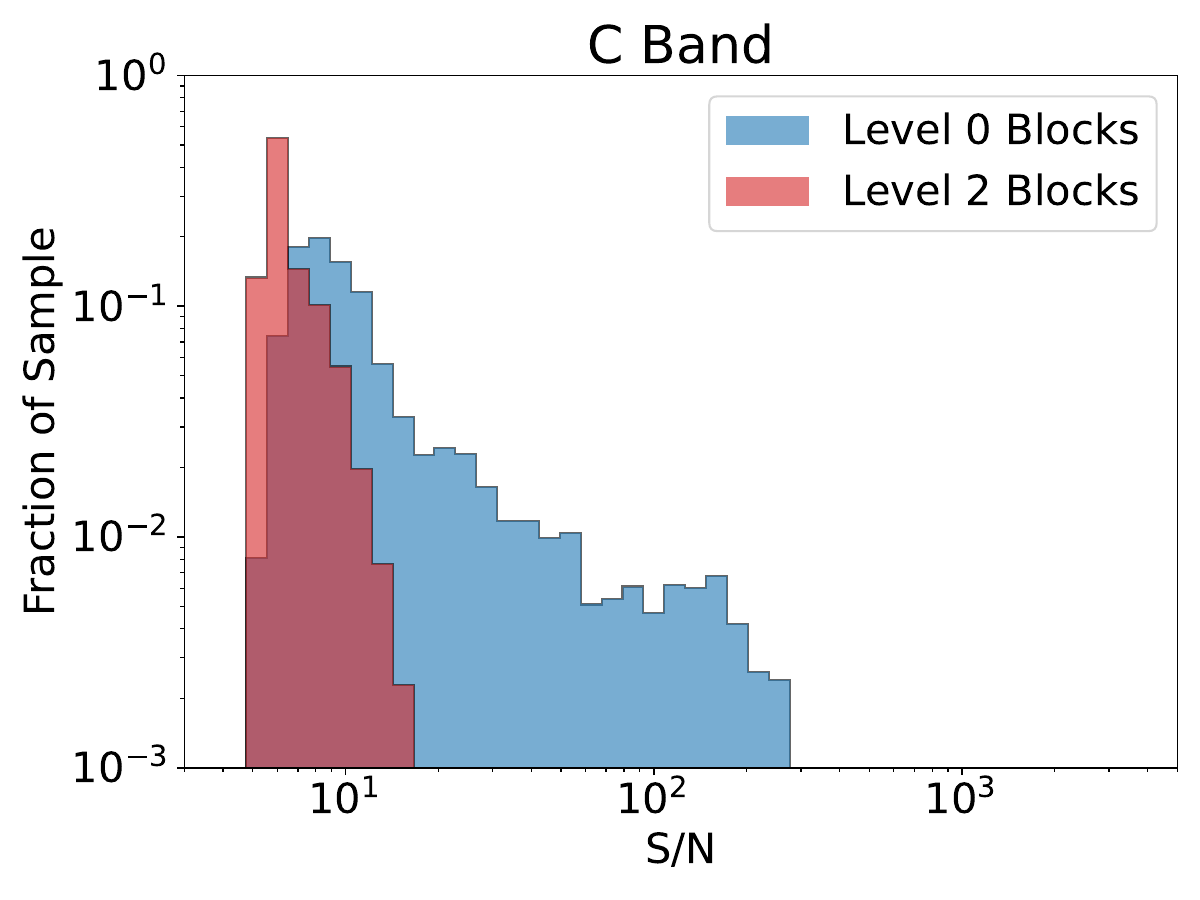}
    \includegraphics[width=.45\textwidth]{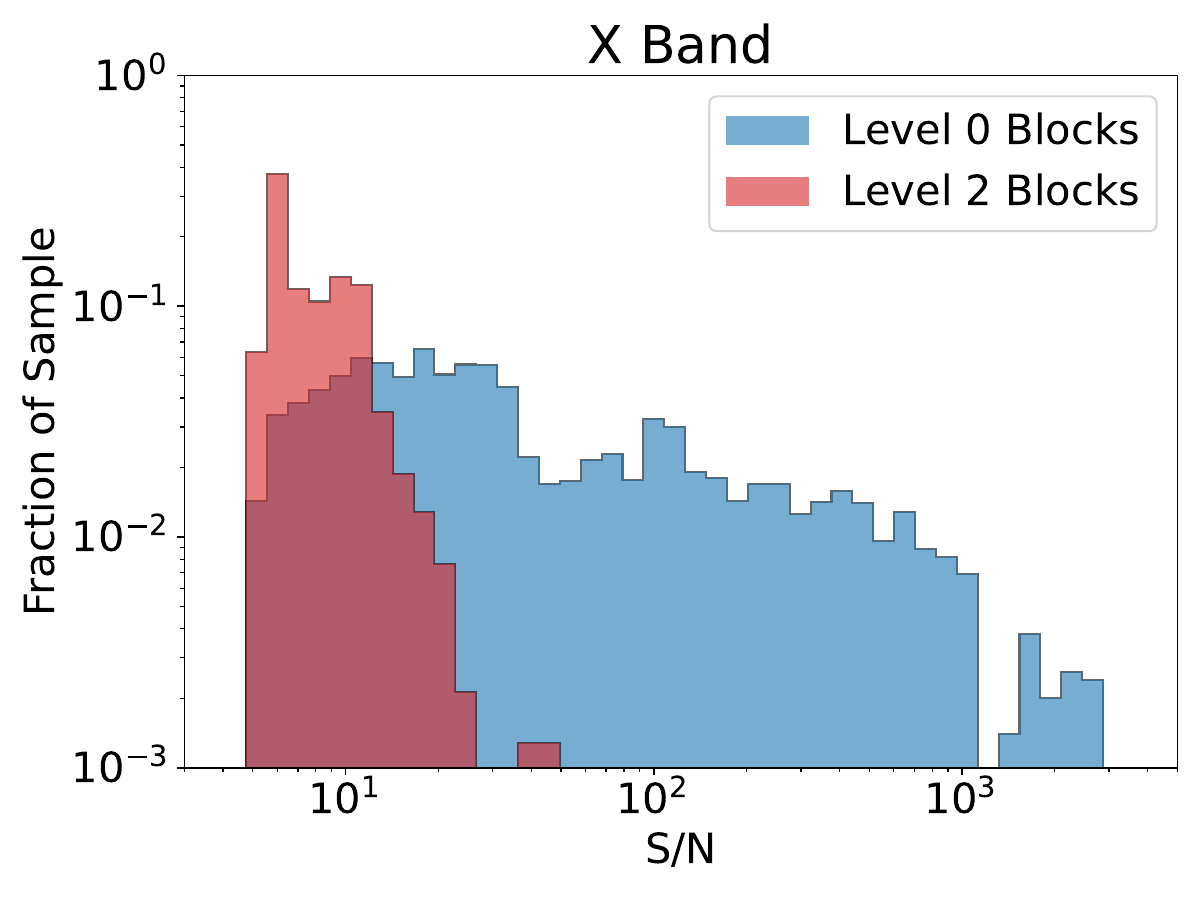}

    \caption{Fractional histograms of SNR for a subsample of all Level 0 blocks (blue) and the full sample of all Level 2 blocks (red) for each GBT receiver band.}
    \label{fig:snr_hist}
\end{figure*}

\section{Discussion}\label{sec:discussion}

\subsection{Comparisons Across bands}

We present a summary of the Level 0, Level 1, and Level 2 blocks across L-band, S-band, C-band, and X-band in Table \ref{tab:pipeline_outputs}. We normalize for observation time across each band, and calculate a block density for each of our levels and each band. This is the number of blocks of a certain level per unit frequency per hour, as shown in Table \ref{tab:densities_hits}. 

A higher block density corresponds to a higher density of RFI. The sensitivity of a telescope band to RFI, and thus the amount of blocks our algorithm detects, is dependent on factors such as the system temperature and the pointing of the telescope. We assume most of these factors are fairly similar across receivers in order to make some basic comparisons between the block densities in each of our bands. 

\begin{table}[t]
\centering
\caption{Summary of Pipeline Results\label{tab:pipeline_outputs}}
\begin{tabular}{ccccc}
\hline
\hline
band & Freq Processed & Level 0  & Level 1  & Level 2 \\
& [MHz] & Blocks & Blocks & Blocks\\
\hline
L & 455 & 4,196,680 & 101,568 & 6,131 \\
S & 710 & 3,392,299 & 33,308 & 3,382\\
C & 3280 & 8,423,757 & 21,751  & 2,838\\
X & 2600 & 8,355,979 & 18,533 & 1,817\\
\hline
Total & 1.10--11.20 & 24,368,715 & 175,160 & 14,168\\
\hline
\end{tabular}
\end{table}

The highest block densities occur at L-band. This is primarily attributed to the higher number of satellites transmitting at 1.1 -- 1.9\,GHz. The density of Level 1 blocks is more than 3 times as high as in S band, and $\sim 20$ times as high as the density of Level 2 blocks at C and X band. The fraction of Level 0 blocks that make it to Level 2 is $1.46 \times 10^{-3}$ for L band, $1.0 \times 10^{-3}$ for S band, $3.4 \times 10^{-4}$ for C band, and $2.1 \times 10^{-4}$ for X band. A much higher percentage of the RFI signals at L band weaken significantly in their respective OFF scans than in other bands.  This may be due to the areal density and motion of satellites at L band, particularly GPS, having a higher tendency to cause RFI with a cadence around 10 minutes, and thus potentially appearing in successive ON scans.

\begin{table}[t]
\centering
\caption{Band Level Density\label{tab:densities_hits}}
\begin{tabular}{ccccc}
\hline
\hline
 & L & S  & C  & X  \\
\hline
Level 0 Density & 8.7 & 6.3 & 4.6 & 4.1 \\
$[$\textrm{Blocks/Hz/Hr]} &  &  &  &  \\
Level 1 Density & 209.4 & 61.6 & 11.8 & 9.2  \\
$[$\textrm{$10^{-3}$ Blocks/Hz/Hr}$]$ &  &  &  &  \\
Level 2 Density & 12.6 & 6.3 & 1.5 & 0.9\\
$[$\textrm{$10^{-3}$ Blocks/Hz/Hr}$]$ &  &  &  &  \\

\hline
\end{tabular}
\end{table}

We also examine the spread in signal strength among the Level 0 and Level 2 blocks in each band. In L, S, and X band, the spread in signal-to-noise for signals in Level 0 blocks ranges from 5 up to $\sim 10^3$, whereas the signal-to-noise of signals only goes up to $\sim 3 \times 10^2$ in C band. The spread of signal-to-noise for signals in Level 2 blocks follows the spread of Level 0 blocks closely in L and S band. In C and X band, the majority of Level 2 blocks have signals with SNR $<20$. We note that while we assume an SNR threshold of 10 when calculating the sensitivity of our search in Section \ref{sec:fom}, in practice our pipeline appears sensitive to signals with signal-to-noise ratios as low as $\sim 6$, as shown by Figure \ref{fig:snr_hist}. However, we do not pick up on signals with SNR $<10$ with high completeness. We calculate the SNR of a signal by taking its mean over all time bins. While these signals need SNR $\geq 10$ in their eighth time bin in order to be detected by our pipeline, many of these signals are weaker in the other time bins of an ON scan, leading to an overall lower SNR for these signals.

\subsection{Parameter Space Probed and Common Signal Morphologies}

We examine the distribution in parameter space that the signal blocks we record fall in. While we record many parameters for each frequency block, the most important ones in our algorithm are the median kurtosis of the ON scans, and the max kurtosis of the OFF scans (as discussed in Section \ref{sec:results}). The values of each of these parameters for all of the frequency blocks we analyze is shown in Figure \ref{fig:all_ks}. We estimate the SNR of the signals found in these frequency blocks, and plot them in Figure~\ref{fig:snr_hist}.

\begin{figure}
\centering
    \includegraphics[width=0.45\textwidth]{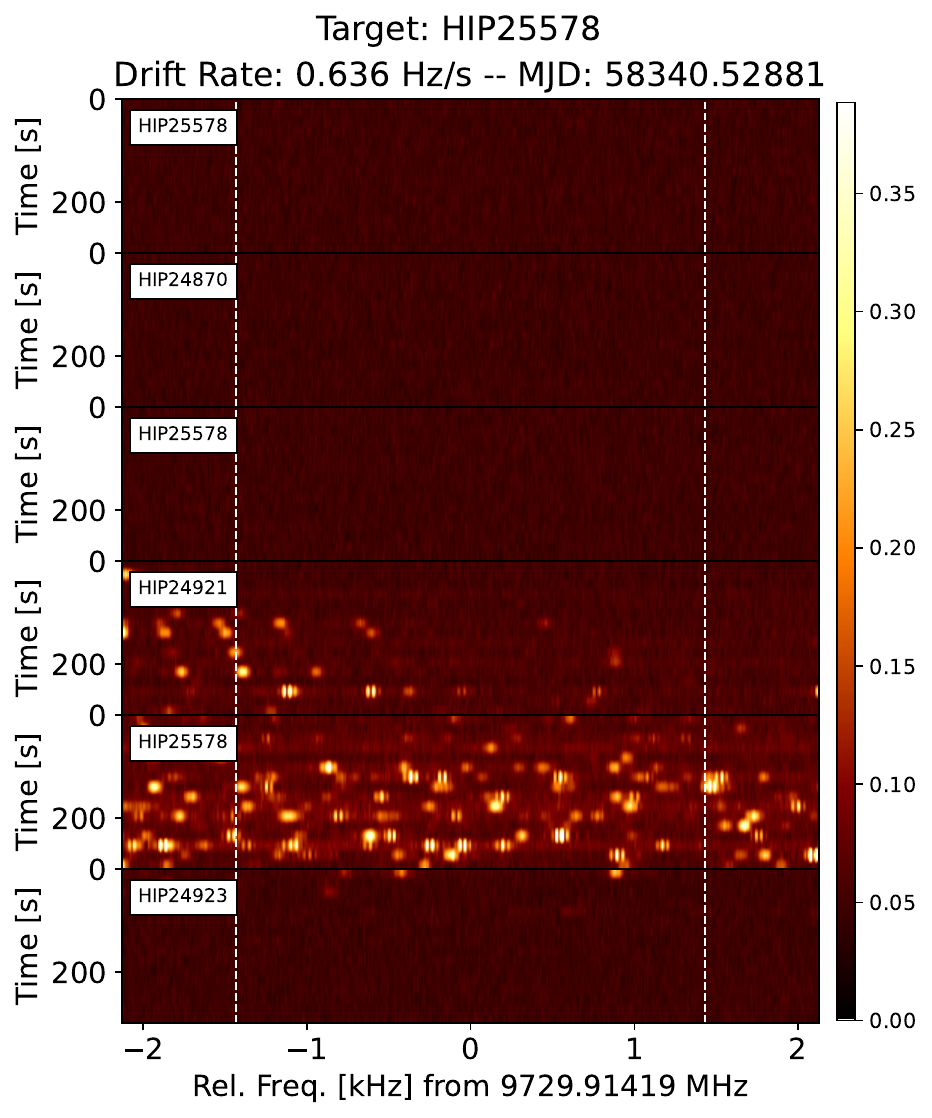}
    \caption{Example Level 0 Block.}
    \label{fig:common_morphs}
\end{figure}

There are some common features across the kurtosis-space frequency block distribution for each band. The vast majority of frequency blocks fall along the $k_{\rm ON, med} = k_{\rm OFF, max}$ line. This is what is to be expected, as the vast majority of frequency blocks contain RFI signals that are equally strong throughout both the ON and OFF scans. Another common feature visible in each band, and especially clear in X band, is a horizontal cluster around $k_{\rm ON, med} \approx 10^{-1}$. Frequency blocks in this region are typically those with some signal in only one of the ON observations, and varying amounts of that signal present in the OFF scans. As a result, the median ON scan kurtosis is very low, and the OFF scan kurtosis can vary depending on how much of the signal contaminates the OFF scans. An example of such a signal can be seen in Figure \ref{fig:common_morphs}. Another common type of block contains some signal in one of the ON scans (commonly a broadband pulse or blip) and no signal in the rest of the ON or OFF scans. As a result, both $k_{\rm ON, med}$ and $k_{\rm OFF, max}$ are very low, producing the small clump in the lower left of each plot. 

We also examine the  distribution of the pointing directions of the observations containing our candidates and Level 2 blocks. We plot the altitude and azimuth of of these pointings in Appendix \ref{app:A}, and find that many of our Level 2 blocks originate from L band observations of sources located at an azimuth of $210\degr \pm 35\degr$. A large fraction of our candidates also originate from this region. This could correspond to a large cluster of geosynchronous satellites as viewed from the GBT. We carry out a search to match our candidate signals to known satellites whose paths pass near to the observation pointing, but are unable to succesfully confirm that any of our candidate signals come from specific satellites.

\subsection{Figures of Merit}\label{sec:fom}

It is difficult to quantify the thoroughness of a SETI search, due to the large parameter space covered and both known and unknown biases in target sampling. Various figures of merit (FoM) have been developed to assess the rigor of searches and allow for comparison between them. For calculations of all these FoMs in this section, we count only individual stars as targets, and do not include observations of many-star systems, such as globular clusters or galaxies. We also do not include the background fields encompassed by the GBT beam. As a result, the total stellar number is likely underestimated, although still a good representation of the number of targets we observe at the set distances we choose in Table {\ref{tab:current_searches}}.   

One of the earliest attempts at such a FoM is the Drake Figure of Merit, given by

\begin{figure*}
    \centering
\centering
    \includegraphics[width=1\textwidth]{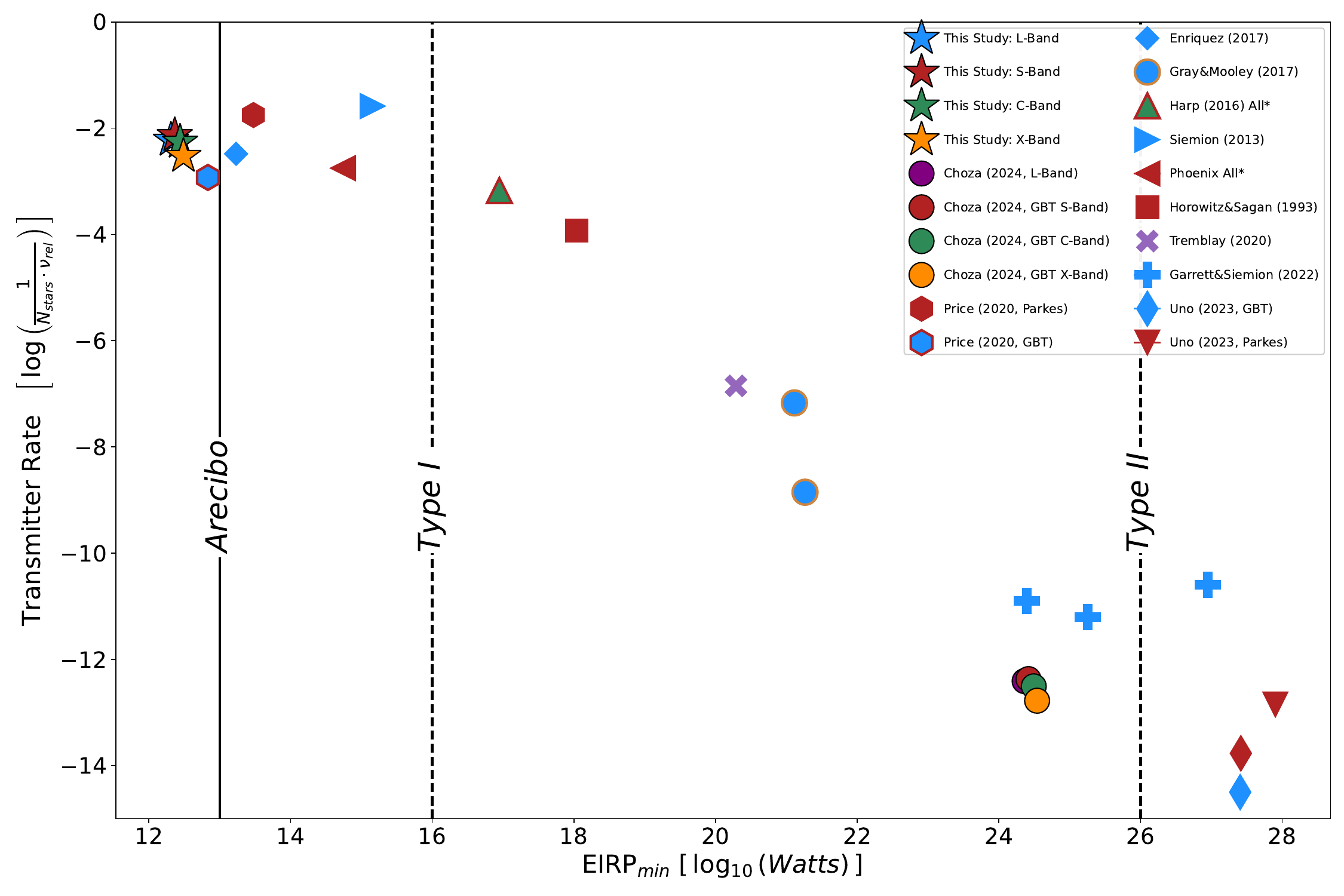}
    \caption{A comparison of transmitter rate vs. EIRP$_{\rm min}$ for this work with previous searches, where blue, red, green, and yellow stars correspond to searches of the L, S, C,
and X bands respectively. The three vertical lines mark the EIRP of the Arecibo S-band radar, the power level of a Kardashev Type I civilization, and the total power
budget of a Kardashev Type II civilization. The results of \citet{Enriquez},  \citet{Price:2020}, \citet{GarrettSiemion:2022}, and \citet{Uno_2023} are corrected for findings by \citet{choza:2024} that a \texttt{turboseti} search run with a SNR threshold of 10 in fact is only sensitive to signals with a SNR $\gtrsim 33$. Other results shown come from \citet{Tremblay:2020}, \citet{Harp:2016}, \citet{Siemion:2013}, \citet{Phoenix:2004}, \citet{Horowitz:1993}. }
    \vspace{.5cm}
    \hspace{-.8cm}
    \label{fig:eirp}
\end{figure*}
\begin{equation}
    {\rm DFM} = \frac{n \Delta \nu_{\rm tot} \Omega}{F_{\rm min}^{3/2}}
\end{equation}

where $n$ is the number of sky pointings, $\Delta \nu_{\rm tot}$ is the observing bandwidth, $\Omega$ is the sky coverage, and $F_{\rm min}$ is the minimum detectable flux in W m$^{-2}$. The -3/2 index comes out of the distance-to-volume scaling ($d^3$) and sensitivity scaling ($d^{-2}$). Since our search covers multiple bands each with their own $F_{\rm min}$ and DFM, we also calculate the Total Drake Figure of Merit, defined as 

\begin{equation}
    {\rm DFM}_{\rm tot} = \sum _i^4 {\rm DFM}_{\rm i},
\end{equation}

where we sum the respective DFM of each band. The minimum detectable flux, $F_{\rm min}$, is a key metric for the sensitivity of a radio-frequency SETI experiment, and is set primarily by the system noise and effective collecting area of the telescope. These properties are encapsulated in the system equivalent flux density (SEFD):

\begin{equation}
    {\rm SEFD} = \frac{2k_{\rm B}T_{\rm sys}}{A_{\rm eff}},
\end{equation}

where $k_{\rm B}$ is the Boltzmann constant, $T_{\rm sys}$ is the system
temperature, and $A_{\rm eff}$ is the effective collecting area. The effective collecting area is a scaling of the physical collecting
area of the telescope by some efficiency factor $\eta$ between 0 and 1. The SEFDs of GBT's L, S, C, and X bands are reported in Table \ref{tab:current_searches}. 

The SEFD can be used to calculate the minimum detectable flux density for a given receiver. In the case of a  narrowband signal (one where the transmitter signal bandwidth
is narrower or equal to the observing spectral resolution, as is the case in SETI technosignature searches), the minimum detectable flux density $F_{\rm min}$ is then given by

\begin{equation}
    F_{\rm min}= \mathrm{SNR}_{\min }\mathrm{SEFD} \sqrt{\frac{B}{\mathrm{n}_{\mathrm{pol}} \tau_{\mathrm{obs}}}},
\end{equation}

where $B=\delta \nu$ is the channel bandwidth,   $n_{\rm pol}$ is the number of polarizations, and $\tau_{\rm obs}$ is the observing time.  The minimum detectable flux $F_{\rm min}$ is related to the minimum detectable flux density $F_{\rm min}$ = $S_{\rm min}/\delta \nu_{\rm t}$ , where $\delta \nu_{\rm t}$ is the bandwidth of the transmitting
signal. In this work we have chosen a value of unity for $\delta \nu_{\rm t}$, in keeping with previous studies \citep{Price:2020}. 

Since our search used four different GBT receivers with varying fields of view, system temperatures, and bandwidths, we compute a combined $\rm{DFM}_{\rm tot}=\sum_i^4\rm{DFM}_{\rm i}$ which is simply the sum of each band's DFM. A larger DFM in theory represents a more sensitive search. The $\rm{DFM}_{\rm tot}$ for this study is $4.8 \times 10^8$ Hz Jy$^{-3/2}$, $3.1\times$ as large as that of \citealt{Price:2020}.

As has been noted in previous SETI searches, one issue with the DFM as a comparison tool is that it fails to take into account the distance to survey targets. An alternative heuristic was proposed by \citet{Enriquez} to take this into account; the Continuous Waveform Transmitter Rate Figure of Merit (CWTFM) which is defined as

\begin{equation}
    \mathrm{CWTFM}=\eta \frac{\mathrm{EIRP}_{\min }}{N_{\mathrm{star}}} \frac{\nu_c}{\Delta \nu_{\mathrm{tot}}}.
\end{equation}

In this formulation, $\nu_c$ is the central observing frequency, $\Delta \nu_{\rm tot}$ is the observing bandwidth, $N_{\rm star}$ is the number of stars observed, $\rm{EIRP}_{\rm min}$ is the minimum detectable equivalent isotropic radiated power in watts, and $\zeta_{\rm AO}$ is a normalization factor set such that the CWTFM is 1 for an EIRP equal to that of Arecibo. The value $\Delta \nu_{\rm tot}/N_{\rm star}\nu_c$ encompasses both the fractional bandwidth and the number of sources, and is referred to as the transmitter rate. The value of EIRP$_{\rm min}$ can be calculated as 

\begin{equation}
    {\rm EIRP}_{\rm min} = 4\pi d^2 F_{\rm min}
\end{equation}

Lower TFM scores represent more sensitive and more complete surveys. Our TFM values for each band are listed in Table \ref{tab:current_searches}. The total TFM score of this four-band survey is $3.2\times$ smaller than the two-band search undertaken by \citet{Price:2020}.

\begin{table*}
\renewcommand{\arraystretch}{1.2}
\setlength{\tabcolsep}{12pt}
\caption{Search Parameters and Figures of Merit}\label{tab:current_searches} 
\makebox[\textwidth]{%
\hspace{-2cm}
\begin{tabular}{|c|c|c|c|c|}
\hline 
Band & L band & S band & C band & X band \\
\hline 
\multicolumn{5}{|c|}{TELESCOPE PARAMETERS} \\
\hline 
Antenna Diameter ( $D$ ) & \multicolumn{4}{|c|}{100} \\
\hline 
Beam Width $(\theta)$  & 8 & 4.5 & 1.7 & 1  \\
\hline 
Aperture Efficiency $(\eta)$ & 0.72 & 0.68 & 0.65 & 0.63  \\
\hline 
System Temperature ( $T_{\text {sys }}$ ) [K] & 20 & 22 & 25 & 27  \\
\hline 
\multicolumn{5}{|c|}{SEARCH PARAMETERS} \\
\hline 
Number of stars $< 50$ pc & 798 & 464 &  328 & 695  \\
\hline 
Number of stars $< 1000$ pc& 1484 & 1046  & 821 & 1239  \\
\hline 
Stellar Spectral Types & All & All & All & All  \\
\hline 
SNR Threshold & 10 & 10 & 10 & 10 \\
\hline 
Spectral Resolution $(\delta \nu)[\mathrm{Hz}]$ & 2.8 & 2.8 & 2.8 & 2.8 \\
\hline 
Band Frequency [GHz] & $1.1 - 1.9$  & $1.8 - 2.8$ & $4.0 - 7.8$ & $7.8 - 11.2$ \\
\hline
Coverage [MHz] [\% of band] & 455 (57\%) & 710 (71\%) & 3280 (86\%) & 2600 (76\%) \\
\hline 
\multicolumn{5}{|c|}{FIGURES OF MERIT} \\
\hline 
EIRP$_{\rm min}$(50 pc) & $2.05 \times 10^{12}$ & $2.32 \times 10^{12}$ & $2.76 \times 10^{12}$ & $3.07 \times 10^{12}$  \\
\hline
EIRP$_{\rm min}$(1000 pc) & $8.2 \times 10^{14}$ & $9.3 \times 10^{14}$ & $1.1 \times 10^{15}$ & $1.2 \times 10^{15}$  \\
\hline
CWTFM (50 pc) & 0.61 & 0.85 & 0.76 & 0.46 \\
\hline
DFM & $2.15 \times 10^8$ & $9.52 \times 10^7$ & $9.8 \times 10^7$&  $6.8 \times 10^7$ \\
\hline 
\end{tabular}
}
\begin{flushleft} 
\end{flushleft}
\end{table*}

\subsection{Transmitter Limit}

We find no signs of narrowband transmitters from
observations of the stars in our sample above the EIRP$_{\rm min}$ values of $2-3\times10^{12}$ W for stars within 50\,pc, and EIRP$_{\rm min}$ values of $8-12\times10^{14}$\,W for stars within 1000\,pc.  

Determining limits on the existence of possible transmitters aimed at star targets is challenging due to factors like RFI, potential transmission intermittency or periodicity, and our pipeline's limitations (see Section \ref{sec:limitations}). However, we are able to establish a probabilistic upper limit on the likelihood of continuous narrowband transmitters above EIRP$_{\rm min}$, assuming such transmitters are rare. Specifically, we can calculate a conditional probability of detecting a signal, should it exist above EIRP$_{\rm min}$ and within the observed band, treating each star target as an independent trial in a Poisson distribution. 

We take the number of stars in each band closer than 50\,pc from Table \ref{tab:current_searches}. Following \citet{choza:2024}, we find upper limits on the fraction of stars with narrowband transmitters above the EIRP$_{\rm min}$. We calculate upper limits of $0.38\%$ for L band (EIRP$_{\rm min}=2.1\times10^{12}$),  $0.64\%$ for S band (EIRP$_{\rm min}=2.3\times10^{12}$),  $0.91\%$ for C band (EIRP$_{\rm min}=2.8\times10^{12}$), and $0.43\%$ for X band (EIRP$_{\rm min}=3.1\times10^{12}$). We note that these transmitter limits only apply to the portion of each band that we include in our analysis. 

We also calculate upper limits for transmitter strengths corresponding to a distance of 1000\,pc. We use the EIRP$_{\rm min}$ values reported in Table \ref{tab:current_searches}. We find upper limits of $0.20\%$ for L band,  $0.29\%$ for S band,  $0.37\%$ for C band, and $0.24\%$ for X band.

\subsection{Comparison to Previous Work}\label{sec:comparisons}

Multiple previous studies have analyzed GBT data as part of SETI surveys. The largest star-sample survey as of this work is \cite{Price:2020}, which analyzed 1,327 nearby stars across GBT's L and S band, and the 10-cm receiver on the 64-m
CSIRO Parkes telescope. Not all of the observations in \cite{Price:2020}'s sample are also in ours. Specifically only 981 of their L band observations make it into our sample. However, all of the signals they highlight as interesting come from stars whose observations we also analyze. In total, they showcase 8 of their best candidates, coming from 7 different stars. Of these 7 stars, our pipeline also identifies all of these sources as containing Level 2 blocks. 

However, we are not able to pick up on all of their specific top ranked signals. Our pipeline only classifies 3 of their 8 signals as Level 2. For two of these candidates, this is because they share their frequency block with contaminating RFI (HIP100064 at 1595.87757\,MHz and HIP1444 at 1449.98475\,MHz). This increases the kurtosis of the OFF observations, and can also cause the frequency block to be thrown out if the RFI is not drifting. Another two signals (in HIP91699 at 2200.05036\,MHz and HIP22845 at 2200.040568\,MHz) occur at frequencies we deem RFI dense and do not analyze. Another signal (HIP109176 at 2414.52441\,MHz) exhibits broadband behavior, which our algorithm filters out. While we do not pick up on all of the signals highlighted by \citet{Price:2020}, our pipeline does capture several dozen Level 2 blocks which contain equally if not more interesting candidate signals. In fact, all of the signals shown in Figures \ref{fig:candidates} and \ref{fig:candidates2} come from observations in \citet{Price:2020}'s survey. This further highlights the need for novel search methods in picking up strong candidate signals.    

\cite{Ma_2023} carried out a SETI search on 820 stars observed with the GBT L-band. They used a $\beta$-Convolutional Variational Autoencoder as opposed to traditional pipelines such as \texttt{turboseti}. They trained their autoencoder on synthetic signals generated with \texttt{setigen}. They identified eight candidate signals from five different targets. All five of these targets are picked up on by our pipeline as containing Level 2 blocks in their observations. 

We do not detect three of the specific candidates found by \citet{Ma_2023}: MLc1 (in HIP13402 at 1188.53923\,MHz), MLc4 (in HIP54677 at 1372.98759\,MHz), and MLc5 (in HIP54677 at 1376.988694\,MHz). MLc1 has $k_{\rm ON, med}=0.1$, which is not strong enough to meet the criteria we imposed. MLc4 and MLc5 are located at frequencies that we filter out due to being in RFI dense environments, but meet the kurtosis requirements for a Level 2 signal. While we do not find these three specific signals, we find $\sim20$ similar quality and morphology candidates in both of the sources these signals come from that. Though these were likely also found by \citet{Ma_2023}, they were not specifically highlighted. 

While we do not recover $100\%$ of the signals highlighted by previous searches, we find many signals of comparable quality from the same sources. In addition, we believe that our pipeline would be successful at finding many of these signals with a slightly modified set of tuning parameters. Due to the size of our search and the quantity of data we analyzed, we imposed strict restrictions on the strength of the signal, the drift rates we searched over, and the the frequency ranges we chose to cut out. Our current false positive rates are $<1\%$, and weaker signals could be picked up on at the expense of a slightly higher false positive rate if the $k_{\rm ON, med}$ cutoff is lowered. Additionally, if a smaller number of observations is being analyzed, a smaller fraction of the frequency coverage could be blocked out. The primary reason we remove it is to speed up computation time for the large number of files we process. We find that we successfully recover the signals found by \citet{Price:2020} and \citet{Ma_2023} in regions we deem RFI dense if we do not cut these regions out preliminarily. However, we also note that a large fraction of these signals that occur in RFI dense regions deserve to be cut. For example, two of \citet{Price:2020}'s candidates occur at 2200\,MHz, a common frequency for satellite communications. Additionally, our pipeline uncovers additional high-quality candidate signals in the dataset used by \citet{Price:2020} that they did not find using \turboseti. 

\subsection{Pipeline Limitations}\label{sec:limitations}

While we detect many candidates that appear interesting upon visual inspection, we acknowledge several limitations in our current pipeline. We also suggest potential ways to address these limitations, although leave the implementation of these approaches to future work.

Our algorithm cannot find ET signals if they occur in especially dense RFI regions. Specifically, if there is persistent RFI within $\sim 1.5$\,kHz of a candidate signal, the frequency block with the candidate will not pass from Level 0 to Level 1, as the kurtosis of the OFF observations will be large due to the RFI. In these cases, a more sophisticated approach to comparing the intensity distributions of the ON and OFF scans could be developed. One potential method would be to compute the Pearson cross-correlations between the ON scans and also between the OFF scans, and highlight frequency blocks with high correlations between all of their ON scans but low correlation between their ON and OFF scans. Another way to address this problem would be to run successive iterations of the pipeline using varying block sizes. Smaller block sizes would make it more sensitive to signals in RFI dense regions, provided they are not drifting too strongly. 

Another limitation is that we have excluded a signficant portion of the GBT's frequency bandwidth from analysis. Only 70.5\% of the total frequency coverage from $1.1 - 11.2$\,GHz is passed into our pipeline. We remove these high-RFI regions as our pipeline performs poorly in especially dense RFI regions (as noted above), and because this $30\%$ of the total frequency coverage accounts for $\sim95\%$ of all signals which would be computationally prohibitive to analyze. These regions can easily be included for analysis, though at the expense of compute time, which may be a worthwhile trade-off for smaller surveys. 

A third limitation is bias towards certain types of signals. While our kurtosis criteria do not strongly preference any specific signal morphology, our additional vetting filters do to some extent. Our broadband filter tends to select signals that do not vary dramatically over time. While this is necessary to reduce the large number of broadband false positives, it could also reject genuine signals that are intermittent. Despite this, our algorithm finds a wider variety of signal morphologies than previous SETI searches focused on narrowband Doppler-drifting signals, while still detecting many such signals. Our pipeline is also less likely than \turboseti to find ``candidate'' events that clearly contain persistent signals in both ON and OFF scans on visual inspection of spectrograms.

A final limitation of our approach is the strong dependence on the size of the frequency block analyzed. A large frequency block increases the ability of detecting a rapidly-drifting signal in multiple ON scans of a source. However, as the frequency block grows too large two issues become more severe. Weak signals in wide frequency blocks may get drowned out by the noise in the rest of the observation, and hence subsequent kurtosis values will fall below our detection threshold. Additionally, a wider frequency block increases the chance of RFI, which would lead to a high kurtosis in the OFF scans. This could also lead to the frequency block getting rejected, even if it contains a genuine candidate signal. 

While we use kurtosis as a metric, there are multiple other statistical measures that can be used to evaluate whether a spectrogram differs significantly from Gaussian noise. As one example, the Kullback–Leibler divergence could be calculated instead to determine the deviation of a frequency block's intensity distribution from the background noise distribution. We leave the implementation of these other approaches to future work.

\section{Conclusion}

As part of the Breakthrough Listen program, we searched 2,623 stars, using data from the GBT. We used four receivers, spanning a combined range of 1.10–11.2 GHz, and found no compelling candidates that are
not attributable to RFI. We present and implement a novel pipeline, designed to search for signals present in the ON scans of observation cadences and not in the OFF scans. This is a search method less biased towards any specific signal type than many previous SETI searches. We analyze 6,630 unique cadences, representing 3,165 hours of on-sky observation time. We identify $14,168$ frequency blocks 2.86\,kHz wide deemed likely to contain a candidate signal. We visually inspected all of these candidates, and determined that all can likely be attributed to RFI. This work highlights the wide variety of RFI that hinders radio SETI searches, and we hope the techniques developed may prove useful in future studies dealing with increasingly heavy RFI environments.  

This study constitutes the most comprehensive survey yet for radio evidence of advanced life around nearby stars. We improve on the results of \cite{Price:2020} in terms of sensitivity and number of stars. Our novel search approach also makes us sensitive to a larger range of signals than previous surveys. The Drake Figure of Merit of our search is $4.8 \times 10^{8}$, the Continuous Waveform Transmittter Rate Figures of Merit of our bands fall between $0.45-0.85$, and we probe down to EIRPs of $2\times10^{12}$\,W for a large numbers of stars (several hundred stars at each band). With respect to each of these metrics, this SETI survey is one of the most sensitive and thorough to date. We add our pipeline to the many algorithms that have been developed to detect technosignatures in radio observations, and hope that this study encourages ensuing SETI surveys to take a more general approach to recovering potential ET signals from complex datasets. 

\section{Acknowledgements}

 The Breakthrough Prize Foundation funds the Breakthrough Initiatives, which manages Breakthrough Listen. The Green Bank Observatory facility is supported by the National Science Foundation, and is operated by Associated Universities, Inc., under a cooperative agreement. CP was funded as a participant in the Berkeley SETI Research Center Research Experience for Undergraduates Site, supported by the National Science Foundation under Grant No.~2244242. The authors thank Joe Bright, Noah Franz, Chenoa Tremblay, and Andrés Luengo for helpful discussions. The authors also thank the anonymous referee and the AAS Journals Statistics Editor for helpful comments which improved the paper.

\newpage

\clearpage
\nocite{Phoenix:2004, Tremblay:2020, Horowitz:1993, Harp:2016, Siemion:2013}

\bibliography{main.bib}{}

\appendix
\section{Pointing Direction of Level 2 Blocks and Candidates} \label{app:A}

Here we plot the RA and declination as well as altitude and azimuth of the sources associated with our final candidates, as well as those associated with our Level 2 Blocks. There appears to be an overdensity of sources located at an azimuth of $210\degr \pm 35\degr$ in L band. This could correspond to a large cluster of geosynchronous satellites as viewed from the GBT. Notably, most of our final candidates come from this cluster. 

\begin{figure}[h]
\centering
    \includegraphics[width=.54\textwidth]{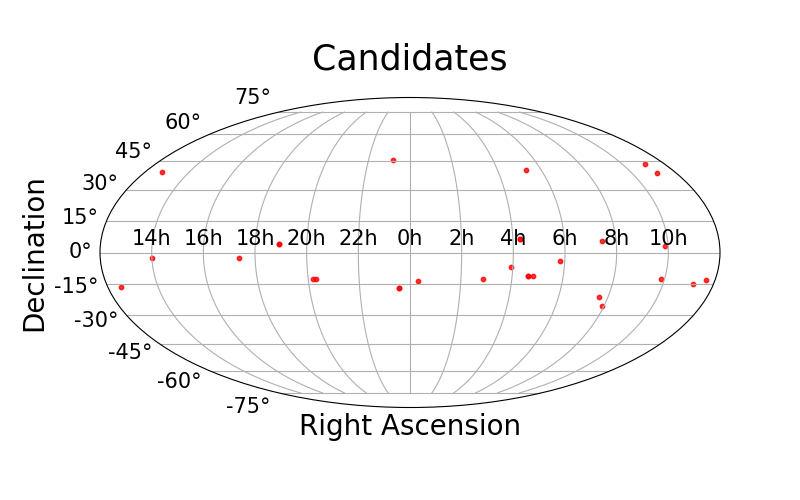}
    \includegraphics[width=.43\textwidth]{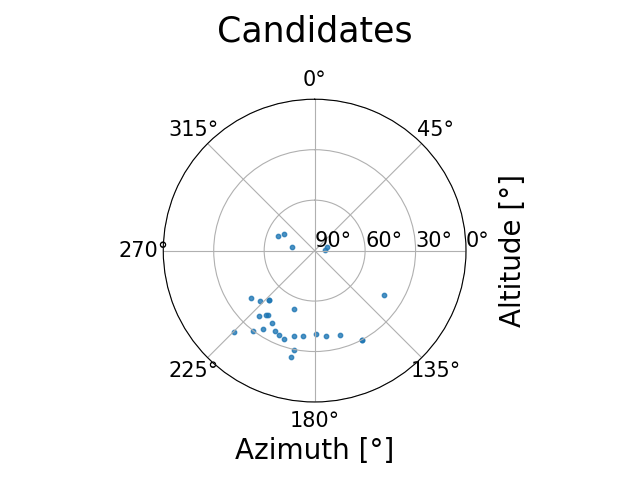}

    \caption{Telescope pointings of the sources containing our candidate signals.}
    \label{fig:candidate_pointings}
\end{figure}

\begin{figure}[h]
\centering
    \includegraphics[width=.54\textwidth]{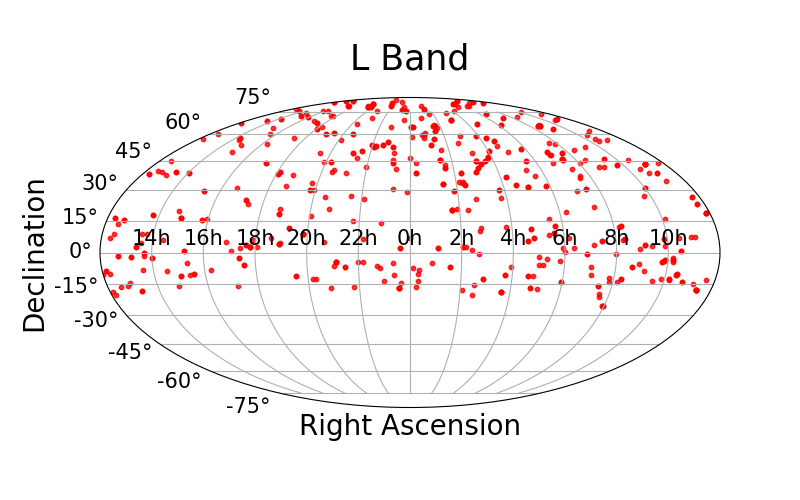}
    \includegraphics[width=.43\textwidth]{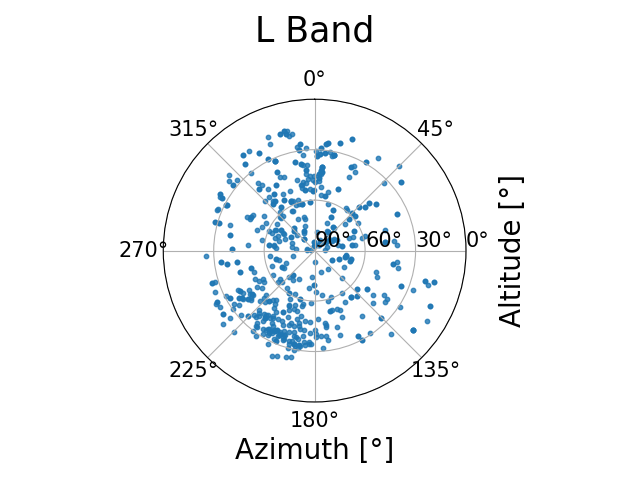}

    \caption{Telescope pointings of the sources containing our Level 2 Blocks in L Band.}
    \label{fig:candidate_pointings}
\end{figure}

\begin{figure}[h]
\centering
    \includegraphics[width=.54\textwidth]{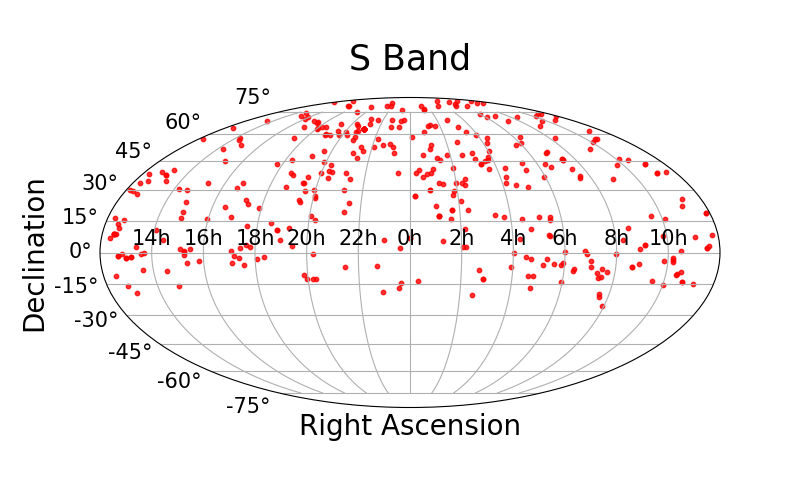}
    \includegraphics[width=.43\textwidth]{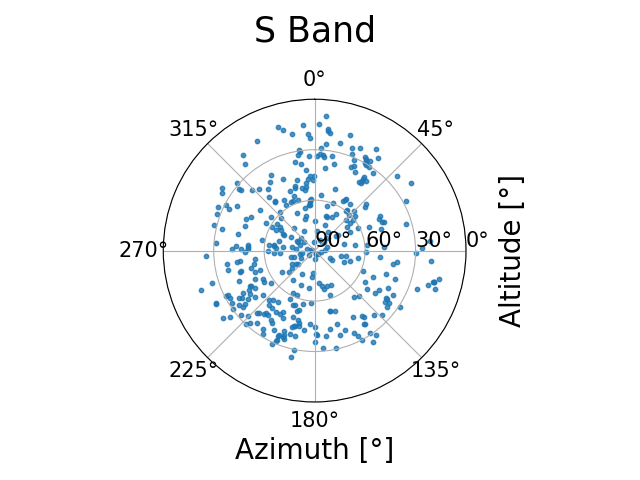}

    \caption{Telescope pointings of the sources containing our Level 2 Blocks in S Band.}
    \label{fig:candidate_pointings}
\end{figure}

\begin{figure}[h]
\centering
    \includegraphics[width=.54\textwidth]{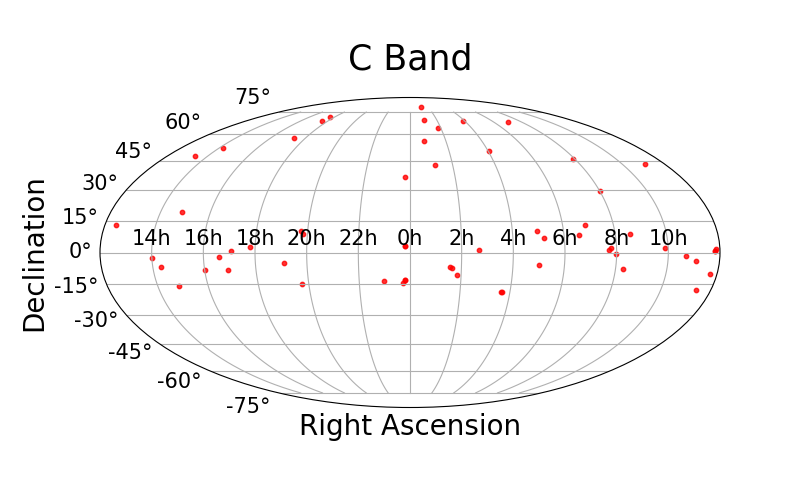}
    \includegraphics[width=.43\textwidth]{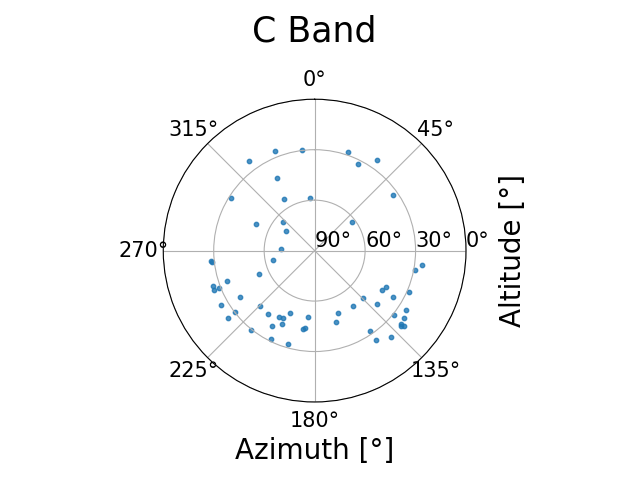}

    \caption{Telescope pointings of the sources containing our Level 2 Blocks in C Band.}
    \label{fig:candidate_pointings}
\end{figure}

\begin{figure}[h]
\centering
    \includegraphics[width=.54\textwidth]{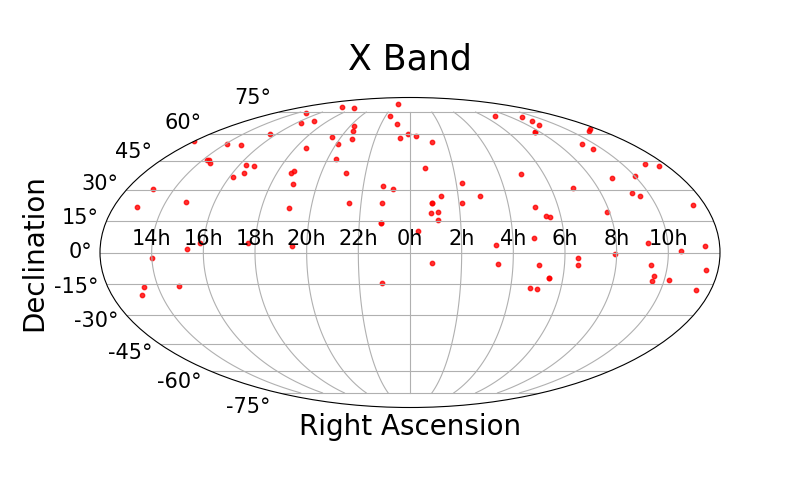}
    \includegraphics[width=.43\textwidth]{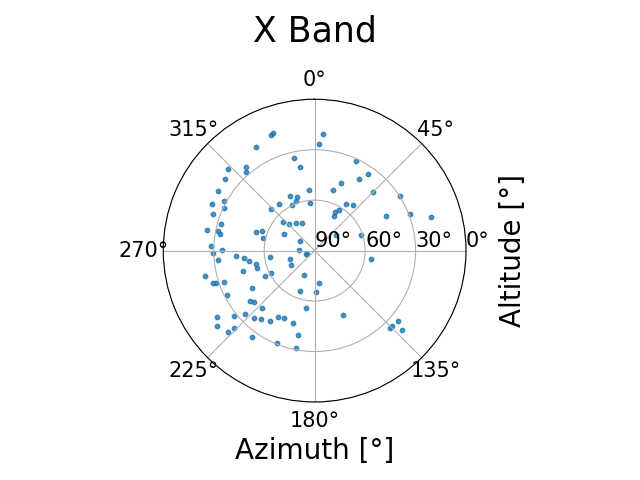}

    \caption{Telescope pointings of the sources containing our Level 2 Blocks in X Band.}
    \label{fig:candidate_pointings}
\end{figure}

\clearpage

\section{Quality Candidates} \label{app:B}

\begin{figure}[!hbtp]
\centering
    \includegraphics[width=.32\textwidth]{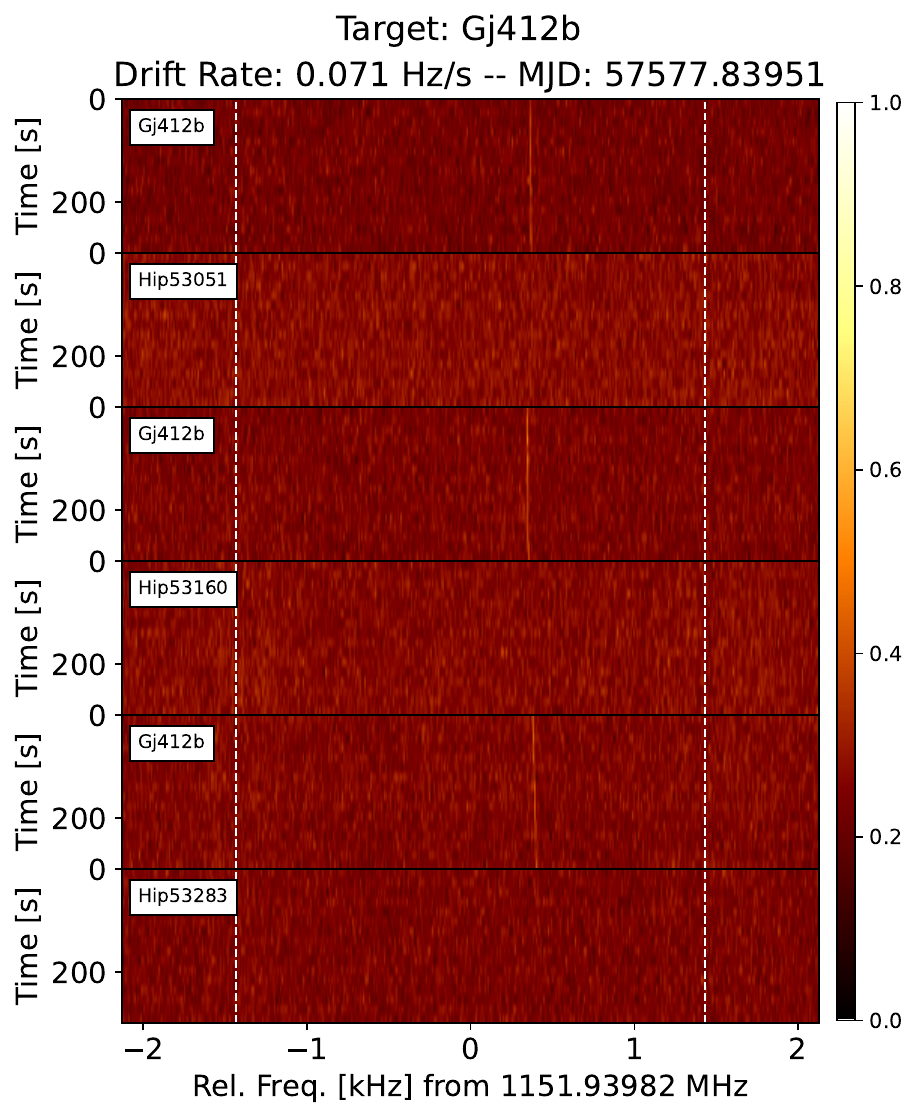}
    \includegraphics[width=.32\textwidth]{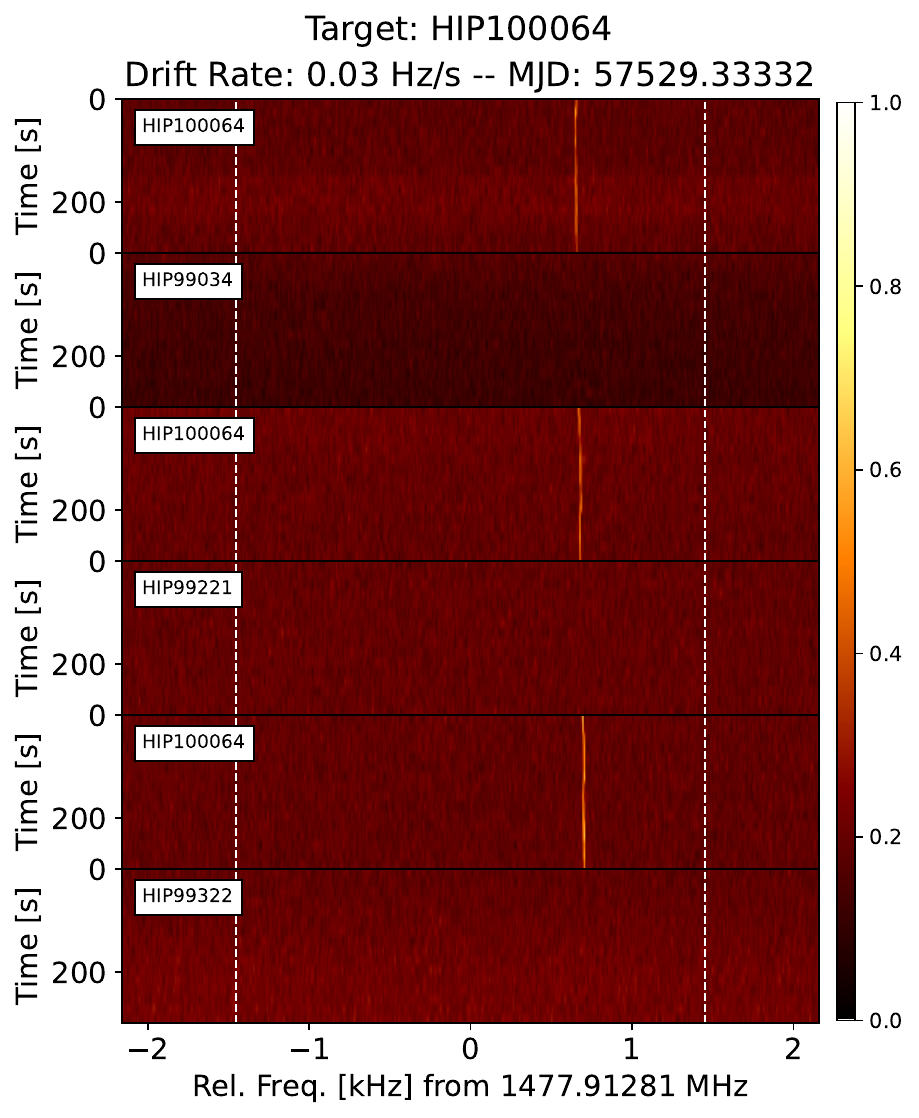}
    \includegraphics[width=.32\textwidth]{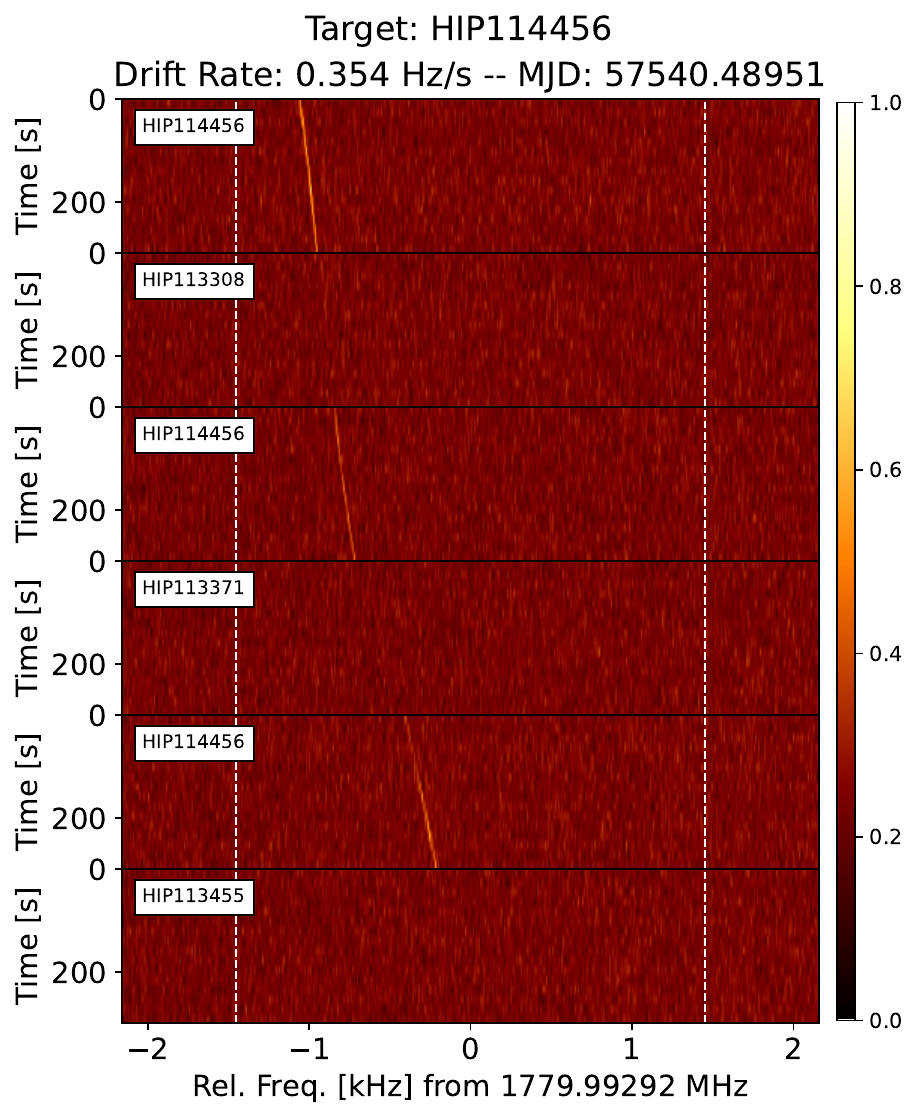}
        
    \includegraphics[width=.32\textwidth]{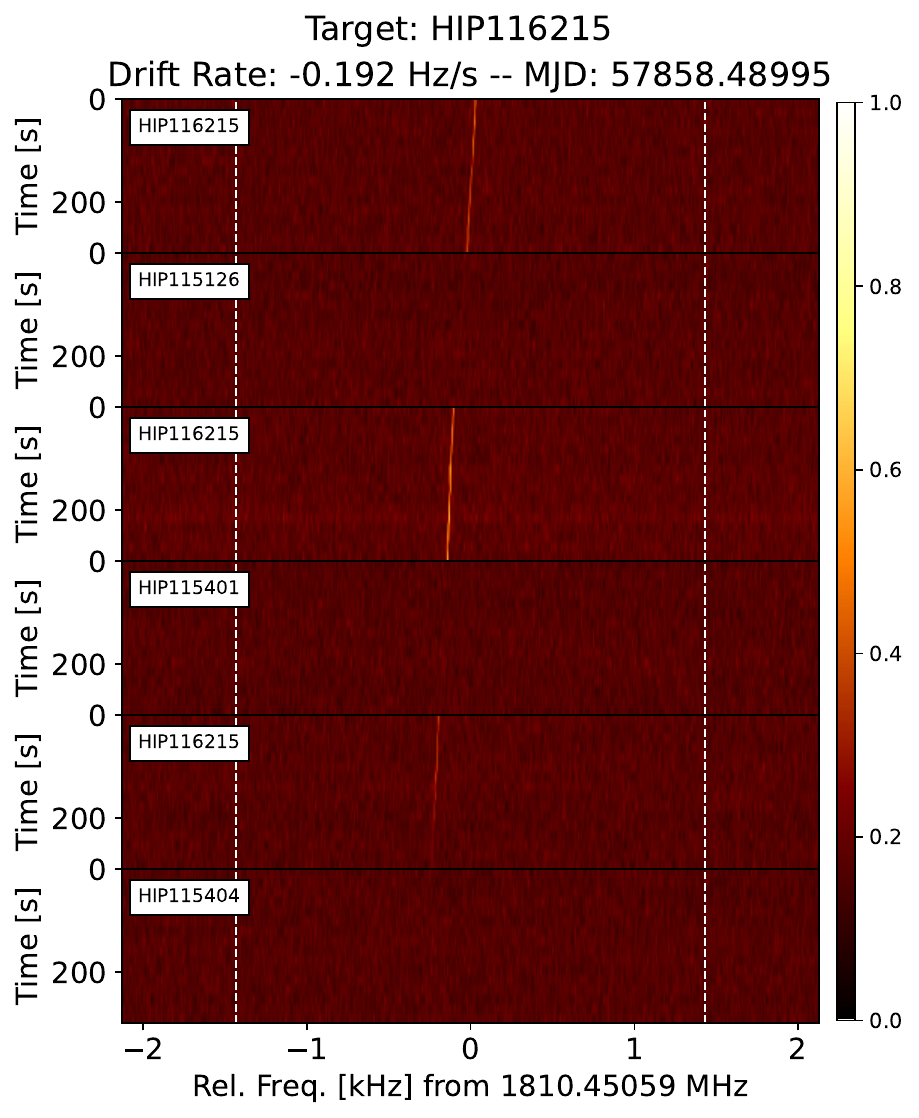}
    \includegraphics[width=.32\textwidth]{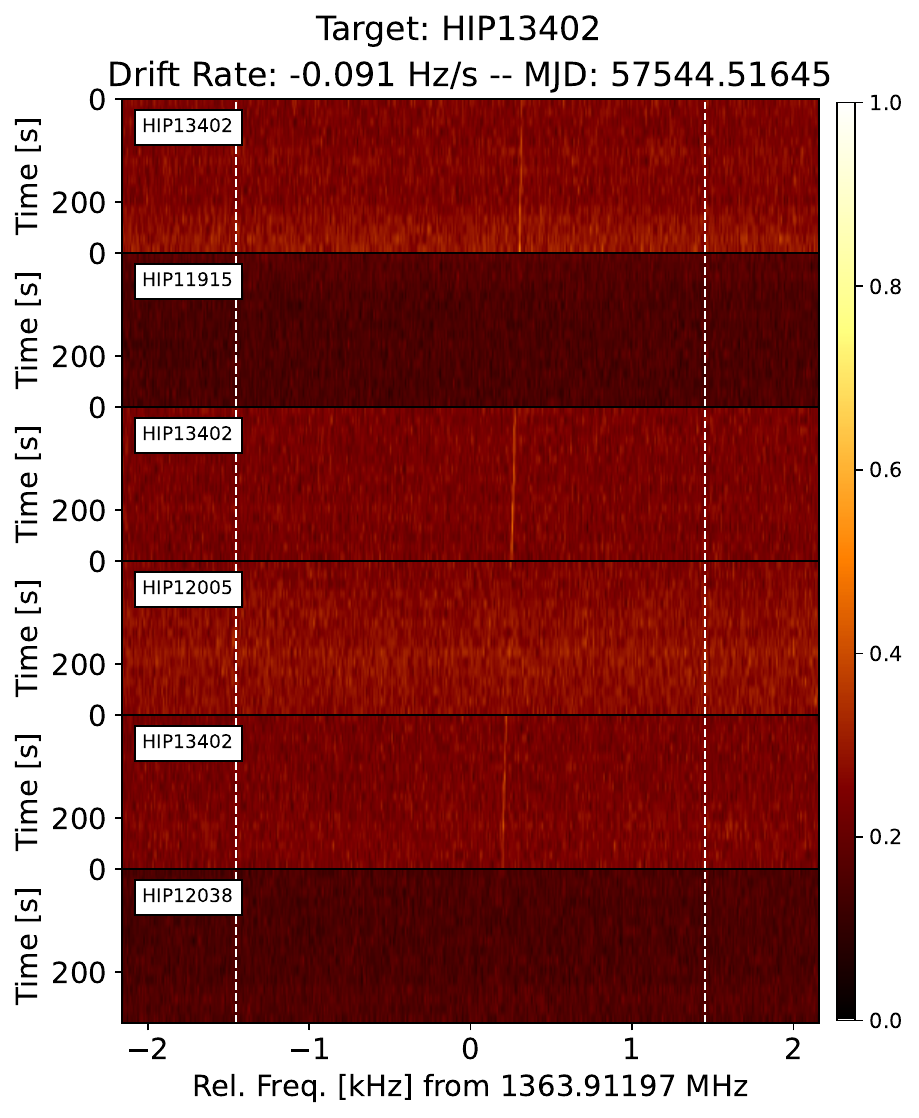} 
    \includegraphics[width=.32\textwidth]{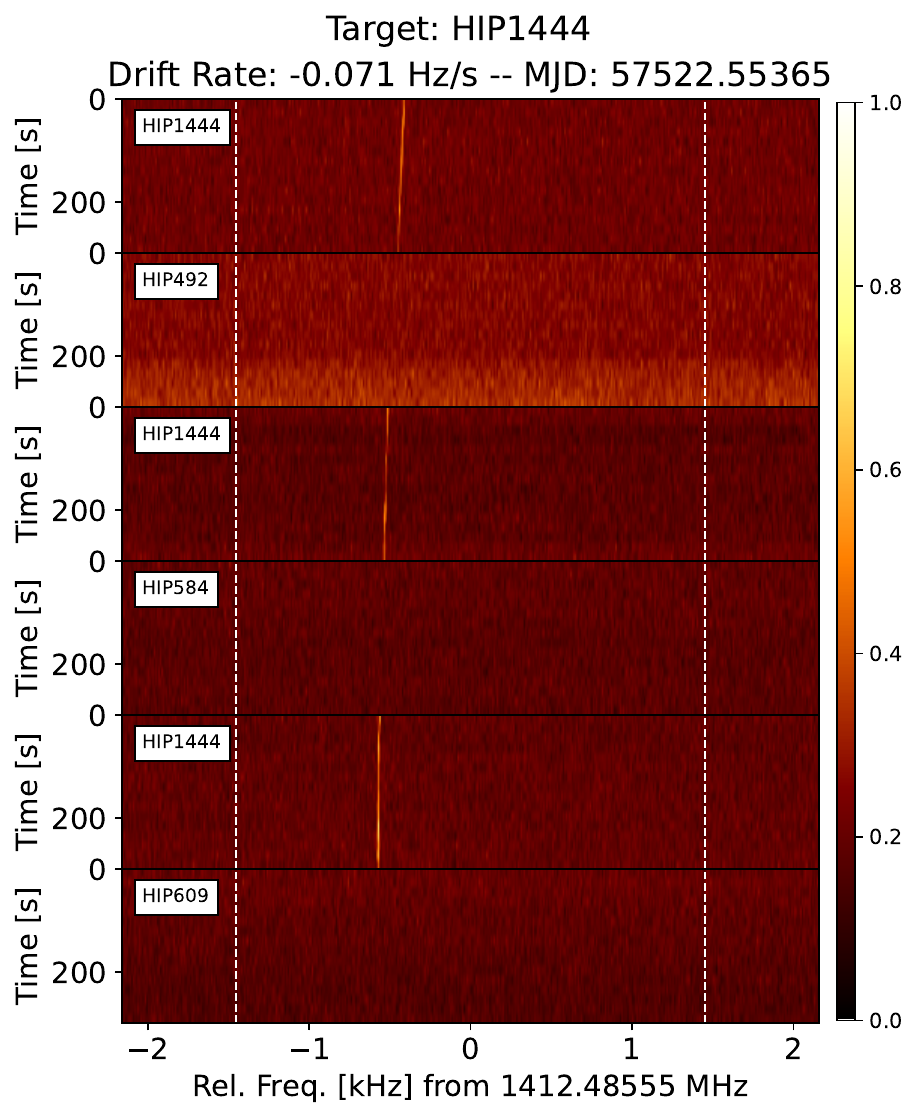}
    
    \includegraphics[width=.32\textwidth]{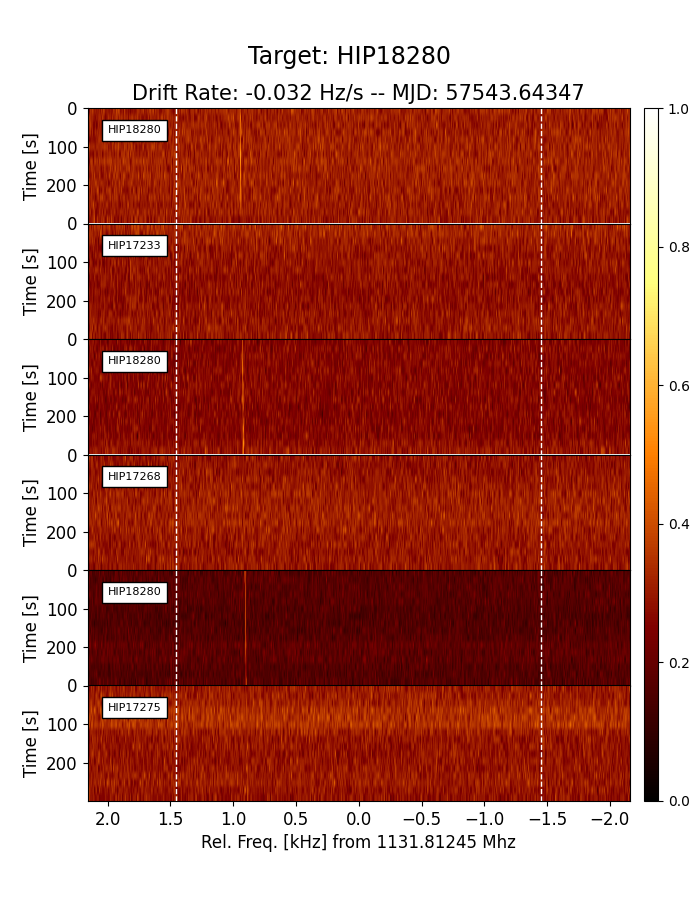}
    \includegraphics[width=.32\textwidth]{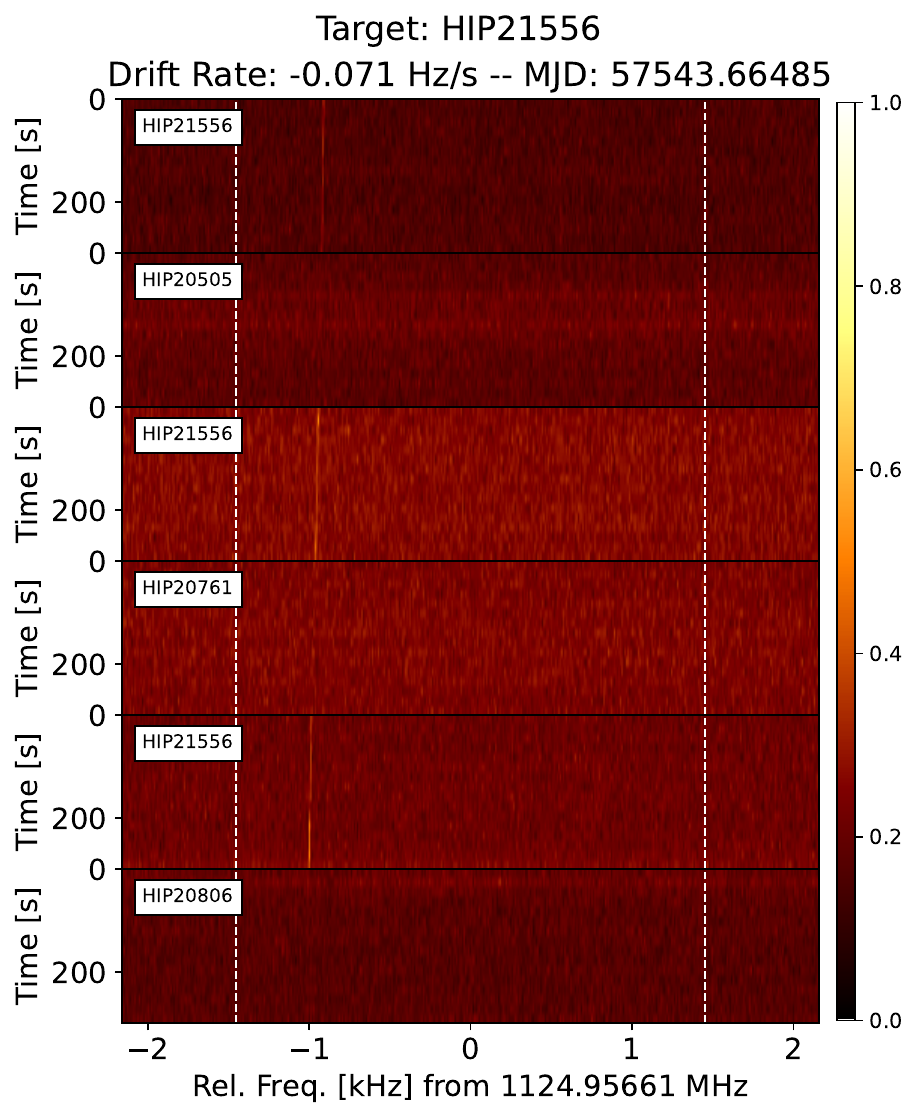}
    \includegraphics[width=.32\textwidth]{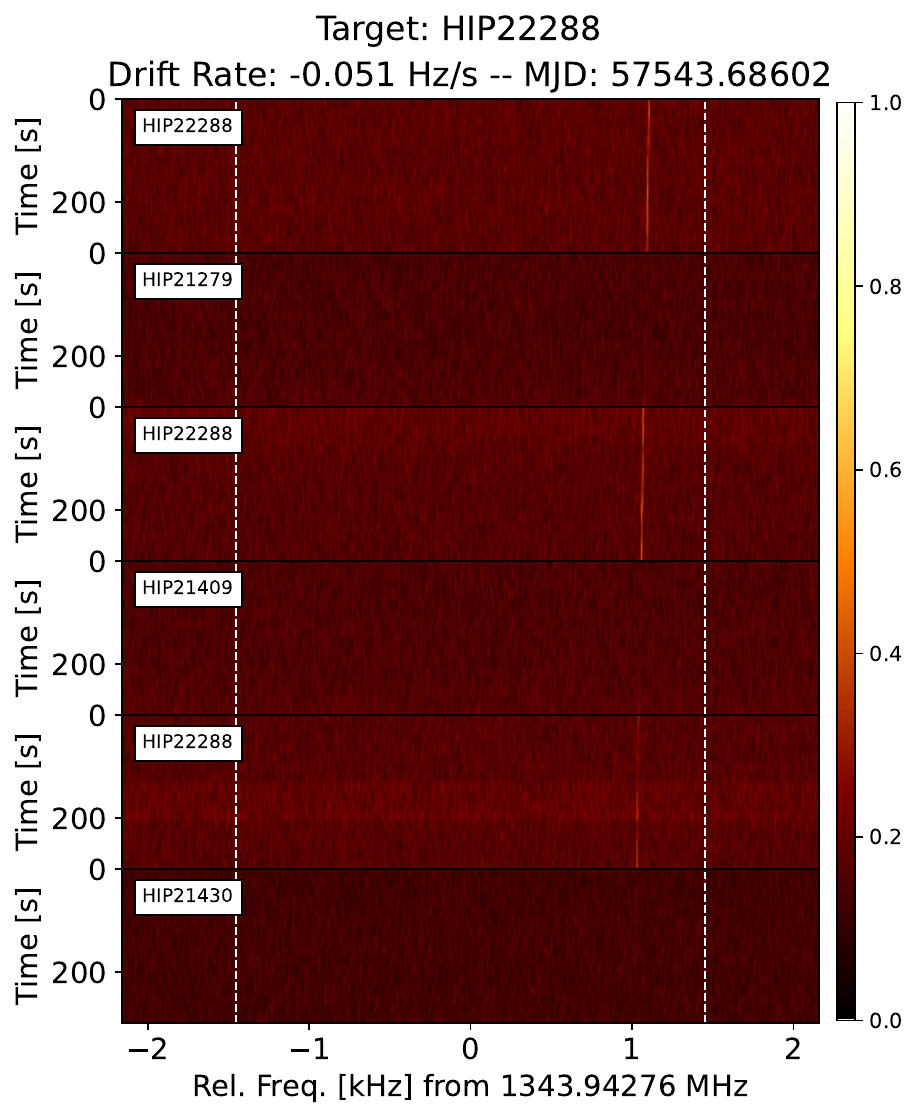}

    \caption{Examples of Level 2 blocks picked up by our pipeline.}
    \label{fig:candidates}
\end{figure}
\vspace{-2cm}

\begin{figure*}
\centering
    \includegraphics[width=.32\textwidth]{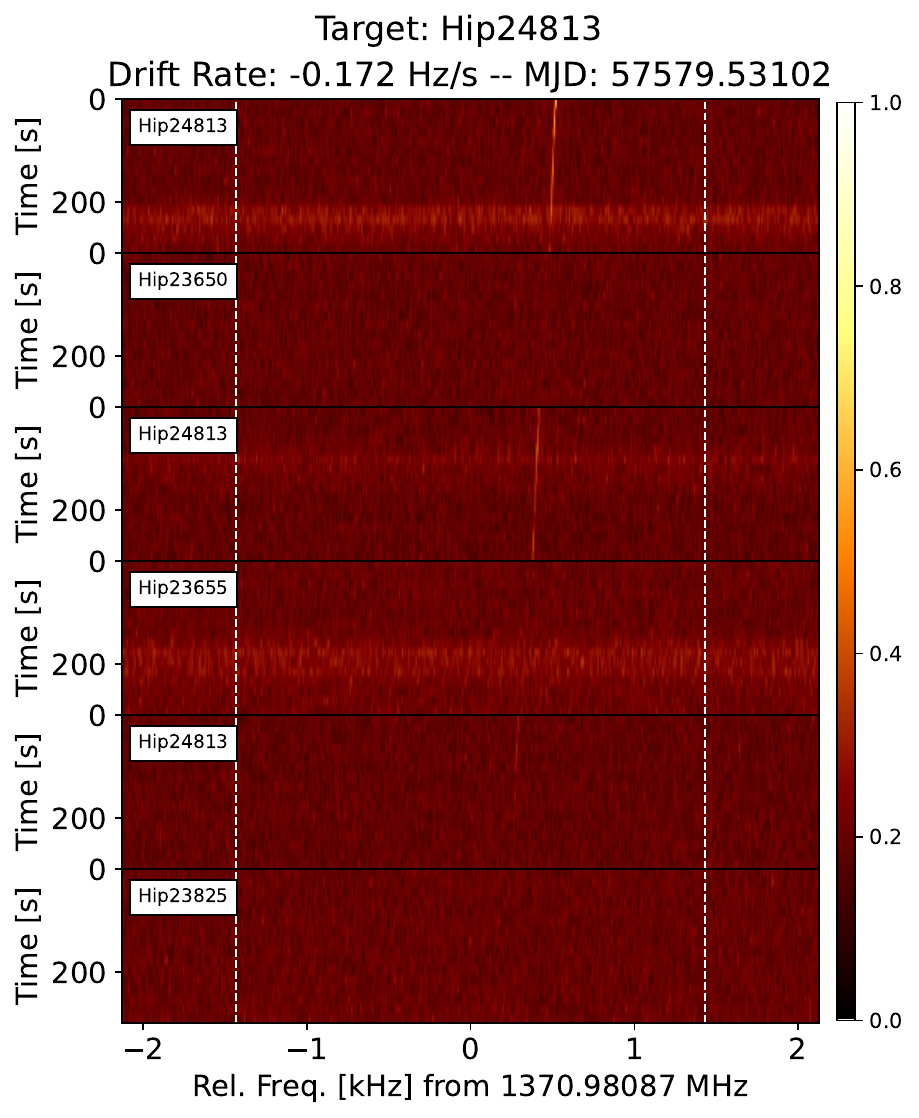}
    \includegraphics[width=.32\textwidth]{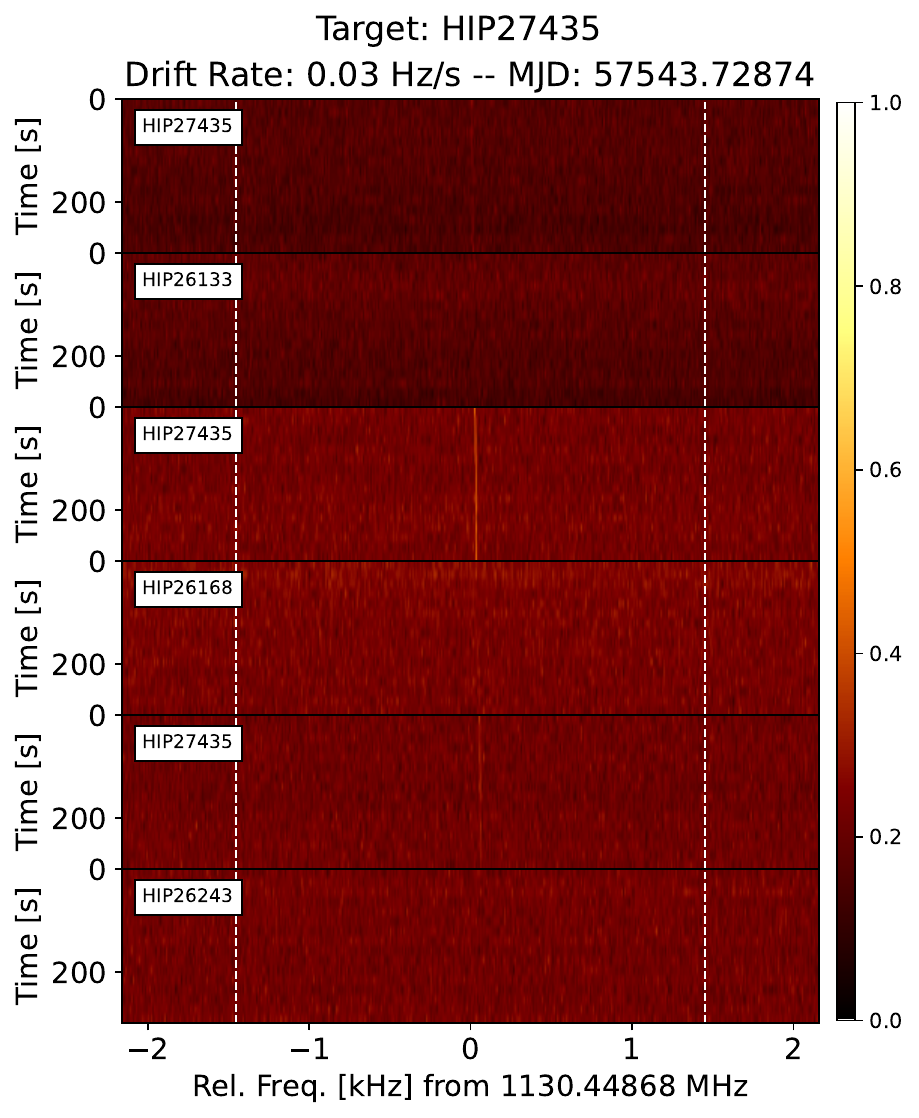}
    \includegraphics[width=.32\textwidth]{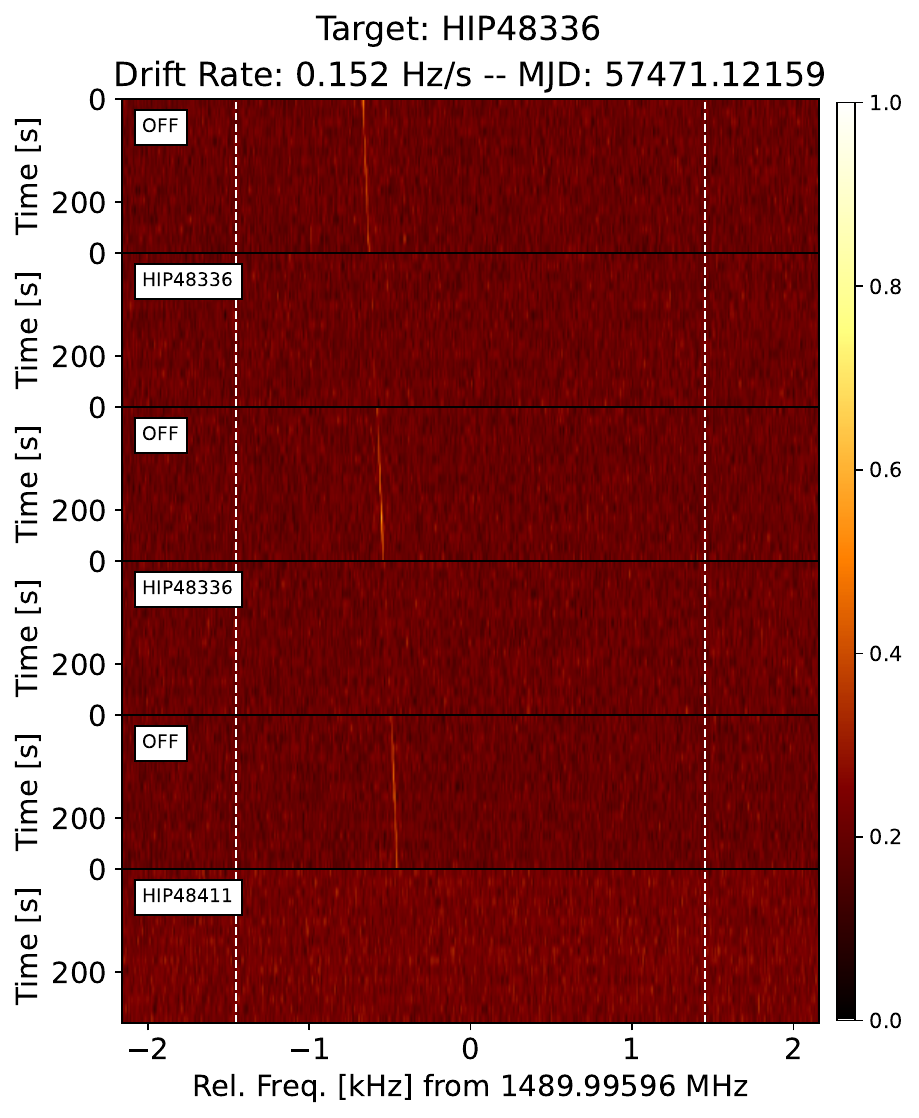}

    \includegraphics[width=.32\textwidth]{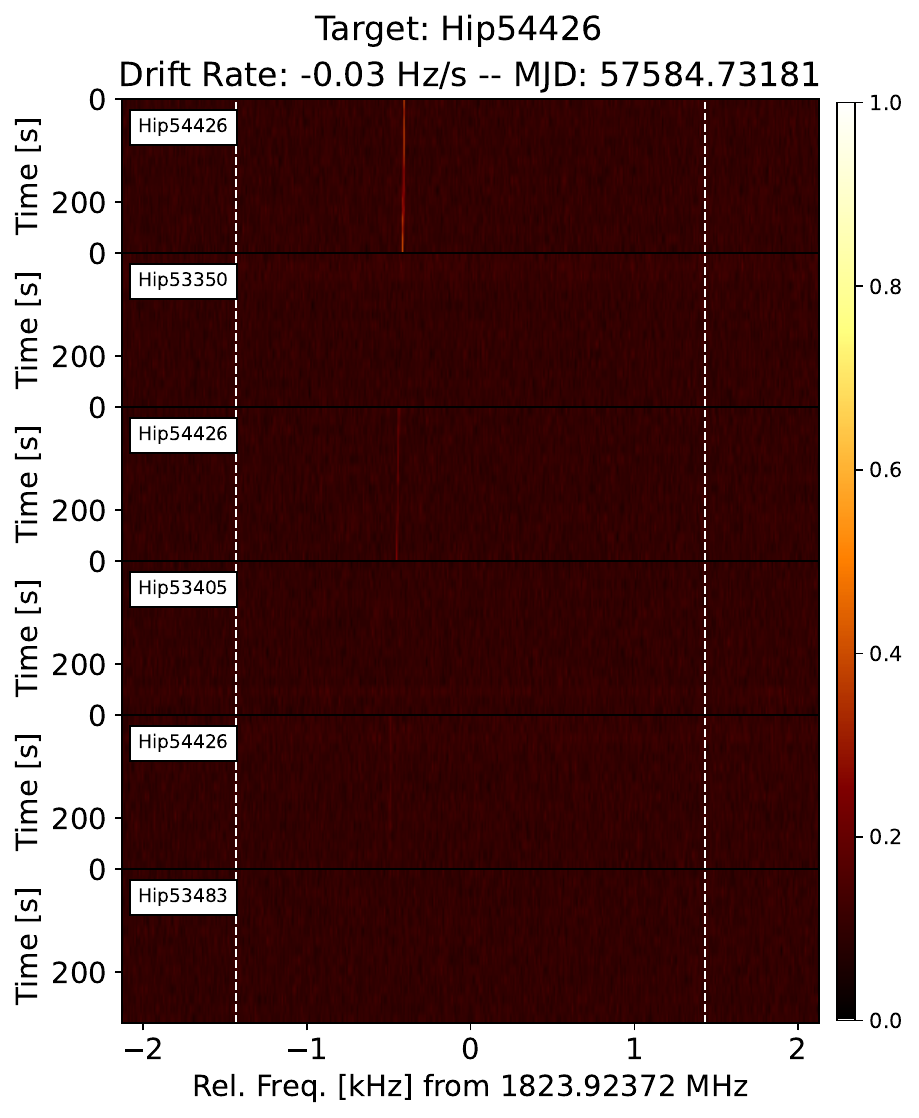}
    \includegraphics[width=.32\textwidth]{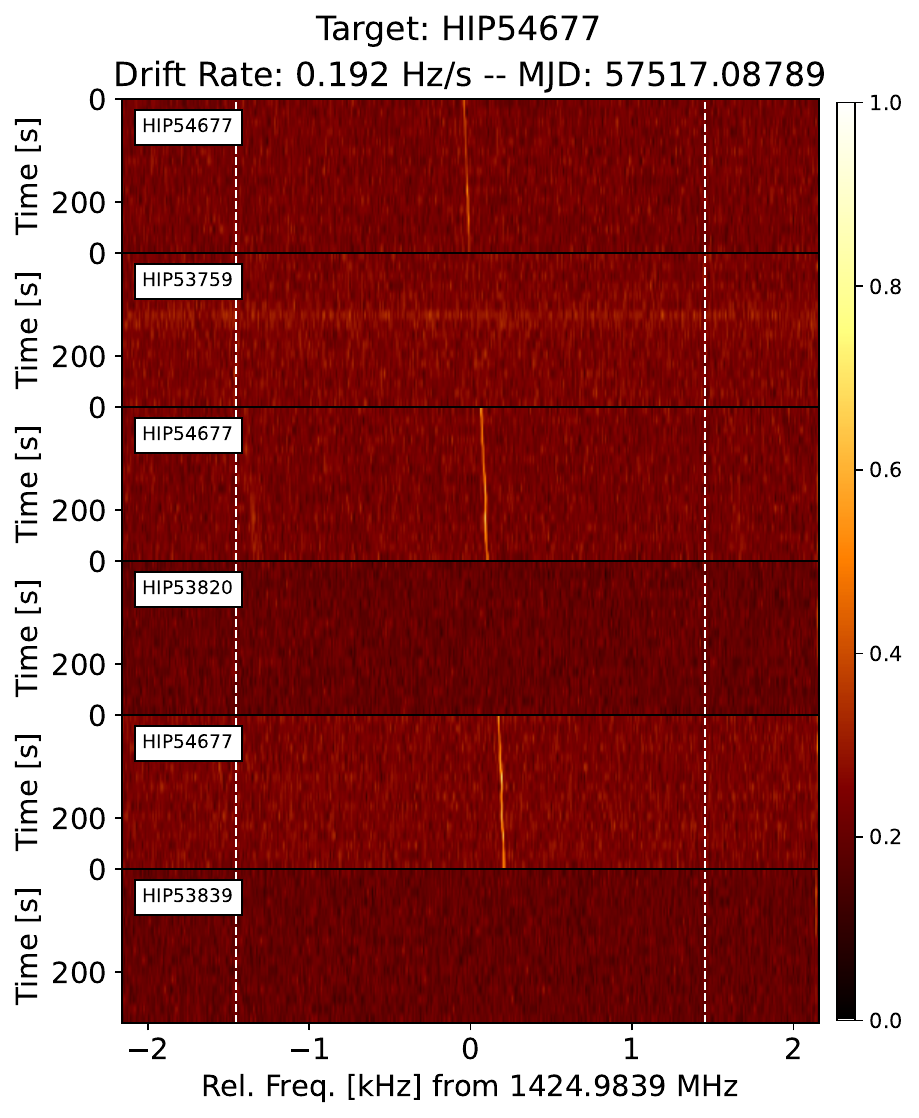} 
    \includegraphics[width=.32\textwidth]{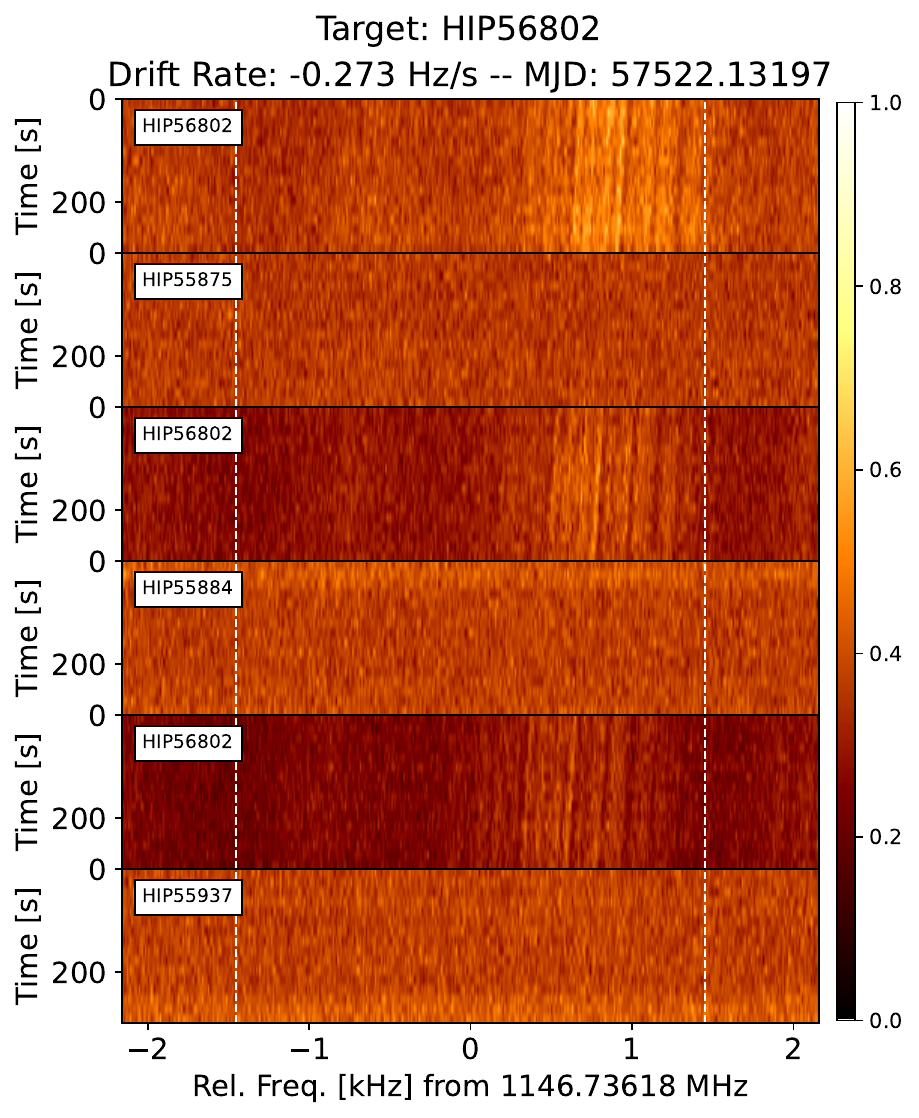}
    
    \includegraphics[width=.32\textwidth]{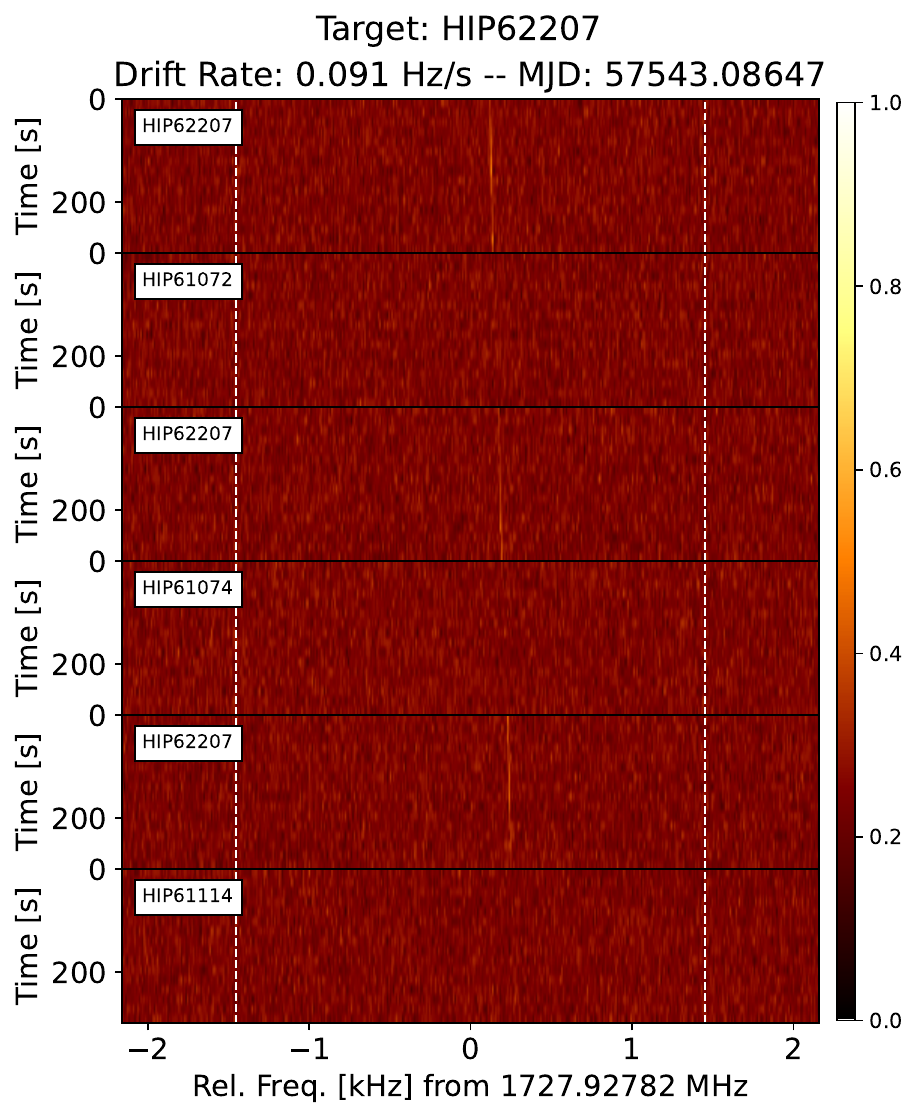}
    \includegraphics[width=.32\textwidth]{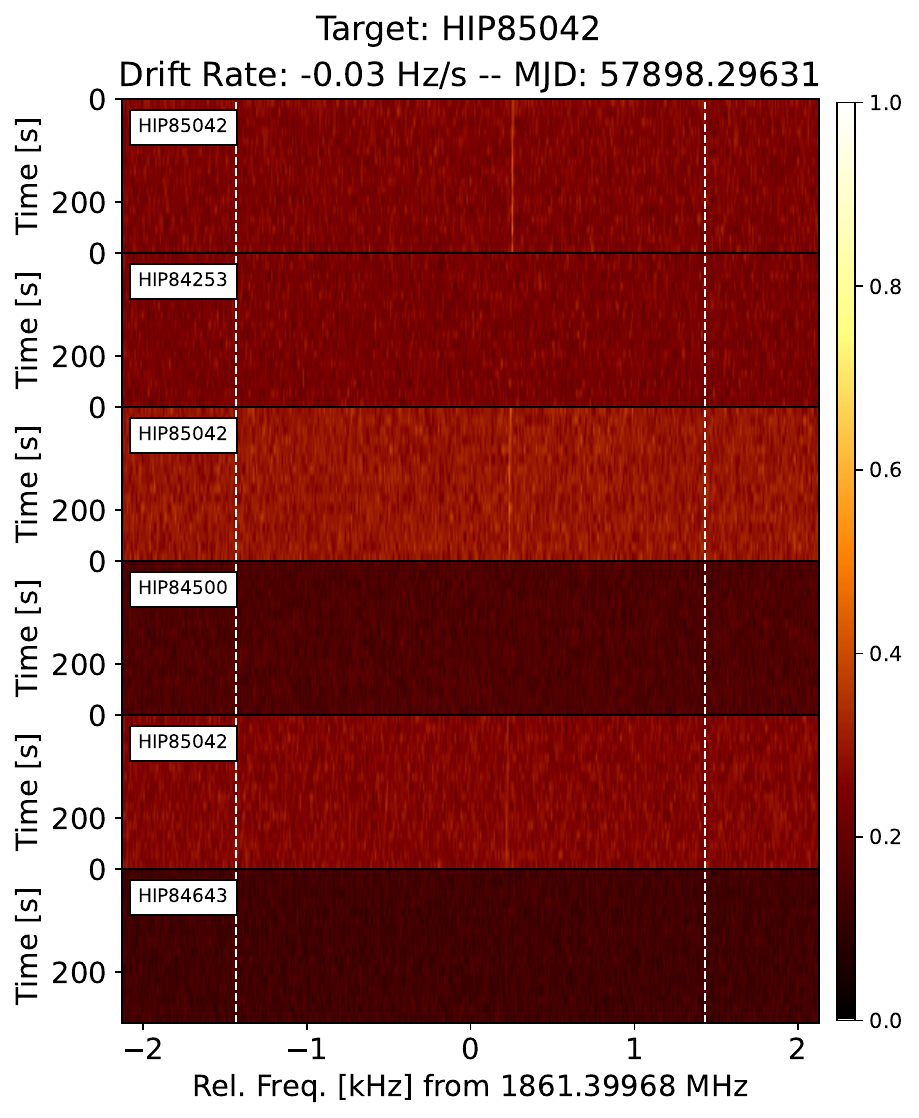}
    \includegraphics[width=.32\textwidth]{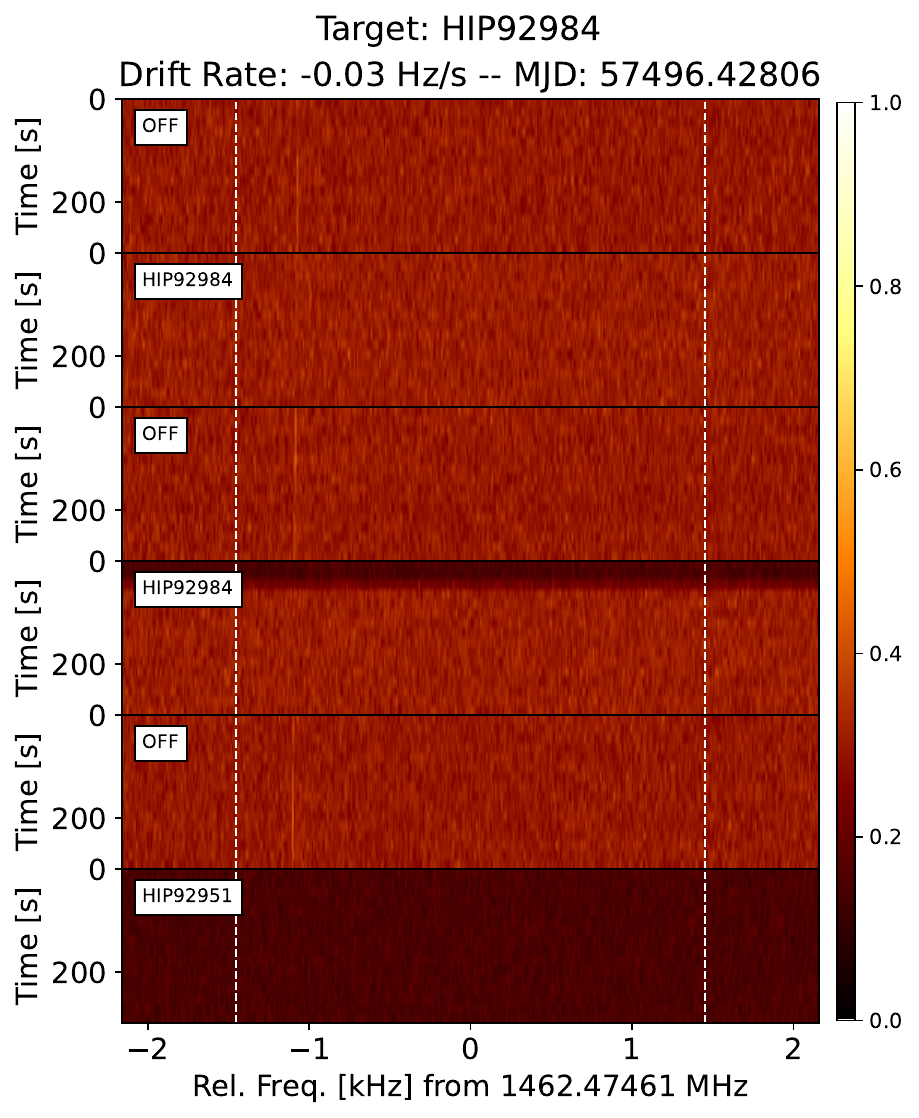}

    \caption{Examples of Level 2 blocks picked up by our pipeline.}
    \label{fig:candidates2}
\end{figure*}

\end{document}